\documentclass[aps,prd,reprint,preprintnumbers,showpacs,floatfix,nofootinbib,superscript address]{revtex4-2}
\usepackage[utf8]{inputenc}
\usepackage{parskip}
\usepackage{amssymb}
\usepackage{stix}
\usepackage{hhline}
\usepackage{xspace}
\usepackage{amsmath}
\usepackage{mathtools}
\usepackage{lipsum}
\usepackage[dvipsnames]{xcolor}
\usepackage{xspace}
\usepackage{multirow,tabularx}
\usepackage{siunitx}
\usepackage{textgreek}
\usepackage{upgreek}
\usepackage{multirow}
\usepackage{graphicx}
\usepackage{xstring}
\usepackage{etoolbox}
\usepackage{notoccite}
\usepackage{natbib}
\usepackage{lineno}
\usepackage{tensor}
\usepackage[thinlines,thicklines]{easybmat}
\usepackage{tikz}
\usepackage{bm}

\newrobustcmd{\planck}{%
  {m_{\text{p}}}%
}
\newrobustcmd{\Planck}{%
  {M_{\text{Pl}}}%
}

\DeclareSIUnit\parsec{pc} 
\DeclareSIUnit\littleh{\mathsf{h}} 
\DeclareSIUnit\ccs{{m_{\text{p}}}^2} 
\DeclareSIUnit\nothing{\relax} 

\newcommand{\og}{\bar{g}} 
\newcommand{\feynman}{\tensor{D}{_{\text{F}}}} 

\input{Macros/will-1.tex}
\usepackage{graphicx}
\usepackage{dcolumn}
\usepackage{bm}

\usepackage{amsmath}

\usepackage{amssymb}

\usepackage{pgfplots}
\pgfplotsset{compat=newest}

\usepgfplotslibrary{external}
\newcommand{\barr}{\begin{array}}
\newcommand{\earr}{\end{array}}
\newcommand{\beqa}{\be \begin{array}{rcl}}
\newcommand{\eeqa}{\end{array} \ee}

\newcommand{\lam}{\lambda}

\newcommand{\gi}{\gamma_{1}}
\newcommand{\gj}{\gamma_{2}}
\newcommand{\gk}{\gamma_{3}}
\newcommand{\go}{\gamma_{0}}

\newcommand{\grad}{\nabla}

\newcommand{\bgrad}{\mbox{\boldmath $\grad$}}

\newcommand{\deriv}[2]{\frac{\partial #1}{\partial #2}}

\newcommand{\kms}{{\rm\,km\,s^{-1}}}

\newcommand{\sfh}{{\sf h}}

\newcommand{\hob}{\bar{\sfh}}

\newcommand{\msun}{M_{\odot}}

\newcommand{\kpc}{{\rm\,kpc}}

\newcounter{bean}
{\begin{list}%
{(\roman{bean})}{\topsep 0in \usecounter{bean}}}%
{\end{list}}

\graphicspath{{figures/}{extra_figures_for_notes/}}

\usepackage{hyperref}
\hypersetup{%
     colorlinks = true,%
     linkcolor = Blue,%
     citecolor = Blue,%
     filecolor = Blue,%
     urlcolor = Blue%
     }%
\usepackage[capitalize]{cleveref} 
\allowdisplaybreaks

\begin{document}

\title{Does gravitational confinement sustain flat galactic rotation curves without dark matter?}
\author{W.E.V. Barker}
\email{wb263@cam.ac.uk}
\affiliation{Astrophysics Group, Cavendish Laboratory, JJ Thomson Avenue, Cambridge CB3 0HE, UK}
\affiliation{Kavli Institute for Cosmology, Madingley Road, Cambridge CB3 0HA, UK}
\author{M.P. Hobson}
\email{mph@mrao.cam.ac.uk}
\affiliation{Astrophysics Group, Cavendish Laboratory, JJ Thomson Avenue, Cambridge CB3 0HE, UK}
\author{A.N. Lasenby}
\email{a.n.lasenby@mrao.cam.ac.uk}
\affiliation{Astrophysics Group, Cavendish Laboratory, JJ Thomson Avenue, Cambridge CB3 0HE, UK}
\affiliation{Kavli Institute for Cosmology, Madingley Road, Cambridge CB3 0HA, UK}

\begin{abstract}
The short answer is \emph{probably no}. Specifically, this paper considers a recent body of work which suggests that general relativity requires neither the support of dark matter halos, nor unconventional baryonic profiles, nor any infrared modification, to be consistent after all with the anomalously rapid orbits observed in many galactic discs. In particular, the gravitoelectric flux is alleged to collapse nonlinearly into regions of enhanced force, in an analogue of the colour-confining chromoelectric flux tube model which has yet to be captured by conventional post-Newtonian methods. However, we show that the scalar gravity model underpinning this proposal is wholly inconsistent with the nonlinear Einstein equations, which themselves appear to prohibit the linear confinement-type potentials which could indicate a disordered gravitational phase. Our findings challenge the fidelity of the previous Euclidean lattice analyses: we propose that the question of confinement demands a gauge-invariant lattice implementation. We confirm by direct calculation using a number of perturbation schemes and gauges that the next-to-leading order gravitoelectric correction to the rotation curve of a reasonable baryonic profile would, in fact, be imperceptible. The `gravitoelectric flux collapse' programme was also supported by using intragalactic lensing near a specific galactic baryon profile as a field strength heuristic. We recalculate this lensing effect, and conclude that it has been overstated by three orders of magnitude. As a by-product, our analysis suggests fresh approaches to (i) the fluid ball conjecture and (ii) gravitational energy localisation, both to be pursued in future work. \emph{In summary, whilst it may be interesting to consider the possibility of confinement-type effects in gravity, such an investigation should be done thoroughly, without relying on heuristics: that task is neither attempted in this work nor accomplished by the key works referenced. Pending such analysis, we may at least conclude here that confinement-type effects cannot play any significant part in explaining flat or rising galactic rotation curves without paradigmatic dark matter halos.}
\end{abstract}


\maketitle

\section{Introduction}\label{introduction}

A substantial body of work has recently
accumulated~\cite{Deur:2009ya,Deur:2013baa,Deur:2014kya,Deur:2016bwq,Deur:2017aas,Deur:2019kqi,Deur:2020wlg,Deur:2021ink,Deur:2022ooc},
asserting that overlooked effects -- otherwise native to quantum chromodynamics
-- are nonlinearly implied by general relativity, and are in fact
manifest among the observed astrophysical and cosmological phenomena.  It is
proposed that non-Abelian graviton-graviton interactions at next-to-leading
order can qualitatively reshape (yet still be adequately described by) the
weak-field regime.  The effect is best seen in the dominant
\emph{gravitoelectrostatic} portion of the weak field, i.e. the familiar Newtonian part sourced by
static mass-energy in a manner analogous to electrical charge in
Maxwell's theory.  If the gravitoelectric flux lines collapse under their
own `weight', regions of appreciably rarefied and enhanced force will appear. 
In what follows, we will use the broad term `\emph{gravitoelectric flux collapse}' (GEFC) to refer to this effect, and to flag the associated literature (e.g. `\gefc{Deur:2009ya}', etc.)\footnote{We stress that \emph{GEFC} is merely a convenient label, and not intended to fairly capture all the effects proposed in~\gefc{Deur:2009ya,Deur:2013baa,Deur:2014kya,Deur:2016bwq,Deur:2017aas,Deur:2019kqi,Deur:2020wlg,Deur:2021ink,Deur:2022ooc}, which might equally well be termed `\emph{overlooked nonlinearity}' (ON), etc.}.

That there may well be a GEFC effect follows reasonably from our current understanding of the strong force~\cite{confinement}. In stark contrast to the photons and leptons of quantum electrodynamics (QED), the gluons and quarks of quantum chromodynamics (QCD) do not appear in the mass spectrum as asymptotic states: these species are believed instead to be \emph{confined}.
The precise mechanism of colour confinement is not yet established~\cite{Roberts:2012sv}, but we can summarise certain principles which are thought to be involved.
The effect is well understood to be tied to the \emph{non-Abelian} nature of the QCD gauge group $\suthreec$, a quality which is not shared by the QED counterpart $\uoneem$.  Gluons -- like photons -- are massless, but the potential between a quark-antiquark pair is only Coulombic at short (i.e. asymptotically-free) distances~\cite{PhysRevLett.30.1343,PhysRevLett.30.1346}. At intermediate distances the potential rises \emph{linearly}, before flattening off due to the energetically favourable pair-production of light quarks in a process called \emph{string-breaking}~\cite{Philipsen:1998de,2000NuPhS..83..310D}. In classical terms, the intermediate regime is associated with the collapse of the Coulombic quark-to-antiquark field lines into a \emph{chromoelectric flux tube} of nearly constant cross-section, which exerts a $\sim\SI{}{\giga\electronvolt\per\femto\meter}$ restoring force aginst orbital angular momentum in favour of a confined meson~\cite{Bali:2000gf}. 
That confinement is a non-Abelian phenomenon may be recognised at various levels. Photons are neutral, but gluons carry the colour charge and so the chromoelectric field lines are \emph{themselves} subject to the strong force. Classically, the structure constants introduce nonlinearities into the strong force $\frac{1}{2}\mathrm{tr} \tensor{G}{_{\mu\nu}}\tensor{G}{^{\mu\nu}}$ Maxwell term, which are absent in the electromagnetic counterpart $\frac{1}{4}\tensor{F}{_{\mu\nu}}\tensor{F}{^{\mu\nu}}$. The effects of these nonlinearities are most conveniently probed on a lattice, where gauge invariant Wilson loops are indeed found to be suppressed by an area law in non-Abelian gauge theories~\cite{Shibata:2017ixa,Greensite:2003bk}. This area law indicates a thermally preferred phase of magnetically disordered gauge links or plaquettes: infinitely heavy probes, whose tips are charged under the gauge group and inserted into a disordered phase, would attract each other with a constant, \emph{confining} force~\cite{confinement,Rothkopf:2009pk}.

Does gravity exist in a state of magnetic disorder? In contrast to the asymptotic freedom of QCD, Coulombic (i.e. \emph{Newtonian}) gravity appears to dominate at \emph{long} distances.
As the length-scale associated with the curvature decreases, nonlinearities certainly emerge, as evidenced by the precession observed in Mercury's orbit~\cite{1916AnP...354..769E}. As with QCD, gravitational nonlinearity is evident in the Lagrangian (from the dependence of $-\frac{1}{2\kappa}R$ on the metric inverse), but we are puzzled that the quadratic Maxwell structure seen above is \emph{missing}. While GR might not therefore qualify as a Yang--Mills theory on these grounds, it \emph{is} undeniably a gauge theory, and we observe moreover that the various natural gauge-theoretic reformulations of GR~\cite{1998RSPTA.356..487L,blagojevic2002gravitation} (and gravitational theory as a whole~\cite{lasenby-hobson-2016,blagojevic2002gravitation}) all point to a local symmetry of the Poincar\'e group $\poincare$, which is non-Abelian\footnote{Although again, since $\poincare$ is not compact, it may seem less suitable than $\mathrm{SU}(n)$ as a basis for Yang--Mills theory.}.

So far, so good, in likening GR to a theory of `classical chromodynamics'. The classical electrodynamic comparison, swapping a \emph{chromoelectric} for a \emph{gravitoelectric} field, seems even more encouraging. Gravitoelectromagnetism (GEM) constitutes a well established, Lorentz-invariant and classical correspondence between linear GR and electromagnetism, in which mass-energy is interpreted as the electrical charge~\cite{landau,Hobson:2006se}. GEM thus offers a convenient, linear foundation --- perhaps less naturally present in QCD~\cite{Catani:1989fe} --- upon which the inherent nonlinearities of the theory can be turned up and examined.

But do gravitons carry the GEM mass-energy charge, as gluons carry colour? They \emph{do}, but only at the expense of general covariance. In developing a GEFC picture of `heavy flux', it seems hard to avoid an appeal to \emph{gravitational energy}, for which there is no preferred, generally-applicable localisation scheme~\cite{landau,jiri_1}. One is in fact spoiled for a choice of gravitational stress-energy pseudotensor with which to perform the self-coupling~\cite{xulu,Barker:2018ilw}. It is not obvious that this will be fatal to a speculative GEFC programme. For example, flux collapse could be interpreted covariantly (e.g. through the curvature as a natural field strength), while some `compensating' gauge-dependence is implicated in the details of how the chosen pseudotensor acts as a source.

So, a classical confinement model for GR does not seem out of the question. What about quantum effects? QCD is \emph{strictly} a quantum theory, but a complete quantum theory of gravity is currently missing~\cite{2016arXiv161008744B}. We are puzzled again, but it is helpful to remember that quantum field theory in curved spacetime is nonetheless very well understood. Experience of the Unruh effect then suggests GEFC will not be so susceptable to string-breaking as QCD~\cite{PhysRevD.14.870,PhysRevD.7.2850,Davies_1975}. Although it is not limited by the $\sim\SI{}{\mega\electronvolt}$ quark mass, suppression by the Planck density means that conditions for appreciable gravity-induced particle production are met only in the most extreme scenarios, such as inefficient inflationary reheating~\cite{Ford:1997hb,PhysRev.183.1057}, and the Hawking effect near small (i.e. hot) black holes~\cite{PhysRevD.15.2738}. If the linear regime is less fragile, we might expect long-range, linear gravitational potentials to be commonplace in nature.

If that were true, GEFC ought to apply quite intuitively over the galactic plane: radial field lines (as sourced by a typical galactic baryon profile) should drawn downwards to be embedded in the disc, where their bunching would enhance the centripetal force on the stars at the periphery and hence --- in a grand astrophysical analogue of the meson Regge trajectories, \emph{accelerate galactic orbits}.

By now this begins to sound potentially exciting. 
As established by the seminal \SI{21}{\centi\meter} observations of Rubin et al~\cite{1970ApJ...159..379R,1978ApJ...225L.107R,1980ApJ...238..471R} and, later, others~\cite{1981AJ.....86.1825B,1985ApJ...295..305V}, we know that most spiral galaxies exhibit flat or rising rotation curves which are inconsistent with the (traditionally modelled~\cite{will_2018,will_2018_2}) weak gravitation of their optically determined baryon content.
A missing or \emph{dark} matter component which might account for this was earlier proposed by Zwicky~\cite{1933AcHPh...6..110Z,1937ApJ....86..217Z}, based on the motions of seven galaxies in the Coma cluster\footnote{See English and Spanish translations in~\cite{2017arXiv171101693A}.}. The current paradigm of course stipulates that most late-type baryonic discs are sitting at the centre of a heavier, more extensive dark matter halo~\cite{Li:2020iib,2017ApJ...836..152L,2016AJ....152..157L}, whose presence may also be inferred by lensing~\cite{Massey:2010hh}.

Current GEFC models promise an alternative to this paradigm. By accounting for rotation curves within the strict context of general relativity (GR),~\gefc{Deur:2019kqi,Deur:2016bwq,Deur:2017aas} is supposed to be supported by lensing calculations~\gefc{Deur:2020wlg}, an observed correlation between missing matter effects and ellipticity~\gefc{Deur:2013baa}, and an extension of its principle to galaxy clusters (specifically the Bullet cluster~\cite{2006ApJ...648L.109C}) \gefc{Deur:2009ya,Deur:2017aas,Deur:2020wlg}, in promising to \emph{eliminate} the original need for the dark matter, whose particle composition continues to remain so elusive~\cite{2010ARA&A..48..495F}.

The r\^ole of (\emph{cold}) dark matter (CDM) on cosmological scales is also central to the prevailing \textLambda CDM cosmic concordance model~\cite{2018arXiv180401318S,2018arXiv180706209P,2016A&A...594A..13P,2014A&A...571A..16P} --- but here too, GEFC is put forward to restore consistency. The enhanced force effect is apparently shown in~\gefc{Deur:2021ink} to be adequate in driving structure formation without the need for CDM. The rarefied force effect is moreover suggested in~\gefc{Deur:2017aas} as an alternative to dark \emph{energy} (viz the cosmological constant $\Lambda$), as suggested by the relevant SNIa observations~\cite{supnv1,supnv2,supnv3}. Most recently~\gefc{Deur:2022ooc} concludes that the effects are also consistent with the observed power spectrum of temperature anisotropies in the cosmic microwave background (CMB), without the need for any dark ingredients~\cite{2018arXiv180706209P,2016A&A...594A..13P,2014A&A...571A..16P}.

\vspace{5pt}

Notwithstanding the theoretical appeal as we have motivated it above, the \emph{extent} to which~\gefc{Deur:2009ya,Deur:2013baa,Deur:2014kya,Deur:2016bwq,Deur:2017aas,Deur:2019kqi,Deur:2020wlg,Deur:2021ink,Deur:2022ooc} credits confinement-like effects with the observed phenomena would seem to warrant a level of skepticism. In particular, it is not clear how such significant behaviours can have been consistently missed in the long history of numerical relativity~\cite{Lehner:2014asa,Lehner:2001wq}, or in the well-developed post-Newtonian formalism~\cite{will_2018,will_2018_2}. If we are not too concerned with string-breaking effects, it seems prudent to pay closest attention to the \emph{onset} of confinement, and ask how this can come about in the weak-field environment of the galactic disc.

\vspace{10pt}

In this paper, we will attempt to refute the main structural elements of the proposed GEFC programme, as it is currently presented.
Much of our commentary follows from a close reading (and sincere attempts at reproduction) of~\gefc{Deur:2016bwq} and~\gefc{Deur:2020wlg}, but it will become clear that our findings also disallow the better part of the techniques which support the broader literature~\gefc{Deur:2009ya,Deur:2014kya,Deur:2017aas,Deur:2019kqi,Deur:2021ink,Deur:2022ooc}\footnote{We note a certain parallel with a previous attempt to explain rotation curves using purely GR effects, by Cooperstock and Tieu~\cite{Cooperstock:2006dt} -- that model was cogently shown to be non-viable by Korzy\'nski~\cite{Korzynski:2007zz}.}.

In particular we note that GEFC has, hitherto and from its inception (see~\gefc{Deur:2016bwq,Deur:2014kya,Deur:2021ink,Deur:2009ya,Deur:2019kqi,Deur:2017aas}), used a \emph{scalar} model of gravitation as a proxy for GR. This model is described in greatest detail in~\gefc{Deur:2016bwq}, from which we understand the scalar to be a nonlinear extension of the gravitoelectrostatic potential.
Now, post-Newtonian scalar models of gravity have been proposed in the past -- most notably by Nordstr\"om and later Einstein -- but they are not faithful to the phenomena~\cite{noord}.
We will show that the GEFC scalar is no different in this regard: it does not descend from GR by any principled means, and has no clear redeeming feature beyond the attractive force law expected of an even-spin representation of the Lorentz group.
In~\gefc{Deur:2016bwq} the scalar model is implemented on a Euclidean lattice to produce remarkable -- ostensibly gravitational -- effects, such as linear potentials between point masses. These results are recapitulated in~\gefc{Deur:2014kya,Deur:2019kqi}.
However, notwithstanding the non-relation between the GEFC scalar and GR, we are not convinced that the specific lattice techniques used in that work have a physical grounding.

We also address the outstanding phenomenological claims of GEFC, insofar as they pertain to galactic rotation curves in~\gefc{Deur:2020wlg,Deur:2019kqi,Deur:2016bwq,Deur:2017aas}.
We attempt to `steel-man'\footnote{We use the term `steel-man' in contrast to the more commonplace `straw-man', to mean that the strongest or most promising interpretation of the GEFC proposal should be considered, where possible.} the GEFC hypothesis by discarding the faulty scalar model, and directly probing nonlinear GR for the claimed phenomena in the presence of reasonable, lenticular baryon profiles.
We are dissapointed to find \emph{no such phenomena} at next-to-leading order, though we consider a range of gauges and perturbation schemes.
In~\gefc{Deur:2020wlg} the effect of graviton self-interaction on rotation curves is actually modelled
by considering the gravitoelectric field lines as the trajectories of
massless gravitons, which are then gravitationally lensed by the galactic
density distribution in the same way as photon trajectories; that the paths of
electric field lines in GR follow precisely those of null geodesics has been
discussed previously by Padmanahban~\cite{padm}. The modified gravitoelectric
field at any point, and hence the force on a test particle, is then determined
by calculating the flux of the lensed field lines through a small surface at
that point. Based on this interesting method, however, our own calculations will indicate lensing effects \emph{three orders of magnitude smaller} than those claimed in~\gefc{Deur:2020wlg}.

The remainder of this paper is organised as follows. We conclude this section by introducing some conventions in~\cref{setupcon,oldgauge}. In~\cref{anmod} we consider the physical meaning of the scalar gravity model which underpins~\gefc{Deur:2016bwq,Deur:2014kya,Deur:2021ink,Deur:2009ya,Deur:2019kqi,Deur:2017aas}, and which is specifically studied using lattice methods in~\gefc{Deur:2016bwq}. We speculate as to how substantial differences may arise between the phenomenology of this model, and of GR. We also try, using standard parameterised post-Newtonian (PPN) methods, to account for how the correct Einstein--Infeld--Hoffman potential comes to be produced by this model. Our attempt to understand the meaning of the lattice method itself is confined to~\cref{ess}. 
In~\cref{gem2} we attempt to `steel-man' the GEFC effect by studying next-to-leading order GR. We tackle the nonlinear gravitoelectric effect directly, by constructing the leading nonlinear correction to the gravitoelectromagnetic (GEM) formalism. This is ostensibly equivalent to the level of approximation used in~\gefc{Deur:2016bwq}. We consider the consistency of our GEM formulation with the PPN formalism in~\cref{bipartite}.
Finally in~\cref{lensing} we recalculate the lensing effects presented in~\gefc{Deur:2020wlg}. We consider both the profile of~\gefc{Deur:2020wlg} and the independently motivated Miyamoto--Nagai profile of \cite{Miyamoto:1975zz}. Conclusions follow in~\cref{conclusions}.

\subsection{Setup and conventions}\label{setupcon}

We will use the `West coast' signature $(+,-,-,-)$.
We sometimes use an overbar for background quantities, and the exact (dimensionless) metric perturbation will be
\begin{equation}
  \tensor{g}{_{\mu\nu}}\equiv \tensor{\og}{_{\mu\nu}}+\tensor{h}{_{\mu\nu}}, \quad
  \tensor{g}{^{\mu\nu}}\equiv \tensor{\og}{^{\mu\nu}}-\tensor{h}{^{\mu\nu}}+\mathcal{  O}(h^2).
  \label{pert}
\end{equation}
Greek indices on perturbations refer strictly to the background metric (and to the metric in any non-perturbative context): we will prefer a flat background $\tensor{\og}{_{\mu\nu}}=\tensor{\eta}{_{\mu\nu}}$ for most purposes, and introduce a timelike vector field $\tensor{\bar{u}}{^\mu}\tensor{\bar{u}}{_\mu}\equiv 1$ to define `static' on the background. Indices from the middle of the alphabet, e.g. $\mu$, $\nu$, run from $0$ to $3$ and from the beginning, e.g. $\alpha$, $\beta$, run from $1$ to $3$ over the spatial coordinates $\tensor{x}{^\alpha}$. The time coordinate is $t\equiv \tensor{x}{^0}$. For a flat background our coordinates are usually Cartesian Lorentz coordinates, and we also use vector notation such as $[\bm{x}]\equiv\tensor{x}{^\alpha}$ so that $|\bm{x}|^2\equiv-\tensor{\eta}{_{\alpha\beta}}\tensor{x}{^\alpha}\tensor{x}{^\beta}$, and $\dot{\bm{x}}\equiv\tensor{\partial}{_t}\bm{x}$, but $[\vect{\nabla}]\equiv\tensor{\partial}{_\alpha}$. We also use standard cylindrical coordinates with radius $R$ (not to be confused with the Ricci scalar), azimuthal angle $\upvarphi$, and $z$ anchored to $z\equiv\tensor{x}{^3}$, and spherical polar coordinates sharing the azimuth $\upvarphi$ but with polar angle $\upvartheta$ and radius $r$. Occasionally, we also suppress indices on four-vectors, e.g. $[x]\equiv\tensor{x}{^\mu}$.

The total Einstein--Hilbert action, with matter added, is taken to be
\begin{equation}
  S_{T}\equiv \int \mathrm{d}^4x \sqrt{-g}
  \left[ 
    -\frac{1}{2\kappa}R+L_{{M}}
  \right],
  \label{eha}
\end{equation}
with $g\equiv\det \tensor{g}{_{\mu\nu}}$.
We can divide up the total action and Lagrangian into (kinetic) gravitational and matter parts ${S_T\equiv S_G+S_M}$ and ${\mathcal{L}_{{T}}\equiv\mathcal{L}_{{G}}+\mathcal{L}_{{M}}}$, where
\begin{equation}
	\mathcal{L}_{G}\equiv-\frac{1}{2\kappa}\sqrt{-g}R, \quad
	\mathcal{L}_{M}\equiv\sqrt{-g}L_M,
	\label{gravmat}
\end{equation}
with the Ricci tensor $\tensor*{R}{^\mu_\nu}\equiv \tensor{R}{^{\mu\sigma}_{\nu\sigma}}$ and scalar $R\equiv \tensor*{R}{^\mu_\mu}$, derived from the Riemann tensor and Christoffel symbols according to
\begin{subequations}
	\begin{gather}
		\tensor{R}{_{\rho\sigma\mu}^\nu}\equiv 2\big(
    \tensor{\partial}{_{[\sigma}}\tensor*{\Gamma}{^{\nu}_{\rho]\mu}}
  +\tensor*{\Gamma}{^\lambda_{[\rho|\mu}}\tensor*{\Gamma}{^\nu_{|\sigma]\lambda}}\big),\\
		\tensor*{\Gamma}{^\mu_{\nu\sigma}}\equiv  \frac{1}{2}\tensor{g}{^{\mu\lambda}}\big(
    \tensor{\partial}{_\nu}\tensor{g}{_{\sigma\lambda}}
    +\tensor{\partial}{_\sigma}\tensor{g}{_{\nu\lambda}}
    -\tensor{\partial}{_\lambda}\tensor{g}{_{\sigma\nu}}
  \big).
  \label{metrical_riemann}
	\end{gather}
\end{subequations} 
The Einstein field equations (EFEs) which follow from a variation of~\eqref{eha} with respect to $\tensor{g}{^{\mu\nu}}$ equate the Einstein ${\tensor{G}{_{\mu\nu}}\equiv \tensor{R}{_{\mu\nu}}-\frac{1}{2}\tensor{g}{_{\mu\nu}}R}$ and energy-momentum tensors
\begin{equation}
  \tensor{G}{_{\mu\nu}}=\kappa\tensor{T}{_{\mu\nu}}, \quad
  \tensor{T}{_{\mu\nu}}\equiv\frac{2}{\sqrt{-g}}\frac{\delta S_M}{\delta\tensor{g}{^{\mu\nu}}}.
  \label{efe}
\end{equation}
The Newton and Einstein constants and the Planck mass are related by ${8\pi G\equiv\kappa\equiv 1/\Planck^2}$, and we naturally take the fundamental speed $c=1$.

The nature of the sources considered in~\gefc{Deur:2016bwq,Deur:2014kya,Deur:2021ink,Deur:2009ya,Deur:2019kqi,Deur:2017aas} suggests that we can confine ourselves to perfect fluid lagrangia for which the stress-energy tensor takes the form
\begin{equation}
  \tensor{T}{^{\mu\nu}}=\left( \rho+P \right)\tensor{u}{^\mu}\tensor{u}{^\nu}-P\tensor{g}{^{\mu\nu}}.
  \label{masterset}
\end{equation}
In~\eqref{masterset} we define the rest-mass energy density $\rho$ of the fluid, including all internal chemical, kinetic and thermal contributions, $P$ is the rest pressure, and $\tensor{u}{^\mu}$ is the four-velocity $\tensor{u}{^\mu}\equiv\mathrm{d}\tensor{x}{^\mu}/\mathrm{d}s$ of the fluid. Note that we are encouraged in~\gefc{Deur:2016bwq} and~\gefc{Deur:2020wlg} to assume $P=0$ to all perturbative orders.

A particularly convenient way to label the perturbative gravitational effects of such a fluid is via the velocity, assuming that velocity is non-relativistic, the source has suitably virialised under its own gravitation and that a variety of other reasonable statistical conditions are met~\cite{will_2018}. If the coordinate velocity is denoted $\tensor{v}{^\alpha}\equiv\tensor{u}{^\alpha}/\tensor{u}{^0}$ and we assume ${\gamma_v \equiv\left(1-|\bm{v}|^2\right)^{-1/2} \approx 1}$, we will accordingly refer to $\mathcal{O}\left( |\bm{v}|^{2n} \right)$ as $\mathcal{O}\left( \varepsilon^{n} \right)$, in keeping with the conventions of the PPN formalism~\cite{will_2018_2}. Of course, the coordinate velocity in this case need not be that expressed in~\eqref{masterset}, if we relax $P=0$. 
Indeed, careful arguments have shown~\cite{will_2018} that $\mathcal{O}\left( P/\rho \right)$ is generally \emph{synonymous} with $\mathcal{O}\left( \varepsilon \right)$, and we will briefly recall in~\cref{oldgauge,smokemirror} that the same is true of $\mathcal{O}\left( h \right)$.

An additional assumption suggested in~\gefc{Deur:2016bwq,Deur:2014kya,Deur:2021ink,Deur:2009ya,Deur:2019kqi,Deur:2017aas} is that of \emph{staticity}. Any number of flagrant inconsistencies are readily seen to arise when we try to combine $\bm{v}=0$ with $P=0$ at nonlinear orders. On these grounds such assumptions ought to be disqualifying, and a reasonable approach to verifying the claims of~\gefc{Deur:2016bwq,Deur:2014kya,Deur:2021ink,Deur:2009ya,Deur:2019kqi,Deur:2017aas} might be to include pressure, or rotation, or both. In fact, we do not believe such onerous extensions are necessary: we will show over~\crefrange{anmod}{lensing} that GEFC suffers more fundamental problems than susceptability to the Jeans instability.

In demonstrating this, we observe that (i) the `static dust' picture will cause contradictions to arise at various points in the analysis, which must be overlooked if any comparison with~\gefc{Deur:2016bwq,Deur:2014kya,Deur:2021ink,Deur:2009ya,Deur:2019kqi,Deur:2017aas} is to be made, and (ii) the $\mathcal{O}\left( \varepsilon^{n} \right)$ PPN formalism may be formally retained in what follows, even though there are (somehow) no velocities.

\subsection{Linearised general relativity}\label{oldgauge}

Two different types of coordinate transformation connect
quasi-Minkowskian systems to each other: global Lorentz
transformations ${x^\prime}^\mu = {\Lambda^\mu}_\nu x^\nu$ and
infinitesimal general coordinate transformations ${x^\prime}^\mu =
x^\mu + \xi^\mu(x)$, under which $h^\prime_{\mu\nu} =
{\Lambda_\mu}^\rho{\Lambda_\nu}^\sigma h_{\rho\sigma}$ and
$h^\prime_{\mu\nu} = h_{\mu\nu} -\partial_\mu\xi_\nu -
\partial_\nu\xi_\mu+\mathcal{O}\left( \left(  \partial\xi \right)^2\right)$, respectively. This suggests that instead of
considering a slightly curved spacetime to represent the
general-relativistic weak field, one can reinterpret $h_{\mu\nu}$
simply as a special-relativistic symmetric rank-2 tensor field that
represents a weak gravitational field on a Minkowski background
spacetime and possesses the gauge
freedom 
\be
h_{\mu\nu} \to h_{\mu\nu} -\partial_\mu\xi_\nu -
\partial_\nu\xi_\mu+\mathcal{O}\left( \left(  \partial\xi \right)^2\right).
\label{ch17:eqn17.5}
\ee

Expanding the Einstein equations~\eqref{efe}
to first-order in $h_{\mu\nu}$ to yield $G^{(1)}_{\mu\nu} \equiv
R^{(1)}_{\mu\nu}-\frac{1}{2}\eta_{\mu\nu}R^{(1)} = \kappa
T_{\mu\nu}+\mathcal{  O}\left( \varepsilon^2 \right)$, one obtains the linearised field equations
\be
\begin{aligned}
G^{(1)}_{\mu\nu} &\equiv -\frac{1}{2}\left(\dalembertian \bar{h}_{\mu\nu} + \eta_{\mu\nu} \partial_\rho
\partial_\sigma \bar{h}^{\rho\sigma} - \partial_\nu\partial_\rho
\bar{h}_\mu^\rho-\partial_\mu\partial_\rho \bar{h}_\nu^\rho\right) 
\\
&
=\kappa T_{\mu\nu}+\mathcal{O}\left( \varepsilon^2 \right),
\label{ch17:eqn17.11}
\end{aligned}
\ee
where $\bar{h}_{\mu\nu} \equiv h_{\mu\nu}-\frac{1}{2}\eta_{\mu\nu} h$
is the trace reverse\footnote{The trace reverse should not be confused with background quantities: we will apply it only to symbols denoting perturbative quantities.} of $h_{\mu\nu}$, with $h \equiv
\eta_{\mu\nu}h^{\mu\nu}$, and $\dalembertian \equiv
\eta^{\mu\nu}\partial_\mu\partial_\nu$ is the d'Alembertian
operator. As expected, the LHS of~\eqref{ch17:eqn17.11} is invariant
under the gauge transformation~\eqref{ch17:eqn17.5}. By
choosing $\xi^\mu(x)$ to satisfy $\dalembertian \xi^\mu = \partial_\rho
\bar{h}^{\mu\rho}$, one may impose the Lorenz gauge condition
$\partial_\rho \bar{h}^{\mu\rho}=0$; note that this gauge condition is
preserved by any further gauge transformation of the form~\eqref{ch17:eqn17.5} provided that the functions $\xi^\mu$ satisfy
$\dalembertian\xi^\mu=0$.  In the Lorenz gauge, the linearised field
equations~\eqref{ch17:eqn17.11} reduce to the simple form
\be
\dalembertian \bar{h}^{\mu\nu} = -2\kappa T^{\mu\nu}+\mathcal{  O}\left( \varepsilon^2 \right).
\label{ch17:eqn17.13}
\ee

The general solution to the inhomogeneous wave equation~\eqref{ch17:eqn17.13} is most easily obtained by using a Green's
function approach, in a similar manner to that employed for solving
the analogous problem in electromagnetism. Denoting spatial 3-vectors
by $\vect{x}$, this yields
\be
\begin{aligned}
\bar{h}^{\mu\nu}(t,\vect{x}) &= -4G \int
\frac{T^{\mu\nu}(t-|\vect{x}-\vect{x}'|, \vect{x}')}{|\vect{x}-
\vect{x}'|} \,\mathrm{d}^3x'
\\
&\ \ \ \ \ +\ppn{2}.
\label{ch17:eqn17.31}
\end{aligned}
\ee
For a stationary source, $\partial_0 T^{\mu\nu}=0$, such that the time
dependence vanishes and retardation is irrelevant, so~\eqref{ch17:eqn17.31} reduces to
\be
\bar{h}^{\mu\nu}(\vect{x}) = -4G \int
\frac{T^{\mu\nu}(\vect{x}')}{|\vect{x}-\vect{x}'|} \, \mathrm{d}^3x'+\ppn{2}.
\label{ch17:eqn17.46}
\ee
%

Following on from our discussion in~\cref{setupcon}, for a \emph{stationary}, non-relativistic source with $P=0$, we approximate the energy-momentum tensor as
\be
\begin{gathered}
	T^{00} =\rho+\ppn{2},\quad T^{\alpha 0} = \rho v^\alpha+\mathcal{O}\left(\varepsilon^{5/2}\right),
	\\ 
	T^{\alpha\beta} = \mathcal{O}\left(\varepsilon^2\right).
\label{emtensor}
\end{gathered}
\ee
Indeed, this is also consistent
with the Lorenz gauge condition $\partial_\rho
\bar{h}^{\mu\rho}=0$, which implies that
$\partial_\alpha\bar{h}^{\beta\alpha}=-\partial_0\bar{h}^{\beta 0}$, which vanishes for
stationary systems.

In the linearised theory, there is an inconsistency between the field
equations~\eqref{ch17:eqn17.11} and the equations of motion for matter
in a gravitational field. From~\eqref{ch17:eqn17.11}, one quickly
finds that $\partial_\mu T^{\mu\nu} = 0$, which should be contrasted
with the requirement from the full GR field equations that $\vect{\nabla}_\mu
T^{\mu\nu}=0$.  The latter requirement leads directly to the geodesic
equation of motion for the worldline $x^\mu(s)$ of a test particle,
namely
\be
\frac{\mathrm{d}^2\tensor{x}{^\mu}}{\mathrm{d}s^2} + \tensor*{\Gamma}{^\mu_{\nu\sigma}}\frac{\mathrm{d}\tensor{x}{^\nu}}{\mathrm{d}s}\frac{\mathrm{d}\tensor{x}{^\sigma}}{\mathrm{d}s}
= 0,
\label{ch17:eqn17.16}
\ee
whereas the former requirement leads to the equation of motion
$\mathrm{d}^2\tensor{x}{^\mu}/\mathrm{d}s^2 = 0$, which means that the gravitational field has {\it
  no effect} on the motion of the particle and so clearly contradicts
the geodesic postulate. Despite this inconsistency, the effect of weak
gravitational fields on test particles may still be computed by
inserting the linearised connection coefficients into the geodesic
equations~\eqref{ch17:eqn17.16} -- we will make use of this approach in~\cref{axisy}.

\section{Analysis of the scalar model}\label{anmod}

In this opening section, we will attempt to show that the scalar model of gravity, which has been used as a basis for many GEFC calculations (see~\gefc{Deur:2016bwq,Deur:2014kya,Deur:2021ink,Deur:2009ya,Deur:2019kqi,Deur:2017aas}), is not descriptive of GR within the regime of its application.
\subsection{The matter coupling}\label{matcou}

We begin our analysis by considering the incorporation of matter sources into the gravity model.
A starting point of the GEFC approach is a perturbative expansion of~\eqref{eha} in the field ${\tensor{\varphi}{_{\mu\nu}}\equiv \Planck \tensor{h}{_{\mu\nu}}}$, proposed in~\gefc{Deur:2016bwq} to take the specific form
\begin{equation}
	\begin{aligned}
		\mathcal{L}_T&=\left[\partial\varphi\partial\varphi\right]+\frac{\sqrt{2}}{\Planck}\left[\varphi\partial\varphi\partial\varphi\right]+\frac{2}{\Planck^2}\left[\varphi^2\partial\varphi\partial\varphi\right]\\
		&\ \ \ -\frac{\sqrt{2}}{\Planck}\tensor{\varphi}{_{\mu\nu}}\tensor{\bar T}{^{\mu\nu}}
		-\frac{1}{\Planck^2}\tensor{\varphi}{_{\mu\nu}}\tensor{\varphi}{_{\lambda\sigma}}\tensor{\bar T}{^{\mu\nu}}\tensor{\eta}{^{\sigma\lambda}}+\ldots,
		\label{deur1}
	\end{aligned}
\end{equation}
where the notation $[\cdot]$ with indices suppressed denotes particular contractions following from the Einstein--Hilbert term, and the customary perturbation is in powers of $\tensor{\varphi}{_{\mu\nu}}/\Planck$. More or less equivalent series to~\eqref{deur1} are proposed in~\gefc{Deur:2014kya,Deur:2021ink,Deur:2009ya,Deur:2019kqi,Deur:2017aas}.

The lowest order terms in~\eqref{deur1} are of course the massless Fierz--Pauli theory, coupled to a matter current.
At higher perturbative orders howerver, we are tempted to move from the outset to an adjacent theory with a modified matter sector.
Our reasons for this are illustrated by a formal expression\footnote{see e.g.~\cite{Butcher:2009ta} for a similar approach} which we can construct for the perturbative expansion of~\eqref{eha}
\begin{equation}
  \begin{aligned}
  S_{M}&\equiv
  -\sum_{n=0}^\infty \frac{1}{n!}
  \left[
    \int \mathrm{d}^4x \frac{\tensor{\varphi}{^{\mu\nu}}}{\Planck}\frac{\delta}{\delta\tensor{\og}{^{\mu\nu}}}
  \right]^n
  \bar{S}_{M}
  =\bar{S}_{M}
  \\
	  -\sum_{n=1}^\infty & \frac{1}{n!}
  \left[
    \int \mathrm{d}^4x \frac{\tensor{\varphi}{^{\mu\nu}}}{\Planck}\frac{\delta}{\delta\tensor{\og}{^{\mu\nu}}}
  \right]^{n-1}
  \int \mathrm{d}^4x \frac{\sqrt{-\bar{g}}\tensor{\varphi}{^{\rho\sigma}}}{2\Planck}\tensor{\bar{T}}{_{\rho\sigma}}.
  \label{lukex}
  \end{aligned}
\end{equation}
Even assuming (as sometimes applies) that $L_M$ contains no derivatives of $\tensor{g}{_{\mu\nu}}$, it would then seem from~\cref{lukex} that a perturbative expansion roughly of the form~\eqref{deur1} would require the curious condition on (or off) the background
\begin{equation}
  \left[\frac{\partial}{\partial\tensor{g}{^{\mu\nu}}}\right]^n \left(\sqrt{-{g}}\tensor{ T}{_{\rho\sigma}}\right)= \tensor[^{(n)}]{X}{_{\dots\rho\sigma}^{\kappa\lambda}}\tensor{ T}{_{\kappa\lambda}},
  \label{restriction}
\end{equation}
where $\tensor[^{(n)}]{X}{_{\dots\rho\sigma}^{\kappa\lambda}}$ is some suitably symmetrized and indexed density concommitant of (the undifferentiated) $\tensor{g}{_{\mu\nu}}$. 
Possibly~\eqref{restriction} can be satisfied by the cosmological constant, though without any other matter present this would appear restrictive. 
The other option, that $\tensor{T}{^{\mu\nu}}$ is independent of $\tensor{g}{_{\mu\nu}}$, is also restrictive; it does not apply even for a spin-0 boson. Moreover this option contradicts the expectation that $\mathcal{L}_M$ be a covariant density: the only ansatz in that case is $\mathcal{L}_M=c_1\sqrt{-g}T+c_2\det\tensor{T}{_{\mu\nu}}$, where $T\equiv\tensor*{T}{^\mu_\mu}$, and this ansatz is not consistent with the EFEs in~\eqref{efe}. Accordingly, and without detailed knowledge of the matter sector, we are not wholly confident that the matter coupling in~\gefc{Deur:2014kya,Deur:2021ink,Deur:2009ya,Deur:2019kqi,Deur:2017aas,Deur:2016bwq} is safe\footnote{It is also possible that the stress energy tensor is being re-introduced to the Lagrangian via a solution to the lowest-order field equations: while we are not able to confirm this in the case of~\eqref{deur1}, we use a similar approach in~\cref{newgauge}.}. We will return to this issue in~\cref{smokemirror}, where we do have such detailed knowledge, but select instead a conventional relativistic point particle action under the standard PPN perturbation scheme.

\subsection{The non-relativistic scalar}\label{nonrelsca}

For the moment the representation of the matter sector does not impede our discussion, since~\eqref{lukex} can be used to obtain all the corrections in~\eqref{deur1} to the gravitational sector, and it is this sector which is principally targeted in~\gefc{Deur:2016bwq}. There and elsewhere in~\gefc{Deur:2014kya,Deur:2021ink,Deur:2009ya,Deur:2019kqi} it is argued that for static spacetimes, the gravitational field may be represented by the \emph{single} degree of freedom
\begin{equation}
  \tensor{\varphi}{^{\mu\nu}}=2\left(2\tensor{\bar{u}}{^\mu}\tensor{\bar{u}}{^\nu}-\tensor{\bar{g}}{^{\mu\nu}}\right)\varphi,
  \label{pertan}
\end{equation}
which in the static, perturbative context is the Newtonian scalar potential $2\Planck\vect{\nabla}^2\varphi=\rho+\mathcal{O}\left(\varepsilon^2\right)$.
However it is made clear in~\gefc{Deur:2016bwq} that $\varphi$ also dominates in some non-perturbative static regime of physical relevance, and so we are effectively being invited to promote~\eqref{pertan} to an isotropic Cartesian line element 
\begin{equation}
  \mathrm{d}s^2=\left(1+\frac{2\varphi}{\Planck}\right) \mathrm{d}t^2
  -\left(1-\frac{2\varphi}{\Planck}\right)\mathrm{d}\bm{x}^2,
  \label{linel}
\end{equation}
which is signature-preserving within the range ${|\varphi/\Planck|<1/2}$.
The ansatz~\eqref{pertan} is apparently \emph{substituted directly into}~\eqref{deur1}, and the static assumption $\dot\varphi\equiv\partial_t\varphi=0$ imposed to obtain a scalar Euclidean lattice action up to the required perturbative order\footnote{Our understanding of the approach is also based on the relevant section in~\cite{Deur:2009ya}.} -- which for much of~\gefc{Deur:2016bwq} entails the equivalent of a $\mathcal{O}\left(\varepsilon^2\right)$ correction, or $\mathcal{O}\left(\varphi^2/\Planck^2\right)$. But if the relevant solutions are indeed non-perturbative, why truncate~\eqref{deur1} at all? We might not be sure for reasons discussed in~\cref{matcou} how~\eqref{deur1} relates to $\mathcal{L}_M$, but we can skip ahead on the gravitational side of the action by substituting the line element~\eqref{linel} \emph{directly into} the fully nonlinear $\mathcal{L}_G$ as it is written in~\eqref{gravmat} to give a novel action $\tilde{S}_G\equiv\int\mathrm{d}t\mathrm{d}^3x \tilde{\mathcal{L}}_G$. Doing so, we find that the lattice calculations in~\gefc{Deur:2016bwq} are really attempting to probe (under the assumption of staticity) the following non-relativistic theory 
\begin{equation}
	  \tilde{\mathcal{L}}_G\equiv
    -
      \frac{
	3\left(
	1
	-\frac{2\varphi}{\Planck}
	\right)\dot{\varphi}^2
	+
	\left(
	1
	-\frac{6\varphi}{\Planck}
	\right)|\vect{\nabla}\varphi|^2
      }{
	\left(
	1
	-\frac{2\varphi}{\Planck}
	\right)
	\sqrt{
	1
	-\frac{4\varphi^2}{\Planck^2}
	}
      }.
    \label{thoughttrain}
\end{equation}

\begin{figure*}[t]
  \center
  \includegraphics[width=\linewidth]{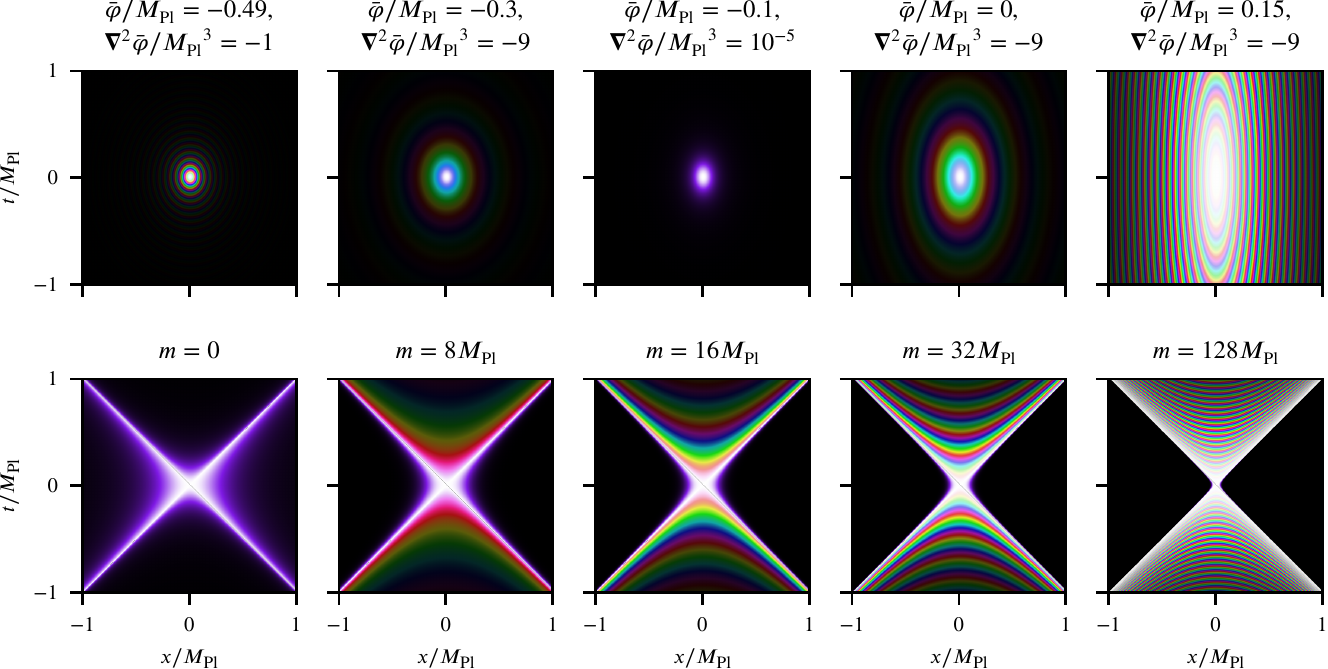}
	\caption{\label{coeffs} 
	Propagator of the scalar gravity model which underpins~\gefc{Deur:2016bwq,Deur:2014kya,Deur:2021ink,Deur:2009ya,Deur:2019kqi,Deur:2017aas}, with a variety of static background potentials $\bar\varphi$, and with the healthy Klein--Gordon propagator (mass $m$) shown below for comparison.
	The scalar model is obtained by discarding all d.o.f in the nonlinear Einstein--Hilbert action beyond the isotropic metric perturbation $\varphi$ leads to a non-relativistic theory~\eqref{thoughttrain} which does not appear to be healthy, though it may seem so on a Euclidean lattice under the assumption of staticity.
      }
\end{figure*}

We will discover in~\cref{nonrel} that the theory~\eqref{thoughttrain} is essentially arbitrary, and not concretely related to nonlinear gravitostatics. Certainly, it is inconsistent even at lowest order with the linearised EFEs in~\eqref{efe}. But does it even impart stability to the static surfaces identified by the lattice? To answer this we will very briefly consider the quantum mechanical implications of~\eqref{thoughttrain}. Our approach in doing so, whilst providing convenient visualisation of the problem in~\cref{coeffs}, should not be taken too seriously in the context of the strictly classical GEFC proposal.

Bearing this caveat in mind, we imagine that a high-powered Euclidean lattice calculation converges on a static background $\bar{\varphi}$ (not to be confused in this context with $\tensor{\bar{g}}{_{\mu\nu}}=\tensor{\eta}{_{\mu\nu}}$), to which the various interesting solutions in~\gefc{Deur:2016bwq} are presumably approximations. Now~\eqref{thoughttrain} evidently imposes a non-relativistic theory of fluctuations around $\bar{\varphi}$, whose propagator in the representation of momentum $\tensor{p}{^\mu}$, with energy $E\equiv\tensor{p}{^0}$, reads\footnote{To avoid discussion of the second quantization of a ghost, we define the propagator here as being just the Greens function of the equation of motion, or the inverse of the perturbative Lagrangian. The customary shift introduced by the $+i\epsilon$ term should not be necessary in this case, since the poles are rotated. Independent of the GEFC analysis, it is of interest to plot~\cref{propagator} in~\cref{coeffs}, since the ghost propagator is not frequently illustrated in the literature.}
  \begin{align} 
      &D(x_1-x_2)_F=\lim_{\epsilon\to 0}\int\frac{\mathrm{d}E\mathrm{d}^3p}{(2\pi)^4}ie^{-i\tensor{p}{_\mu}\left(\tensor{x}{_1^\mu}-\tensor{x}{_2^\mu}\right)}
\sqrt{
	1
	-\frac{4\bar\varphi^2}{\Planck^2}
      } 
\nonumber\\
&
	      \times\left[
	-6E^2
	-
	2\frac{
	\left(
	1
	-6\bar\varphi/\Planck
      \right)
    }{\left(
	1
	-2\bar\varphi/\Planck
	\right)}|\bm{p}|^2
	-\mu\left(\bar{\varphi},\vect{\nabla}^2\bar{\varphi}\right)^2
	+i\epsilon
      \right]^{-1},
      \nonumber\\
      &
\mu\left(\bar{\varphi},\vect{\nabla}^2\bar{\varphi}\right)^2\equiv 
\frac{4\left(1+\bar\varphi/\Planck+6{\bar\varphi}^2/\Planck^2\right)}{\left(1-2\bar\varphi/\Planck\right)\left(1-4{\bar\varphi}^2/\Planck^2\right)}
\frac{\vect{\nabla}^2\bar\varphi}{\Planck}.
      \label{propagator}
  \end{align}

The pathologies in~\eqref{propagator} appear to be quite severe. For weak fields we might expect $\vect{\nabla}^2\bar\varphi$ to be small outside the matter source on the lattice, however it may be coupled, suppressing the effective mass $\mu$. In that case the residue about $|\bm{p}|=0$ suggests that no unitary quantum theory lives on the portion of the background in which the light cone structure of~\eqref{linel} is preserved. This ghost is invisible on the lattice, which appears furthermore to be shielded from gradient instabilities within the range $-1/2<\bar\varphi/\Planck<1/6$. By analogy to the Newtonian potential, we might expect the lattice to prefer $\bar\varphi/\Planck<0$ anyway, possibly accounting for the excellent numerical results of~\gefc{Deur:2016bwq}. There follows a brief window where the lattice solutions would not be destabilised by classical waves of the dynamical theory, before such waves exit the light cone of $\tensor{\bar{g}}{_{\mu\nu}}$. In general we expect sources and the nonlinear aspects of~\eqref{thoughttrain} to sometimes induce a substantial $\mu$, which could be analysed for any tachyonic character. These observations are illustrated in~\cref{coeffs}.

It is clear from the above analysis that the theory~\eqref{thoughttrain} diverges wildly from the GR phenomena if the assumption of staticity is relaxed. In the context of the wider literature, the possibility of a \emph{linear} gravitational potential would ordinarily lead to the same conclusion in the static case; however that phenomenon is actually proffered in~\gefc{Deur:2016bwq} as being innate to GR, and so in the static case we must proceed more carefully.
\subsection{The non-relation to GR}\label{nonrel}
How can we decide whether static extrema of~\eqref{thoughttrain} are really representing the gravitostatic limit of GR? 
That action entails only one vacuum equation of motion 
\begin{equation}
	\tensor{c}{_j}\tensor{q}{_j}=0,
\label{singfeld}
\end{equation}
where
where $[\tensor{q}{_j}]\equiv\left(\ddot{\varphi},\dot{\varphi}^2/\varphi,\vect{\nabla}^2\varphi,|\vect{\nabla}\varphi|^2/\varphi\right)$ and $\tensor{c}{_{j}}$ is a  rational function in $\varphi/\Planck$ whose components are
\begin{subequations}
\begin{align}
  c_1&\equiv 6\left(1-2\varphi/\Planck\right)\left(1-4\varphi^2/\Planck^2\right),\\
  c_2&\equiv 12\left(\varphi/\Planck\right)^2\left(1-2\varphi/\Planck\right),\\
  c_3&\equiv 2\left(1-6\varphi/\Planck\right)\left(1-4\varphi^2/\Planck^2\right),\\
  c_4&\equiv 4\left(\varphi/\Planck\right)\left(1+\varphi/\Planck+6\varphi^2/\Planck^2\right).
\end{align}
\end{subequations}
However, once a gauge such as~\eqref{linel} is chosen the EFEs in~\eqref{efe} can impose up to \emph{six} such equations, in addition to \emph{four} constraints on the initial data: these had better all be consistent with~\eqref{singfeld}, otherwise we will no longer be studying gravity in any regime whatever. Taking for example the line element~\eqref{linel} substituted into the vacuum equation ${\tensor*{G}{^\mu_\mu}=0}$ and accompanying constraint ${\tensor*{G}{^{\mu\nu}}\tensor{\bar{u}}{_{\mu}}\tensor{\bar{u}}{_{\nu}}=0}$, we obtain the system
\begin{equation}
\tensor{c}{_{aj}}\tensor{q}{_j}=0,
\end{equation}
where $\tensor{c}{_{aj}}$ is a $2\times 4$ matrix which can be diagonalised over the first $2\times 2$ block. Now if we assume staticity, so $\dot{\varphi}=\ddot{\varphi}=0$, we need retain only the second $2\times 2$ block $\tensor{c}{_{ab}}$, writing $\tensor{c}{_{ab}}\tensor{q}{_{b}}=0$ where $[\tensor{q}{_{b}}]\equiv\left(\vect{\nabla}^2\varphi,|\vect{\nabla}\varphi|^2/\varphi\right)$. However direct calculation yields
\begin{equation}
\det \tensor{c}{_{ab}}\propto\frac{\left(\varphi/\Planck\right)\left(7+6\varphi/\Planck\right)}{\left(1-2\varphi/\Planck\right)\left(1-4\varphi^2/\Planck^2\right)^3},
\label{finaldet}
\end{equation}
so the EFEs do not actually seem to admit any nontrivial static solutions under the GEFC ansatz. On this basis it would appear that~\eqref{linel} is too restrictive for nonlinear gravity, and this is not surprising. In the perturbative case the field $\varphi$ essentially corresponds to the principal PPN potential, which cannot be considered in isolation at any PN order~\cite{will_2018}. Note that $\det \tensor{c}{_{ab}}\to 0$ as $\varphi/\Planck \to 0$. In this limit we recover the only link between the GEFC scalar and gravity, namely the static vacuum condition $\vect{\nabla}^2\varphi=0$.
We can now attempt to diagnose a potential issue in~\eqref{thoughttrain}, and so also in the scalar approach of~\gefc{Deur:2016bwq,Deur:2014kya,Deur:2021ink,Deur:2009ya,Deur:2019kqi,Deur:2017aas}. The discrepancy between~\eqref{singfeld} and the EFEs in~\eqref{efe} appears to occur because the GEFC ansatz is substituted before variations (or equivalently lattice path integrals) are performed. These are in general non-commuting operations: in reverse order they may well yield the phenomena described in~\gefc{Deur:2016bwq,Deur:2014kya,Deur:2021ink,Deur:2009ya,Deur:2019kqi,Deur:2017aas} which, however colourful, seem less likely to be gravitational in origin.

\subsection{Why the two-body potential \emph{looks} correct}\label{smokemirror}
As a final check on the scalar model which underpins~\gefc{Deur:2016bwq,Deur:2014kya,Deur:2021ink,Deur:2009ya,Deur:2019kqi,Deur:2017aas}, we consider the significance of the observation in~\gefc{Deur:2016bwq}, that the perturbative interpretation of~\eqref{thoughttrain} recovers the parameterised post-Newtonian (PPN) energy of a system of (proper) point masses, $\tensor*{m}{^*_n}$ at $\bm{x}_n$.
We firstly point out that the perturbative ansatz~\eqref{pertan} is already consistent with the standard PPN gauge~\cite{will_2018} at $\mathcal{O}\left(\varepsilon\right)$, for which $\tensor{\bar{g}}{_{\mu\nu}}=\tensor{\eta}{_{\mu\nu}}$. In that gauge the components of the (dimensionless) metric perturbation $\tensor{h}{_{\mu\nu}}\equiv\tensor{\varphi}{_{\mu\nu}}/\Planck$ in~\eqref{pert}, with the PPN parameters fixed to those of GR, are
\begin{subequations}
  \begin{align}
	  \tensor{h}{_{00}}&=-2U+2\left\{\Phi_2+U^2\right\}+\mathcal{O}\left(\varepsilon^3\right),
  \label{ppn_pert}
	  \\
	  \tensor{h}{_{\alpha\beta}}&=2U\tensor{\eta}{_{\alpha\beta}}+\mathcal{O}\left(\varepsilon^2\right),
  \label{ppn_pert_2}
	  \\
	  \tensor{h}{_{0\alpha}}&=-4\tensor{V}{_\alpha}+\frac{1}{2}\tensor{\partial}{_\alpha}\dot X+\mathcal{O}\left(\varepsilon^{5/2}\right),
  \label{ppn_pert_3}
  \end{align}
\end{subequations}
where we use $\{\cdot\}$ to signify those contributions which originate in the $\mathcal{O}\left(\varepsilon^2\right)$ correction to $\tensor{h}{_{00}}$, and with PPN potentials defined by adapting the PPN conventions in~\cite{will_2018} to our choice of signature
\begin{equation}
  \begin{aligned}
	  U&\equiv\frac{\kappa}{8\pi}\int\frac{\mathrm{d}^3 x' {\rho^*}'}{|\bm{x}-\bm{x}'|},
  \quad
	  X\equiv\frac{\kappa}{8\pi}\int\mathrm{d}^3 x' {\rho^*}'|\bm{x}-\bm{x}'|,\\
	  \Phi_2&\equiv\frac{\kappa}{8\pi}\int\frac{\mathrm{d}^3 x' {\rho^*}'U'}{|\bm{x}-\bm{x}'|},
  \quad
	  \tensor{V}{^\alpha}\equiv\frac{\kappa}{8\pi}\int\frac{\mathrm{d}^3 x' {\rho^*}'\tensor{{v'}}{^\alpha}}{|\bm{x}-\bm{x}'|},
	  \label{ppnpot}
  \end{aligned}
\end{equation}
where the source fluid has total rest mass $M$, rest mass density $\rho$ and we introduce the conserved density $\rho^*\equiv\rho\sqrt{-g}\tensor{u}{^0}$. Recall the coordinate velocity is $\tensor{v}{^\alpha}\equiv\tensor{u}{^\alpha}/\tensor{u}{^0}$ and the four-velocity is $\tensor{u}{^\mu}$. It is clear that the standard PPN gauge~\eqref{ppn_pert} may deviate from~\eqref{pert} above $\mathcal{O}\left(\varepsilon\right)$. To put this another way, careful (and well-tested) consideration of nonlinear effects suggests departure from~\eqref{pert} and~\gefc{Deur:2016bwq}. Let us now consider the PN energy associated with a collection of point sources, i.e. 
\begin{equation}
\rho^*=\sum_{n=1}^N \tensor*{m}{^*_n}\delta^3(\bm{x}-\bm{x}_n),
  \quad
  M\equiv\int\mathrm{d}^3x\rho^*=\sum_{n=1}^N \tensor*{m}{^*_n}.
  \label{manybod}
\end{equation}
No compressional energy is involved, and so the relativistic matter Lagrangian in~\eqref{gravmat} will simply be 
\begin{equation}
  \mathcal{L}_M=-\rho^*/\tensor{u}{^0}=-\rho^*\left[\tensor{g}{_{00}}+2\tensor{g}{_{0\alpha}}\tensor{v}{^\alpha}+\tensor{g}{_{\alpha\beta}}\tensor{v}{^\alpha}\tensor{v}{^\beta}\right]^{\frac{1}{2}}.
	\label{sqrtlag}
\end{equation}
We recall from~\eqref{sqrtlag} why the $\mathcal{O}\left(\varepsilon^2\right)$ part of $\tensor{h}{_{\alpha\beta}}$ is allowed to be suppressed in~\eqref{ppn_pert_2} when considering the first nonlinear corrections to the dynamics, but why the same is not (immediately) true for $\tensor{h}{_{00}}$. In other words, we could imagine that $\tensor{h}{_{00}}=\tensor{h}{_{\alpha\alpha}}+\mathcal{O}\left(\varepsilon^3\right)$, in line with~\eqref{pertan}, and retain only $\tensor{h}{_{00}}$ to next-to-leading order through the calculations.
Accordingly, expanding~\eqref{sqrtlag} to $\mathcal{O}\left(\varepsilon^2\right)$ under the scheme~\crefrange{ppn_pert}{ppn_pert_3}, we find quite directly
  \begin{equation}
  \begin{aligned}
    \mathcal{L}_M &= \frac{1}{2}\rho^*\left(1+3U\right)|\bm{v}|^2
    +\frac{1}{8}\rho^*|\bm{v}|^4
    \\ 
    &
    \ \ \ 
    -\rho^*\bm{v}\cdot\left(4\bm{V}+\frac{1}{2}\vect{\nabla} \dot X\right)
    \\
    &
    \ \ \ 
    -\rho^*\left(1-U-\frac{1}{2}U^2+\left\{\Phi_2+U^2\right\}\right)+\mathcal{O}\left(\varepsilon^3\right).
    \label{matpert}
  \end{aligned}
  \end{equation}
  The $\mathcal{O}\left(\varepsilon^2\right)$ expansion of $\mathcal{L}_G$ as it is defined in~\eqref{gravmat} is more challenging, but by shaking out all surface terms and reducing the potentials (see~\cref{tracking}) we eventually find that the Einstein--Hilbert contribution has the form
\begin{equation}
\begin{aligned}
  \mathcal{L}_G &= 
  \frac{1}{2}\rho^*\bm{v}\cdot\left(4\bm{V}
  +\frac{1}{2}\vect{\nabla} \dot X\right)
  \\
  &
  \ \ \ 
  +\rho^*\left(\frac{1}{2}-\frac{1}{2}U-U^2+\left\{\Phi_2+U^2\right\}\right)+\mathcal{O}\left(\varepsilon^3\right),
  \label{ehpert}
\end{aligned}
\end{equation}
where again we track the $\mathcal{O}\left(\varepsilon^2\right)$ contribution to $\tensor{h}{_{00}}$ through the calculation and retain it in braces. By comparing~\cref{matpert,ehpert} we now see that the gravitational and matter corrections stemming from the $\mathcal{O}\left(\varepsilon^2\right)$ correction to the gravitaional field have an \emph{equal and opposite} effect on the total action. In other words, the leading PN correction to the GEFC scalar \emph{does not survive} in the $\mathcal{O}\left(\varepsilon^2\right)$ corrections to the phenomena when they are properly calculated, which are instead quadratic in the $\mathcal{O}\left(\varepsilon\right)$ Newtonian potential. 

What are these phenomena in the context of~\gefc{Deur:2016bwq}, i.e. static point sources? The total Lagrangian obtained from adding~\cref{matpert,ehpert} is
\begin{equation}
\begin{aligned}
    \mathcal{L}_T &= \frac{1}{2}\rho^*\left(1-3U\right)|\bm{v}|^2
    +\frac{1}{8}\rho^*|\bm{v}|^4
    \\
    &
    \ 
    -\frac{1}{2}\rho^*\bm{v}\cdot\left(4\bm{V}+\frac{1}{2}\vect{\nabla} \dot X\right)
    -\frac{1}{2}\rho^*\left(1-U+U^2\right)
\\
	&
	\
	+\mathcal{O}\left(\varepsilon^3\right).
  \label{finlag}
\end{aligned}
\end{equation}
To discover the true potential energy associated with the two-static-point-source setup in~\gefc{Deur:2016bwq}, we can substitute~\eqref{manybod} into~\eqref{finlag} with $N=2$. Integration over the Cauchy surface to remove the Dirac functions, once suitably regularised self-energies have been discarded, leads to a reduced Lagrangian over the time coordinate
\begin{align}
	\int&\mathrm{d}^3x\mathcal{L}_T=
  \frac{1}{2}\left(\tensor*{m}{^*_1}|\bm{v}_1|^2+\tensor*{m}{^*_2}|\bm{v}_2|^2\right)
  +\left[v^4~\text{corrections}\right]\nonumber\\
  &+\frac{\kappa \tensor*{m}{^*_1}\tensor*{m}{^*_2}}{8\pi|\bm{x}_1-\bm{x}_2|}\left(1-\frac{\kappa (\tensor*{m}{^*_1}+\tensor*{m}{^*_2})}{16\pi|\bm{x}_1-\bm{x}_2|}\right)
+\mathcal{O}\left(\varepsilon^3\right).
  \label{eih}
\end{align}
The post-Newtonian kinetic corrections, which we suppress, are such that variation of~\eqref{eih} with respect to $\bm{x}_1$ and $\bm{x}_2$ yields the two-body Einstein--Infeld--Hoffman (EIH) equations~\cite{1938AnMat..39...65E}. The final term in~\eqref{eih}, with its own post-Newtonian correction, is the negative of the static two-point potential that we sought.

This potential is also put forward in~\gefc{Deur:2016bwq} as evidence in favour of the scalar model discussed in~\cref{nonrelsca,nonrel}. 
The steps by which it is extracted from~\cref{deur1,pert} appear to run as follows. The field $\varphi$ is calculated up to `some' $\mathcal{O}\left(\varepsilon^2\right)$ correction by adding the $\mathcal{O}\left(\varepsilon^2\right)$ expansion of~\eqref{thoughttrain} to the $\mathcal{O}\left(\varepsilon\right)$ (i.e. Fierz--Pauli) matter current in~\eqref{deur1}, and solving the resulting field equation. The influence of the higher-order couplings in~\eqref{deur1} appears to be neglected, and it is the $\mathcal{O}\left(\varepsilon\right)$ Fierz--Pauli coupling term which is finally recycled (up to self energies) to give a statement of the potential complete with a $\mathcal{O}\left(\varepsilon^2\right)$ correction. 
As discussed in~\cref{matcou}, we lose confidence in the Fierz--Pauli coupling ${\mathcal{L}_M=-\sqrt{2}\tensor{h}{_{\mu\nu}}\tensor{\bar{T}}{^{\mu\nu}}}$ at PN orders, preferring the well-known Lagrangian formulation for relativistic point particles ${\mathcal{L}_M=-\rho^*/\tensor{u}{^0}}$ (which incorporates PN corrections covariantly). Given that the correct potential is produced in~\gefc{Deur:2016bwq}, something must therefore be compensating for the use of the Fierz--Pauli coupling. The likely culpret is the use of the first PN correction to $\varphi$ within the strict context of the GEFC scalar model~\eqref{pert}. As we have shown in~\cref{nonrel}, the scalar model does not describe gravitostatics at PN orders for reasons which have nothing to do with the proposed matter coupling. In studying the dynamical Lagrangian, we are allowed to imagine that the PN correction to $\tensor{h}{_{\mu\nu}}$ still adheres to the scalar model, but as we have just witnessed that correction \emph{cancels} in the analysis and does not contribute to the PN potential correction, which happens to just comprise squared Newtonian terms. It would appear in summary that the correct PPN potential is produced in~\gefc{Deur:2016bwq} because an even number of physically unsound steps have been introduced.

As a final observation, it may seem better to be cautious about how convincing such a result \emph{could} have been. Any PN model, adjusting for self-energy diagrams, must necessarily correct the Newtonian tadpole with an EIH-like term in~\eqref{eih}, with the only freedom being in the magnitude of that same correction.  

\section{Nothing new at second order}\label{gem2}

We have outlined in~\cref{anmod} some concerns about the specific scalar gravity model used in~\gefc{Deur:2016bwq,Deur:2014kya,Deur:2021ink,Deur:2009ya,Deur:2019kqi,Deur:2017aas}.
However our analysis is insufficient to rule out the broader \emph{principle} of GEFC.
In this section, we would therefore like to `steel-man' the GEFC proposal by discarding the scalar model, whilst still exploring the nonlinear but perturbative regime at the level of rigour just set in~\cref{smokemirror}. We will discuss speculative extensions to the strong gravity regime in~\cref{conclusions}. Our treatment in this section will also facilitate a transition to studying \emph{axisymmetric} GEFC applications, i.e. to galactic rotation curves, which we continue to discuss in~\cref{lensing}.

\subsection{Axisymmetric spacetime}\label{axisy}

 We first set up a metric, without reference to the PPN gauge, that is possibly the most general needed for a static axisymmetric system, viz:
\be
\begin{gathered}
\tensor{g}{^{00}} = \left(1+a_1\right)^2,  \quad
\tensor{g}{^{11}} = -\left(1+b_1\right)^2,  \\
\tensor{g}{^{22}} = -\left(1+d_1\right)^2,  \quad
\tensor{g}{^{33}} = -\left(1+c_1\right)^2, \\
\end{gathered}
\label{eqn:next-h-function}
\ee
where $a_1$ through $d_1$ are functions of (cylindrical) $R$ and $z$. We then set
\be
d_1 = -\frac{a_1}{1+a_1}.
\label{eqn:d1-use}
\ee
The point of this is that it aligns the implied metric with the zero-rotation case of that used by Cooperstock \& Tieu~\cite{Cooperstock:2006dt}, who say that `their metric is in the most general form necessary'. Nothing in the exact Einstein equations appears to call for this particular value of $d_1$, but on the other hand there are no obvious problems that emerge from imposing it, and since it simplifies the exact equations considerably, we use it here.

Next we carry out a $\ppn{1}$ linearisation of the exact Einstein equations implied by this metric. Specifically, we assume each of $a_1$, $b_1$ and $c_1$ and the matter density $\rho$ is $\mathcal{O}\left(\varepsilon\right)$ and then expand the Einstein equations to $\mathcal{O}\left(\varepsilon\right)$. 
Note that in contrast to~\cref{smokemirror} we will choose to work with the covariantly conserved density $\rho$.
This then implies the relations $b_1=c_1=d_1=-a_1$, together with the single remaining relation $\vect\nabla^2 a_1=4\pi G\rho+\ppn{2}$, showing us that $-a_1$ is the Newtonian potential.

The above can be recapitulated, but with the expansions taken to $\ppn{2}$ instead. 
Thus we now set up our metric as 
\be
\begin{gathered}
	\tensor{g}{^{00}}  = \left(1+ a_1^f +  a_1^s \right)^2,   \quad
	\tensor{g}{^{11}}  = \left(1- a_1^f +  b_1^s \right)^2,   \\
	\tensor{g}{^{33}}  = \left(1- a_1^f +  c_1^s \right)^2.
\end{gathered}
\label{eqn:nn-h-function}
\ee
The superscript $f$ and $s$ refer to $\ppn{1}$ (first order) and $\ppn{2}$ (second order) quantities.
Notice that we already substitute for $d_1$ in terms of $a_1$ as given in \eqref{eqn:d1-use} in the exact Einstein and geodesic equations, hence we do not need to specify an ansatz for this part of the metric. %
It is clear that the pressure only enters at $\ppn{2}$, and following from our discussion in~\cref{setupcon} we will briefly consider its presence until~\cref{beginmike} to strengthen the `steel-man' approach to GEFC.
Thus we are assuming the $\rho$ density is specified in advance, and then the pressure $P$, the `potential' $a_1$ and other quantities will be derived from it. In terms of its physical meaning, $\rho$ is the eigenvalue associated with the timelike eigenvector of the matter stress-energy tensor, and hence is physically well-defined and gauge invariant. 
If we were taking another approach to the equations, e.g.\ by assuming a given equation of state, then it would be sensible to have $\rho$'s defined at different orders of solution, but we do not need that here.

Given these choices, the Einstein equations are automatically satisfied at $\ppn{1}$, and at $\ppn{2}$ we are able to reorganise part of them into the following interesting expression:
\be
\bgrad^2 a_1^s=-12\pi\left(a_1^f \rho + P\right)+\left\lvert\bgrad a_1^f\right\rvert^2+\ppn{3}.
\label{eqn:2nd-order-pot}
\ee
This looks like a fully physical equation, and in principle enables us to find the $\ppn{2}$ contribution to the potential, $-a_1^s$, once the $\ppn{1}$ one has been found \emph{exactly} from the Poisson equation
\be
\bgrad^2 a_1^f = - 4 \pi \rho,
\label{eqn:a1f-poisson}
\ee
and also assuming that we can find the pressure (see below for more on the latter). Before continuing, it is worth quickly verifying that the spacetime solution we are constructing in~\cref{eqn:a1f-poisson,eqn:2nd-order-pot} is consistent with the PPN result in~\crefrange{ppn_pert}{ppn_pert_3}. This check can be performed by expanding $\tensor{u}{^0}$ and $\sqrt{-g}$ to find
\begin{subequations}
	\begin{gather}
		\rho=\left(1-3U\right)\rho^*+\ppn{2},\\ 
		\frac{\kappa}{8\pi}\int\frac{\mathrm{d}^3x'\rho'}{|\bm{x}-\bm{x}'|}=U-3\Phi_2+\ppn{3}.
	\end{gather}
\end{subequations}
By (temporarily) substituting $P=0$ into~\eqref{eqn:2nd-order-pot} and applying the useful identity~\eqref{coroll}, we can then recover~\eqref{ppn_pert} by inverting the metric~\eqref{eqn:nn-h-function} to $\ppn{2}$.

If we then want to find out the effect of this $\ppn{2}$ correction on the rotation curve of the galaxy, we need to be careful since the relation between the rotation curve velocity and $a_1$ may itself be modified by $\ppn{2}$ effects. Indeed, now expanding the exact geodesic equations~\eqref{ch17:eqn17.16} for massive particles to $\ppn{2}$, using the ansatz \eqref{eqn:nn-h-function} above, we get the following result for the circular velocity:
\be
\begin{aligned}
	|\bm{v}|^2 &= -R\frac{\partial}{\partial R} \left(a_1^f + a_1^s \right)\\ 
	&\ \ \ +R \frac{\partial a_1^f}{\partial R} \left( a_1^f + R\frac{\partial a_1^f}{\partial R}\right)+\ppn{3}.\label{masspar}
\end{aligned}
\ee
The first term is what we may expect, but the second is new, and would also need to be taken into account.

Finally we discuss the pressure, and whether this can successfully be found from the equations. It is easy to find the following relation from the $\ppn{2}$ equations:
\be
\frac{\partial P}{\partial R} = \rho \, \frac{\partial a_1^f}{\partial R}+\ppn{3}.
\label{eqn:P-so}
\ee
Finding the equivalent relation for the $z$ derivative of the pressure, we need to consider the $R$ and $z$ derivatives of $b_1^s$. We will not go through the details here, but it turns out that once one has fixed $d_1$ to the value in \eqref{eqn:d1-use}, then both $b_1$ derivatives are available explicitly, and we can commute on these to get a constraint. This yields 
\be
\frac{\partial P}{\partial z} = \rho \, \frac{\partial a_1^f}{\partial z}+\ppn{3},
\ee
and then the consistency relation between the these derivatives will imply the result, that
\be
\frac{\partial \rho}{\partial R}\frac{\partial a_1^f}{\partial z}=\frac{\partial \rho}{\partial z}\frac{\partial a_1^f}{\partial R}+\ppn{3},
\label{eqn:par_derivs_consist}
\ee
i.e.\ that the shape of the density distribution has to be the same as the shape of the $-a_1^f$ potential, which does not seem possible, for a realistic distribution. This will be discussed elsewhere, and may be of some relevance to the \emph{fluid ball conjecture}~\cite{2020arXiv200414240C}. In any case, the consistent derivatives would allow us to reconstruct $P$ if we wished to, hence all the elements needed for explicitly calculating the $\ppn{2}$ potential from \eqref{eqn:2nd-order-pot} are available. In general, this would have to be done via numerical evaluation of integrals, but it is of interest to see the machinery working in a completely analytic case, and so in~\cref{axisph} we calculate the $\ppn{2}$ GR correction to the Newtonian potential for a uniform density sphere. Of course, the spherical case is not expected to lead to a GEFC effect: it is the breaking of the spherical symmetry which allows the collapse process, and this is supposed to be indicated in~\gefc{Deur:2013baa} by the correlation between the assumed size of the dark matter halo and the optical ellipticity of the host galaxy. 
By testing the result \eqref{eqn:2nd-order-pot} for the spherical case (for which an exact solution is known), we can show that it does legitimately \emph{correct} the Newtonian approximation, and we illustrate this in~\cref{fig:a1-comp}. We then remember that the formula~\eqref{eqn:2nd-order-pot} should also be valid for a general axisymmetric situation, and notice that there is \emph{no hint} in this expression that cases with extreme flattening will lead to anything special. 

Direct application of~\eqref{eqn:2nd-order-pot} to the axisymmetric and flattened case is of course also possible, but quite involved.
In order to render the analysis tractable, and to reconnect with the chromoelectric analogy in~\cref{introduction}, we will instead address the flattened case in~\cref{secgem} using a heuristic nonlinear extension of the GEM formalism, which we now develop in~\crefrange{beginmike}{newgauge}.

\subsection{Gravitoelectromagnetism}\label{beginmike}

Our initial axisymmetric analysis in~\cref{axisy,axisph} does not suggest GEFC phenomena, but nor is it grounded in any systematic perturbation scheme for \emph{general} spacetimes: for this we might look to the PPN formalism introduced in~\cref{smokemirror}. Alternatively, the GEM formalism would seem to be naturally suited to the GEFC hypothesis. Since GEFC is a proposed graviton self-coupling effect, we may imagine a nonlinear extension of GEM in which the gravitoelectric charge (mass-energy) is augmented by a contribution from the gravitoelectric field strength density.
In keeping with the `steel-man' directive, we therefore now transition to the GEM formalism in the hope that a hidden GEFC effect will become apparent.

GEM provides a useful, notionally-familiar
description of linearised general relativity (GR), by drawing a close
analogy with classical electromagnetism (EM). We will limit our
discussion to non-relativistic stationary matter sources, for which
one may obtain GEM field equations \emph{and} a GEM `Lorentz' force law
that are fully consistent and have forms precisely analogous to their
counterparts in EM, which is not possible for more general
time-dependent scenarios.  In particular, these assumptions regarding
the matter source are appropriate for modelling rotation curves in
galaxies. In the standard approach to such modelling, one assumes the
more restrictive static, Newtonian limit for the matter source, in
which a test particle is subject only to the gravitoelectric force
derived from the usual gravitational potential produced by the
galactic density distribution. This usual approach fails to predict
the flat rotation curves observed in many galaxies in terms of their
visible matter distribution, as discussed in~\cref{introduction}.

The GEM formalism for linear GR with a stationary, non-relativistic
source is based on the simple ansatz of relabelling the components of
$\bar{h}^{\mu\nu}$ as\footnote{Conventions in the literature vary up
  to a multiplicative constant for the definition of the
  gravitomagnetic vector potential $A^\alpha$. These factors variously
  modify the analogues of the EM field equations and the Lorentz force
  law, with no scaling choice allowing all the GEM and EM equations to
  be perfectly analogous. Here, we follow the conventon used in \cite{Hobson:2006se}.}
\be
\begin{gathered}
\bar{h}^{00}=4\Phi+\mathcal{O}\left(\varepsilon^2\right),\qquad \bar{h}^{0\alpha} =
A^\alpha+\mathcal{O}\left(\varepsilon^{5/2}\right),
	\\ 
	\bar{h}^{\alpha\beta}={\cal O}\left(\varepsilon^2\right),
\label{ch17:eqn17.47}
\end{gathered}
\ee
where we have defined the gravitational scalar potential $\Phi$ and
spatial gravitomagnetic vector potential $A^\alpha$. On lowering indices,
the corresponding components of $h_{\mu\nu}$ are $h_{00}=
h_{11}=h_{22}=h_{33}=2\Phi+\mathcal{O}\left(\varepsilon^2\right)$ and $h_{0\alpha}=A_\alpha+\mathcal{O}\left(\varepsilon^{5/2}\right)$.
Thus, the linear GEM potentials in~\eqref{ch17:eqn17.47} can be \emph{approximately} defined in terms of the PPN potentials in~\eqref{ppnpot}
\begin{equation}
	\Phi\equiv -U+\mathcal{  O}\left(\varepsilon^2\right), \quad \tensor{A}{^\alpha}\equiv -4\tensor{V}{^\alpha}+\mathcal{  O}\left(\varepsilon^{5/2}\right).
  \label{GEMtoPPN}
\end{equation}
Just as we resurrected $P$ within~\cref{axisy,axisph}, we see from~\eqref{GEMtoPPN} and~\eqref{ppnpot} that GEM allows us to resurrect the fluid velocity -- though in making a fair comparison to the GEFC proposal we will suppress this velocity again in~\cref{secgem}.
It should be remembered that raising or lowering a spatial (Roman)
index introduces a minus sign with our adopted metric signature. Thus
the numerical value of $A_\alpha$ is minus that of $A^\alpha$, the latter being
the $\alpha$th component of the spatial vector $\vect{A}$.  It is also
worth noting that both $\Phi$ and $A_\alpha$ are dimensionless,
thereby yielding dimensionless components $h_{\mu\nu}$, which is
consistent with our choice of coordinates $\left[x^\mu\right] = (t,x^\alpha)$ having
dimensions of length. 

Indeed, reverting for the moment to the viewpoint in which
$g_{\mu\nu}=\eta_{\mu\nu}+h_{\mu\nu}$ defines the metric of a
(slightly) curved spacetime, one may write the line element in the
limit of a stationary, non-relativistic source in quasi-Minkowski
coordinates as
\be
  \mathrm{d}s^2 = \left(1+2\Phi\right)\mathrm{d}t^2 -2\bm{A}\cdot \mathrm{d}t \mathrm{d}\bm{x}
- \left(1-2\Phi\right)
\mathrm{d}\bm{x}^2
	+\mathcal{O}\left(\varepsilon^2\right).
\label{ch17:eqn17.51}
\ee
Determining the geodesics of this line element provides a
straightforward means of calculating the trajectories of test
particles in the weak gravitational field of a stationary,
non-relativistic source. In particular, one need not
assume that the test particles are slowly moving, and so the
trajectories of photons may also be found by determining
the null geodesics of the line element (\ref{ch17:eqn17.51}).

In~\cref{smokemirror} we chose to work exactly with the density $\rho^*$, which satisfies the Euclidean conservation law $\tensor{\partial}{_\mu}\left(\rho^*\tensor{u}{^\mu}/\tensor{u}{^0}\right)=0$. In~\cref{axisy,axisph} we chose to work instead with the physical density $\rho\equiv\rho^*/\sqrt{-g}\tensor{u}{^0}$, which appears in the perfect fluid stress-energy tensor~\eqref{masterset} and which satisfies the covariant conservation law $\tensor{\vect\nabla}{_\mu}\left(\rho\tensor{u}{^\mu}\right)=0$. Within this section and~\cref{newgauge,secgem}, we will further complicate matters slightly (for later convenience) by introducing the density
\begin{equation}
\begin{aligned}
	\rho^\dagger\equiv\tensor{T}{_{00}}&=\rho^*\left(1-5U\right)+\ppn{3}\\
	&=\rho\left(1-2U\right)+\ppn{3}.
	\label{densx}
\end{aligned}
\end{equation}
With the identifications \cref{ch17:eqn17.47,densx}, we may choose to write the linearised field
equations (\ref{ch17:eqn17.13}) in the Lorenz gauge for a stationary,
non-relativistic source \emph{exactly} in the scalar/vector form
\be
\vect\nabla^2 \Phi  \equiv \frac{\kappa}{2} \rho^\dagger, 
 \quad
\vect\nabla^2 \vect{A}  \equiv 2\kappa \vect{j}^\dagger,
\label{ch17:eqn17.66}
\ee
where we have defined the momentum density (or matter current density)
$\vect{j}^\dagger\equiv \rho^\dagger\vect{v}$, and the (linear) Lorenz gauge condition
$\partial_\rho \bar{h}^{\mu\rho}=0$ itself becomes
$\bg{\nabla}\cdot\vect{A}=0$. Moreover, the general solutions to the
equations (\ref{ch17:eqn17.66}) may be read off directly from
(\ref{ch17:eqn17.46}), (\ref{emtensor}) and (\ref{ch17:eqn17.47}) to yield~\eqref{ppnpot}.
%
%
Clearly, the first equations in (\ref{ch17:eqn17.66}) and
(\ref{ppnpot}) recover, respectively, the Poisson equation and
its solution for the gravitational potential, familiar from Newtonian
gravity, whereas the second pair of equations determine the
gravitomagnetic vector potential that describes the `extra' (weak)
gravitational field predicted in linearised GR, which is produced by
the {\it motion} of the fluid elements in a stationary, non-relativistic source.

One may take the analogy between linearised GR and EM further by
defining the {\it gravitoelectric} and {\it gravitomagnetic} fields
$\vect{E} \equiv -\vect{\nabla}\Phi$ and $\vect{B} \equiv
\vect{\nabla}\times\vect{A}$.  Using the
equations~(\ref{ch17:eqn17.66}), it is straightforward to verify that
the fields $\vect{E}$ and $\vect{B}$ satisfy the {\it
  gravitational Maxwell equations}
  \be
  \begin{gathered}
  \vect{\nabla} \cdot \vect{E} = -\frac{\kappa}{2}\rho^\dagger, 
  \quad 
  \vect{\nabla} \cdot \vect{B} = 0, \\
\vect{\nabla} \times\vect{E} = \vect{0},  
\quad
\vect{\nabla}\times\vect{B}  = -2\kappa \vect{j}^\dagger.
\end{gathered}
\ee
The gravitoelectric field $\vect{E}$ describes the standard
(Newtonian) gravitational field produced by a static mass
distribution, whereas the gravitomagnetic field $\vect{B}$ is the
`extra' gravitational field produced by {\it moving} fluid elements in
the stationary, non-relativistic source.

The equation of motion for a test particle is the geodesic equation
(\ref{ch17:eqn17.16}) in the metric (\ref{ch17:eqn17.51}), from which
one may determine the trajectories of either massive particles,
irrespective of their speed, or massless particles, by considering
timelike or null geodesics, respectively.  In line with the PPN assumptions set out in~\cref{setupcon}, we will assume here,
however, that the test particle is massive and {\it slow-moving},
i.e. its speed $v$ is sufficiently small that we may neglect terms in
$v^2$ and higher. Hence we may take $\gamma_v \equiv \left(1-v^2\right)^{-1/2}
\approx 1$, so that the 4-velocity of the particle may be written
$u^\mu \equiv \gamma_v(1,\vect{v}) \approx (1,\vect{v})$.  This immediately
implies that $\ddot{x}^\sigma=0$ and, moreover, that $\mathrm{d}t/\mathrm{d}s = 1$,
so one may consider only the spatial components of
(\ref{ch17:eqn17.16}) and replace dots with derivatives with respect
to $t$. Expanding the summation in (\ref{ch17:eqn17.16}) into terms
containing respectively two time components, one time and one spatial
component, and two spatial components, neglectng the purely
spatial terms since their ratio with respect to the purely temporal
term is of order $v^2$, expanding the connection coefficients
to first-order in $h_{\mu\nu}$ and remembering that for a stationary field
$\partial_0 h_{\mu\nu}=0$, one obtains
\be
\begin{aligned}
	\frac{\mathrm{d}v^\alpha}{\mathrm{d}t} &=  -
\frac{1}{2} \delta^{\alpha\beta}\partial_\beta h_{00}-
\delta^{\alpha\gamma}\left(\partial_\gamma h_{0\beta}-\partial_\beta h_{0\gamma}\right)v^\beta
\\
	&\ \ \ 
+\mathcal{O}\left(\varepsilon^2\right).
\end{aligned}
\ee
Recalling that one inherits a minus sign on raising or lower a spatial
(Roman) index, this equation of motion may be rewritten in vector form
in terms of GEM fields as
\be
\frac{\mathrm{d}\vect{v}}{\mathrm{d}t} = -\vect{\nabla}\Phi +
\vect{v}\times\left(\vect{\nabla}\times\vect{A}\right) = 
\vect{E} +
\vect{v}\times\vect{B}
+\mathcal{O}\left(\varepsilon^2\right)
,
\label{gemeom}
\ee
which recovers the {\it gravitational Lorentz force law}
for slow-moving massive particles in the gravitational field of a stationary
non-relativistic source. The first term on the right-hand side gives
the standard Newtonian result for the motion of a test particle in
the field of a static non-relativistic source, whereas the second
term gives the `extra' force felt
by a {\it moving} test particle in the presence of the `extra' field
produced by {\it moving} fluid elements in the stationary non-relativistic
source.

\subsection{Second-order general relativity}\label{newgauge}

As mentioned in~\cref{introduction},~\gefc{Deur:2020wlg} has proposed a separate
approach to using GR to model galaxy rotations curves without dark
matter, which neglects gravitomagnetic forces entirely but instead
includes the effects of graviton self-interaction. To include this
effect, at least to leading order in the self-interaction, one must
consider second-order GR. We again closely follow the approach used
in \cite{Hobson:2006se}.

In this approach, one again assumes a weak gravitational field, but now expands the
Einstein equations~\eqref{efe} to second-order in $h_{\mu\nu}$ to yield
$G^{(1)}_{\mu\nu} + G^{(2)}_{\mu\nu} = \kappa T_{\mu\nu}$, where the
second-order Einstein tensor is given by
\be
G^{(2)}_{\mu\nu} = R^{(2)}_{\mu\nu}-\frac{1}{2}\eta_{\mu\nu}
R^{(2)} -\frac{1}{2}h_{\mu\nu}R^{(1)} + \frac{1}{2} \eta_{\mu\nu}
h^{\rho\sigma}R^{(1)}_{\rho\sigma},
\label{ch17:eqn17.56}
\ee
where $R^{(2)}_{\mu\nu}$ denotes the terms in the Ricci tensor that
are second order in $h_{\mu\nu}$, and $R^{(1)}$ and $R^{(2)}$ denote
the terms in the Ricci scalar that are first and second order in
$h_{\mu\nu}$, respectively. One may show, however, that, unlike
$G^{(1)}_{\mu\nu}$, the quantity $G^{(2)}_{\mu\nu}$ is {\em not}
invariant under the gauge transformation (\ref{ch17:eqn17.5}).  Before
addressing this shortcoming, it is useful to perform a trivial
rearrangement of the second-order field equations to yield
$G^{(1)}_{\mu\nu} = \kappa\left(T_{\mu\nu} + t_{\mu\nu}\right)$, where we have
defined $t_{\mu\nu} \equiv -\kappa^{-1}G^{(2)}_{\mu\nu}$, which may
then be interpreted as the energy-momentum of the linearised
gravitational field to leading order in the field
self-interaction. This interpretation prompts one to take
seriously the fact that the energy--momentum of a gravitational field
at a point in spacetime has no real meaning in GR, since at any
particular event one can always transform to a free-falling frame in
which gravitational effects disappear. 

A convenient opportunity to distance oneself from the gauge-dependent nature of gravitational energy arises when one is concerned only with the corrective back-reaction to spacetimes dominated at $\mathcal{O}( \tensor{h}{_{\mu\nu}} )$ by \emph{gravitational radiation}~\cite{landau}.
One can in such cases, at each point in spacetime, average $G^{(2)}_{\mu\nu}$ at a `mezoscale' 
granularity (i.e. between the back-reaction and wavenumber scales) in order to probe the physical curvature of the spacetime,
which gives a gauge-invariant measure of the gravitational field
strength. Denoting this averaging process by $\langle\cdot\rangle$,
the second-order field equations should then read 
$G^{(1)}_{\mu\nu} + \left\langle G^{(2)}_{\mu\nu}\right\rangle = \kappa
T_{\mu\nu}$, or equivalently $G^{(1)}_{\mu\nu} = \kappa\left(T_{\mu\nu} +
\left\langle t_{\mu\nu}\right\rangle\right)$.

Of course, GEFC is not proposed to be radiative in origin; but in order to provide measurable phenomena it \emph{must} be gauge-invariant. For the remainder of this section and in~\cref{secgem} therefore, we will employ the radiative average $\langle\cdot\rangle$ to arrive at heuristic proxies for second-order gravitoelectrostatic corrections of the kind apparently implicated in GEFC. In this way will be able to correct certain model Newtonian galactic rotation curves by means of analytically tractable integrals, and confirm explicitly that such corrections are of no astrophysical significance. 
In particular, this $\ppn{2}$ approach will not be tied in any way to axisymmetry, as was the case with our earlier attempt in~\cref{axisy,axisph}.
Corrections obtained in this way will not be faithful to GR, but they introduce `radiative' errors which are no greater than $\ppn{2}$, and which are therefore too small to conceal the claimed GEFC phenomena. In summary, we are asserting that `\emph{perturbative calculations give perturbative results}' -- a tautology which we are forced to explore directly since it does not appear to be adequately addressed in~\gefc{Deur:2013baa,Deur:2014kya,Deur:2021ink,Deur:2009ya,Deur:2020wlg,Deur:2019kqi,Deur:2017aas,Deur:2022ooc}. In moving forward, we nonetheless take care in~\cref{bipartite} to track the error introduced by the radiative average\footnote{It is
    worth noting that although the radiative average is usually
    envisaged as being taken over some small spacetime `patch' at each
    point, the formalism does not require this interpretation. To
    fulfil its practical usefulness in calculations, it is necessary
    only for the averaging to allow one to assume that the first
    derivatives of any function of spacetime position vanish (at least
    on scales smaller than the averaging scale). For spacetimes with
    particular symmetries, one may thus equally well average over
    larger regions that contain the Killing congruences of the
    spacetime. For example, in a static, spherically-symmetric
    spacetime, one may average over a thin spherical shell at each
    radius. Similarly, in a stationary, axisymmetric spacetime that is
    symmetric about the centre-plane $z=0$ (which is a reasonable
    approximation for galactic systems), each averaging region may
    have the form two `halo'-shaped tubes of small cross-section
    centred on the coordinate curves $R=R_0$, $\upvarphi=\upvarphi_0$
    and $z = \pm z_0$. In both cases, in Cartesian Lorentz coordinates
    the first derivatives of any spacetime function of position will
    average to zero over such regions, as required, and one also is
    prevented from adopting any coordinate system that constitutes a
    free-fall frame over the whole of such a region, thereby yielding
    gauge-invariant results.}.

For now let us assume that the solution to the second-order field equations
has the form $h_{\mu\nu} = \ell_{\mu\nu} + \delta h_{\mu\nu}$,
where $\ell_{\mu\nu}$ is the solution to the first-order (linear) field
equations $G^{(1)}_{\mu\nu} = \kappa T_{\mu\nu}$ and $|\delta
h_{\mu\nu}| \ll |\ell_{\mu\nu}|$ is a small perturbation to it. 
We will assume as described above that $\tensor{\ell}{_{\mu\nu}}$ is `susceptable' to radiative averaging, without itself being radiative, and that this operation has some physical justification.
Since $G^{(1)}_{\mu\nu}$ is linear in
$h_{\mu\nu}$, one may write $G^{(1)}_{\mu\nu}(h) =
G^{(1)}_{\mu\nu}(\ell)+ G^{(1)}_{\mu\nu}(\delta h)$, where the
function arguments are merely a shorthand for the various
gravitational field pseudotensors, rather than denoting their
traces. Since $G^{(2)}_{\mu\nu}$ is non-linear (quadratic) in
$h_{\mu\nu}$, one instead has
\be
\begin{aligned}
	G^{(2)}_{\mu\nu}(h) &= G^{(2)}_{\mu\nu}(\ell) 
+ \left.\pd{G^{(2)}_{\mu\nu}}{h_{\rho\sigma}}\right|_\ell \delta
h_{\rho\sigma} +\mathcal{O}\left(\varepsilon^4\right)
\\
&
= G^{(2)}_{\mu\nu}(\ell) 
+\mathcal{O}\left(\varepsilon^3\right).
\end{aligned}
\ee
Adopting a `mean-field' approach, one ignores the final term on the RHS, and so the second-order field
equations may be written symbolically as
\be
G^{(1)}_{\mu\nu}(\delta h) + \left\langle G^{(2)}_{\mu\nu} (\ell) \right\rangle
=\mathcal{O}\left(\varepsilon^3\right),
\label{master}
\ee
where the two terms on the LHS are of order ${\cal O}(\delta h)$ and
${\cal O}(\ell^2)$, respectively, and hence both second-order small.
Equation (\ref{master}) thus determines the correction $\delta
h_{\mu\nu}$ to the solution $\ell_{\mu\nu}$ of the linearised GR field
equations that occurs due to the leading-order graviton
self-interaction.

It now remains only to determine the form of $\left\langle G^{(2)}_{\mu\nu}
(\ell) \right\rangle$; again our calculation closely follows that in \cite{Hobson:2006se}.
First, since $\ell_{\mu\nu}$ is the solution to the linearised GR
field equations, one may express the last two terms on the right-hand
side of (\ref{ch17:eqn17.56}), which depend on the first-order Ricci
tensor and Ricci scalar, in terms of the matter energy-momentum
tensor as
\be
\begin{aligned}
G^{(2)}_{\mu\nu}(\ell) &= R^{(2)}_{\mu\nu}(\ell)-\frac{1}{2}\eta_{\mu\nu}
R^{(2)}(\ell) 
\\
&\ \ \ 
+\frac{1}{2}\kappa \left(\bar{\ell}_{\mu\nu}T+
\eta_{\mu\nu}\ell^{\rho\sigma}T_{\rho\sigma}\right).
\label{ch17:eqn17.59}
\end{aligned}
\ee
One then requires only an expression for
$R^{(2)}_{\mu\nu}(\ell)$, from which $R^{(2)}(\ell)$ can be found by
contraction. Expanding connection coefficients to second-order in
$\ell_{\mu\nu}$, substituting them into the usual expression for the
Ricci tensor and keep only those terms quadratic in $\ell_{\mu\nu}$,
one finds
\be
\begin{aligned}
R^{(2)}_{\mu\nu}&(\ell)  = \frac{1}{4} \partial_\mu
\ell^{\rho\sigma}\partial_\nu \ell_{\rho\sigma} -
\frac{1}{2}\ell^{\rho\sigma}\big(\partial_\mu\partial_\sigma\ell_{\nu\rho}
\\
&\ \ \ 
  +\partial_\nu\partial_\sigma \ell_{\mu\rho}-
\partial_\mu\partial_\nu \ell_{\rho\sigma}-
\partial_\rho\partial_\sigma \ell_{\mu\nu}\big)  
\\
&\ 
-\frac{1}{2}\partial^\sigma \ell^\rho_\nu\big(\partial_\rho
\ell_{\sigma\mu} -\partial_\sigma \ell_{\rho\mu}\big)
\\
&
-\frac{1}{2}\left(\partial_\sigma \ell^{\rho\sigma} 
 -\frac{1}{2}\partial^\rho \ell\right)\left(\partial_\mu \ell_{\nu\rho} +
\partial_\nu \ell_{\mu\rho} - \partial_\rho \ell_{\mu\nu}\right).
\label{ch17:eqn17.61}
\end{aligned}
\ee
Although the third group of terms on the right-hand side is not
manifestly symmetric in $\mu$ and $\nu$, this symmetry is easy to
verify. In fact, in subsequent calculations it is convenient to
maintain manifest symmetry by writing out this term again with
$\mu$ and $\nu$ reversed and multiplying both terms by one-half.

To evaluate the averaged expression $\left\langle R^{(2)}_{\mu\nu}
\right\rangle$, one merely notes that first derivatives average to zero. Thus, for
any function of spacetime position $a(x)$, one has $\langle
\partial_\mu a \rangle = 0$. This has the important consequence that
$\langle \partial_\mu (ab) \rangle = \langle (\partial_\mu a)b \rangle
+ \langle a(\partial_\mu b) \rangle = 0$, and hence we may swap
derivatives in products and inherit only a minus sign, i.e.  $\langle
(\partial_\mu a)b \rangle = - \langle a(\partial_\mu b) \rangle$. One
first makes use of this result to rewrite products of first
derivatives in (\ref{ch17:eqn17.61}) in terms of second
derivatives. Using the first-order field equations to substitute for
terms of the form $\dalembertian \ell_{\mu\nu}$, and then applying the
averaging result once more to rewrite terms containing second
derivatives as products of first derivatives, one finally obtains
\be
\begin{aligned}
\left\langle R^{(2)}_{\mu\nu}(\ell) \right\rangle & =  -\frac{1}{4}
\Big \langle \partial_\mu \ell_{\rho\sigma}\partial_\nu
\ell^{\rho\sigma} -2\partial_\sigma \ell^{\rho\sigma}\partial_{(\mu}
\ell_{\nu)\rho} 
\\
+2&\partial_\rho \ell \partial_{(\mu} \ell^\rho_{\nu)} -
\partial_\mu \ell\partial_\nu \ell   
\\
+\kappa &\left(2\ell_{\mu\nu} T + 2 \ell T_{\mu\nu}-
\eta_{\mu\nu} \ell T - 4 \ell_{\rho(\mu}T^\rho_{\nu)} \right) 
\Big
\rangle.
\label{ch17:eqn17.62}
\end{aligned}
\ee
Contracting this expression, and once again making use of the
averaging result and the first-order field equations, one quickly
finds that $\left\langle R^{(2)}(\ell) \right\rangle = \frac{1}{2} \kappa
\left\langle \ell^{\rho\sigma} T_{\rho\sigma}\right\rangle$. Combining these
expressions and writing the result (mostly) in terms of the
trace reverse field, one thus finds
\be
\begin{aligned}
\langle G^{(2)}_{\mu\nu}(\ell)\rangle & =  -\frac{1}{4}\Big \langle\vphantom{\frac{a}{2}} (\partial_\mu
\bar{\ell}_{\rho\sigma})\partial_\nu \bar{\ell}^{\rho\sigma} -
2(\partial_\sigma
\bar{\ell}^{\rho\sigma})\partial_{(\mu}\bar{\ell}_{\nu)\rho} 
\\
-
\frac{1}{2}&(\partial_\mu \bar{\ell})\partial_\nu \bar{\ell} -\, \kappa \left ( 4\bar{\ell}_{\rho(\mu}T^\rho_{\nu)} +
\eta_{\mu\nu} \ell^{\rho\sigma} T_{\rho\sigma} \right )
\Big\rangle.
\label{ch17:eqn17.64}
\end{aligned}
\ee
It may be verified by direct substitution that this expression is
indeed invariant under the gauge transformation
(\ref{ch17:eqn17.5}) (with $h_{\mu\nu}$ replaced by $\ell_{\mu\nu}$), as required. 

One may then substitute (\ref{ch17:eqn17.64}) and the expression for
$G^{(1)}_{\mu\nu}$ in the linearised GR field equations
(\ref{ch17:eqn17.11}) (with $\bar{h}_{\mu\nu}$ replaced by
$\delta\bar{h}_{\mu\nu}$) into (\ref{master}) to obtain an equation
for the trace-reversed correction $\delta \bar{h}_{\mu\nu}$ in an
arbitrary gauge. Since both terms in (\ref{master}) are separately
invariant to the gauge transformation (\ref{ch17:eqn17.5}) (with
$h_{\mu\nu}$ replaced by $\delta h_{\mu\nu}$ or $\ell_{\mu\nu}$,
respectively), however, one can impose the separate Lorenz gauge
conditions $\partial_\rho \delta\bar{h}^{\mu\rho}=0$ and
$\partial_\rho \bar{\ell}^{\mu\rho}=0$, which yields
\be
\begin{aligned}
 \dalembertian\delta\bar{h}_{\mu\nu} & = \frac{1}{2}\Big\langle \kappa \left ( 4\bar{\ell}_{\rho(\mu}T^\rho_{\nu)} +
\eta_{\mu\nu} \ell^{\rho\sigma} T_{\rho\sigma} \right )
\\
&\ \ \ 
-\partial_\mu
\bar{\ell}_{\rho\sigma}\partial_\nu \bar{\ell}^{\rho\sigma} 
+\frac{1}{2}\partial_\mu \bar{\ell}\partial_\nu \bar{\ell}
\Big\rangle
+\mathcal{O}\left(\varepsilon^3\right)
	.
\label{masterfinal}
\end{aligned}
\ee

\subsection{Second-order gravitoelectrostatics}\label{secgem}

Equipped with the apparatus from~\cref{newgauge}, we here develop a GEM formalism for {\em second-order} GR, thereby including
the leading-order graviton self-interactions while avoiding the heuristic approach of~\gefc{Deur:2020wlg} which considers the lensing of
gravitoelectric field lines. We will consider the question of lensing separately in~\cref{lensing}.

We will again confine our attention to stationary, non-relativistic
matter sources. By analogy with the GEM ansatz (\ref{ch17:eqn17.47}),
in which we now replace $h_{\mu\nu}$ with $\ell_{\mu\nu}$, one may
make a corresponding identification for the corrections $\delta
h_{\mu\nu}$, such that
\be
\delta\bar{h}^{00}=4\delta\Phi,\qquad \delta\bar{h}^{0\alpha} =
\delta A^\alpha, \qquad \delta\bar{h}^{\alpha\beta}={\cal O}\left(\varepsilon^2\right).
\label{GEM2defs}
\ee
and we again approximate the energy-momentum tensor of a stationary,
non-relativistic source using (\ref{emtensor}).

In this paper we do not consider the general case, which includes matter currents and
the resulting gravitomagnetic field, rather we make contact with the GEFC 
approach by considering the more restrictive case of a static
matter source, for which one instead assumes the space-time components
of the matter energy-momentum tensor to vanish, $T^{\alpha 0} =  0$. In
this case, there exists only the gravitoelectric field derived from
the Newtonian potential and one must strictly also assume the fluid to support
pressure in order to establish an equilibrium configuration for the
galaxy. Following GEFC and our discussion in~\cref{setupcon}, however, we will ignore this pressure
contribution in determining the correction to the Newtonian potential
resulting from leading-order graviton self-interactions.
In this simplified case, one need consider only the $00$-component of
the general result (\ref{masterfinal}) in the absence of any
time-dependence or source motions. Thus, with no sum on $\alpha$, one has that
$\ell_{00} = \ell_{\alpha\alpha} = 2\Phi$, $\bar{\ell}_{00}=4\Phi$,
$\bar{\ell}_{\alpha\alpha} = 0$, $T_{00} = \rho^\dagger $, $T_{\alpha\alpha} = 0$ and time
derivatives $\partial_0$ of any quantity vanish, such that $\dalembertian =
-\vect\nabla^2$. This yields the remarkably simple result
\be
\vect\nabla^2\delta\Phi = -\frac{9\kappa}{4}\Phi\rho^\dagger
+\mathcal{O}\left(\varepsilon^3\right),
\label{electromaster}
\ee
which is the principal fruit of the radiative averaging process.
Thus, in principle, for any specified density distribution $\rho^\dagger$, one
need only determine the Newtonian gravitational potential $\Phi$ using
the first equation in (\ref{ppnpot}) and then subtitute this
result into (\ref{electromaster}), to which, by analogy, the solution
is given by
\be
\delta\Phi = \frac{9\kappa}{16\pi} \int
\frac{\mathrm{d}^3x'\Phi'{\rho^\dagger}'}
{|\vect{x}-
{\vect{x}}'|}
+\mathcal{O}\left(\varepsilon^3\right)
=
-\frac{9}{2}\Phi_2
+\mathcal{O}\left(\varepsilon^3\right),
\label{deltaphisol}
\ee
where we compare again with~\eqref{ppnpot}.
The resulting $\ppn{2}$ solution for the gravitational potential is
then simply $\Phi + \delta\Phi$. Let us pause for a moment to reconnect with the PPN formulation in~\cref{smokemirror}. The correction~\eqref{deltaphisol} and definitions~\eqref{GEMtoPPN} imply
\begin{equation}
  \tensor{\bar h}{_{00}}=-2U-\frac{9}{2}\Phi_2+\mathcal{O}\left(\varepsilon^3\right),
  \label{mikeppn}
\end{equation}
which could be consistent with~\eqref{ppn_pert}, but turns out not to be when the trace $\delta\bar h$ is also calculated from~\eqref{masterfinal}. This discrepancy might arise because~\eqref{ppn_pert} encodes the standard PPN gauge, whilst the gauge choices made en route to~\eqref{masterfinal} are only linearly equivalent to the harmonic gauge and are not, to our knowledge, used beyond this work. As shown in~\cite{will_2018_2}, the $\mathcal{  O}\left(\varepsilon^3\right)$ corrections to $\tensor{h}{_{00}}$ do not differ between the standard PPN and harmonic coordinate gauges when pressures and accelerations are neglected. In fact, we show in~\cref{bipartite} that it is the radiative averaging procedure, rather than the gauge choice, which causes the deviation from PPN. 

Moving forward, as a first example we connect with~\cref{axisy} by calculating in~\cref{axisph} the $\ppn{2}$ gravitational
potential of a sphere of uniform density.
As a result of the radiative average discussed in~\cref{newgauge}, our result in~\eqref{cerdis} is not required to be strictly faithful to the exact results~\cref{axisphr1,axisphr2} which follow from the treatment in~\cref{axisy}, but we are satisfied that the corrections are comparable in magnitude.
An example system for which one may even more straightforwardly derive an
analytical result is two static point particles -- as considered already in~\cref{smokemirror} -- for which
the density is given by~\eqref{manybod} with $N=2$, and for which $\tensor*{m}{^\dagger_n}$ may be solved for in terms of $\tensor*{m}{^*_1}$ via~\eqref{densx}. Subsituting $\rho^\dagger$
into the integral solution in the first equation in
(\ref{ch17:eqn17.66}), one immediately obtains the well-known result
\be \Phi = -\frac{G\tensor*{m}{^\dagger_1}}{|\vect{x}-\vect{x_1}|} -
\frac{G\tensor*{m}{^\dagger_1}}{|\vect{x}-\vect{x}_2|}.  
\ee
Substituting this expression and that for $\rho^\dagger$ into
(\ref{deltaphisol}), and ignoring the infinite self-energy terms, then
gives
\be
\begin{aligned}
	\delta\Phi &= -\frac{9G^2}{2}\frac{\tensor*{m}{^\dagger_1}\tensor*{m}{^\dagger_2}}{|\vect{x}_1-\vect{x}_2|}\left(\frac{1}{|\vect{x}-\vect{x_1}|}+\frac{1}{|\vect{x}-\vect{x}_2|}\right)
\\
&
+\mathcal{  O}\left(\varepsilon^3\right).
\label{cerdo}
\end{aligned}
\ee
The above analysis may be easily extended to an arbitary number $N$ of
point particles. As with~\eqref{cerdis}, we do not really expect a precise agreement between~\eqref{cerdo} and an equivalent exact formula following from our considerations in~\cref{smokemirror} --- once again however, the corrections are of a comparable magnitude.

For modelling galaxy rotation curves while retaining some analytical
simplicity, however, one must consider axisymmetric density
distributions of the kind introduced already in~\cref{axisy}. In this case, the integral solution in the first
equation in (\ref{ppnpot}) may be written in cylindrical polar coordinates with azimuthal symmetry:
\be
\begin{aligned}
\Phi(R,z) = 
-2G \int_0^\infty \mathrm{d}R' \int_{-\infty}^\infty &\mathrm{d}z'\,\rho^\dagger(R',z') 
\\
&
\times R'\sqrt{\frac{m}{RR'}}K(m),
\label{phisolagain}
\end{aligned}
\ee
where $K(m)$ is a complete elliptic integral function of the first
kind and $m \equiv 4RR'/\left[(R+R')^2 + (z-z')^2\right]$. 
Moreover, the derivatives $\partial\Phi/\partial R$ and $\partial\Phi/\partial z$
may also be expressed analytically as
\begin{subequations}
\begin{align}
  \pd{\Phi}{R} & = G\int_0^\infty \mathrm{d}R' \int_{-\infty}^\infty \mathrm{d}z'\,
\rho^\dagger(R',z')
\nonumber
\\
\times&\frac{R'}{R}\sqrt{\frac{m}{RR'}}\left[ K(m) +
  \frac{1}{2}\left(\frac{R}{R'}-\frac{2-m}{m}
  \right)\frac{mE(m)}{1-m}\right], \label{dphidr} 
  \\ 
  \pd{\Phi}{z} & = \frac{G}{2}\int_0^\infty \mathrm{d}R' \int_{-\infty}^\infty \mathrm{d}z'\,
\rho^\dagger(R',z')
\nonumber
\\
& \ \ \ \ \ \ \ \ \ \ \ \ \ \ \ \ \ \ \ \
\times\left(\frac{z-z'}{R}\right)\sqrt{\frac{m}{RR'}}\frac{mE(m)}{1-m},\label{dphidz}
\end{align}
\end{subequations}
where $E(m)$ denotes a complete elliptic integral of the second kind.

By analogy, the
$\ppn{2}$ correction (\ref{deltaphisol}) may immediately be written
as
\be
\begin{aligned}
\delta\Phi(R,z) = 
	9G \int_0^\infty \mathrm{d}R'& \int_{-\infty}^\infty  \mathrm{d}z'\,\Phi(R',z')\,\rho^\dagger(R',z') 
\\
&
\times R'\sqrt{\frac{m}{RR'}}K(m)+\mathcal{  O}\left(\varepsilon^3\right).
\label{deltaphisolaxi}
\end{aligned}
\ee

Some analytical density-potential pair solutions to (\ref{phisolagain})
exist~\cite{ramsey,binney}, most notably for uniform density spheroids (plus some
non-axisymmetric examples, such as a uniform density triaxial
ellipsoid \cite{chand}). In principle, one could model a galaxy using a very
flattened uniform density prolate spheroid to obtain an analytical
expression for $\Phi(R,z)$, or perhaps the closely related
Miyamoto--Nagai density distribution \cite{Miyamoto:1975zz} employed by Ludwig~\cite{Ludwig:2021kea} -- we will consider this distribution further in~\cref{lensing}, in the context of intragalactic lensing. The
resulting expression for $\Phi(R,z)$ may then be substituted into
(\ref{deltaphisolaxi}) to obtain $\delta\Phi(R,z)$, but no analytical
solution exists in this latter case, even if the density is
uniform. Thus, there seems no alternative to evaluating
(\ref{deltaphisolaxi}) numerically.  Since none of the analytical
density-pair solutions to (\ref{phisolagain}) are a particularly good
approximation to a real galaxy, one might instead consider an
alternative form for the specified density distribution that is more
appropriate, but in that case one must perform {\em both} integrals
(\ref{phisolagain}) and (\ref{deltaphisolaxi}) numerically, where the
integrand of the latter is itself described only numerically as output
from the former. Either approach is reasonable depending on the
required level of approximation of realistic galactic density
profiles.

In fact, since we are most interested here in galaxy rotation curves
it is useful before performing any numerical integrations first to
derive a direct expression for $|\bm{v}(R,z)|^2 =
R\partial\Phi_T(R,z)/\partial R$, where we have defined the total
gravitational potential up to $\ppn{2}$ as $\Phi_T(R,z) \equiv \Phi(R,z) +
\delta\Phi(R,z)$ and we follow the approach in~\gefc{Deur:2020wlg} of neglecting the second term in the $\ppn{2}$ massive particle equations of motion~\eqref{masspar}.  Following Ludwig in~\cite{Ludwig:2021kea}, we note that observations of the
rotation velocity are typically measured along the galactic equatorial
plane, so one may consider merely $|\bm{v}(R,0)|^2 =
R\partial\Phi_T(R,0)/\partial R+\mathcal{  O}\left(\varepsilon^3\right)$. In particular, from (\ref{dphidr}),
one has
\be 
\begin{aligned}
	|\bm v(R,0)|^2  =  G&\int_0^\infty \mathrm{d}R' \int_{-\infty}^\infty  \mathrm{d}z'\,
\left[1-\frac{9\Phi(R',z')}{2}\right]
\\
&
\times \rho^\dagger(R',z')\,F(R,R',z')
+\mathcal{  O}\left(\varepsilon^3\right)
	, 
\label{dphitotdr}
\end{aligned}
\ee
where for convenience we have defined the function
\be
\begin{aligned}
F(R,R',z') \equiv R'&\sqrt{\frac{n}{RR'}}\bigg[ K(n) 
\\
&
  +
  \frac{1}{2}\left(\frac{R}{R'}-\frac{2-n}{n}
  \right)\frac{nE(n)}{1-n}\bigg],
\end{aligned}
\ee
in which $n \equiv 4RR'/\left[(R+R')^2 + z^{\prime 2}\right]$.
In principle, one may obtain the rotation curve $|\bm{v}(R,0)|$ in the
equatorial plane of the galaxy for any specified density distribution
$\rho^\dagger(R,z)$ by (numerically) evaluating the double integral
(\ref{dphitotdr}), where $\Phi(R,z)$ in the integrand is itself given
by (\ref{phisolagain}).

Again following Ludwig in~\cite{Ludwig:2021kea}, one may avoid the evaluation of double integrals by
analytically approximating the integrals over $z'$ in
(\ref{phisolagain}) and (\ref{dphitotdr}) under the assumption of a
vertically symmetric galactic density distribution and a thin-disc
approximation of the form 
\be
\rho^\dagger(R,z) \approx \rho^\dagger(R,0)\exp\left(-\frac{z^2}{2\Delta^2(R)}\right),
\label{thindiscagain}
\ee
where $\Delta(R)$ is a radially-dependent characteristic disc
width with some assumed form, and we necessarily lose strict contact with the order of $\varepsilon$, transitioning to the notation $(\approx)$. For small values of $\Delta(R)$, one can
estimate integrals over $z'$ analytically using the Laplace
approximation, which amounts to setting $z'=0$ in the integrand and
multiplying by the volume $\sqrt{2\pi}\Delta(R)$ of the Gaussian
factor in (\ref{thindiscagain}); this yields
\be
\begin{aligned}
|\bm v(R,0)|^2 \approx  \sqrt{2\pi}GR\int_0^\infty&
\left[1-\frac{9\Phi(R',0)}{2}\right]\rho^\dagger(R',0)\,
\\
&
\times \Delta(R')\,F(R,R',0)\,\mathrm{d}R'.
\label{vsqresult}
\end{aligned}
\ee
where the function $F(R,R',0)$ may be written in the simplified form
\be
\begin{aligned}
F(R,R',0) = 
\frac{2R'}{R+R'}\bigg[&K\left(\frac{4RR'}{(R+R')^2}\right)
  \\
  &
  + \frac{R+R'}{R-R'}E\left(\frac{4RR'}{(R+R')^2}\right)\bigg],
\end{aligned}
\ee
and $\Phi(R,0)$ in the integrand of (\ref{vsqresult}) is itself given by
\be
\begin{aligned}
\Phi(R,0) \approx
-4\sqrt{2\pi}G \int_0^\infty & \frac{R' \rho^\dagger(R',0)
  \Delta(R')}{R+R'}
\\
&
\times K\left(\frac{4RR'}{(R+R')^2}\right)\,\mathrm{d}R'.
\label{equiphi}
\end{aligned}
\ee

It is worth noting that~\gefc{Deur:2020wlg} also assumes a vertically symmetric thin
disc approximation for the galactic density distribution, but
of a slightly different form to that in (\ref{thindiscagain}). In
particular,~\gefc{Deur:2020wlg} assumes the fully separable distribution
\be
\rho^\dagger(R,z) = \rho^\dagger_0\exp\left(-\frac{R}{h_R}\right)\exp\left(-\frac{|z|}{h_z}\right),
\label{deurdensity}
\ee
where $\rho^\dagger_0 \equiv \rho^\dagger(0,0)$ and both the radial and vertical
factors are exponentials characterised by the constant scale lengths
$h_R$ and $h_z$, respectively. One may adopt an approach analogous to
the Laplace approximation used above, whereby one sets ${z'=0}$ in the
integrand in (\ref{dphitotdr}) but now multiplies by the volume $2h_z$
of the exponential $z$-dependent factor in (\ref{deurdensity}). In
this case, the expressions (\ref{vsqresult}) and (\ref{equiphi}) are
replaced by
\begin{subequations}
\begin{align}
|\bm v(R,0)|^2 & \approx  2h_zGR\rho^\dagger_0\int_0^\infty
\left[1-\frac{9\Phi(R',0)}{2}\right]\nonumber
\\
&\ \ \ \ \ \ \ \ \   
\times \exp\left(-\frac{R'}{h_R}\right)\,F(R,R',0)\,\mathrm{d}R',\label{deurv2}
\\
\Phi(R,0) & \approx 
-8h_zG\rho^\dagger_0 \int_0^\infty \frac{R'}{R+R'}\exp\left(-\frac{R'}{h_R}\right)\nonumber
\\
&\ \ \ \ \ \ \ \ \   
\times K\left(\frac{4RR'}{(R+R')^2}\right)\,\mathrm{d}R'.\label{deurphi}
\end{align}
\end{subequations}

\begin{figure}
\centerline{\includegraphics[width=\linewidth]{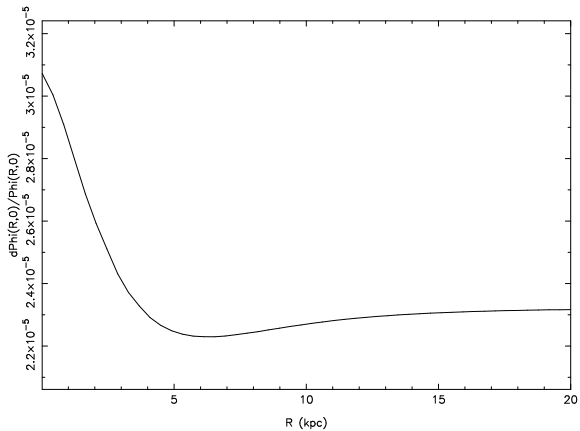}}
\caption{The fractional correction $\delta\Phi(R,0)/\Phi(R,0)$ to the
  Newtonian potential as a function of galactic radius (in kpc), which
  arises from the leading-order graviton self-interaction, for a
  galaxy having the density distribution (\ref{deurdensity}) with
  total (baryonic) mass $M^\dagger = 4\pi h_z h^2_R \rho^\dagger_0 = 3\times
  10^{11}$~M$_\odot$, radial scale length $h_R = 1.5$~kpc and vertical
  scale length $h_z = 0.03 h_R$.
\label{fig:dphionphi}}
\end{figure}

\begin{figure}
\centerline{\includegraphics[width=\linewidth]{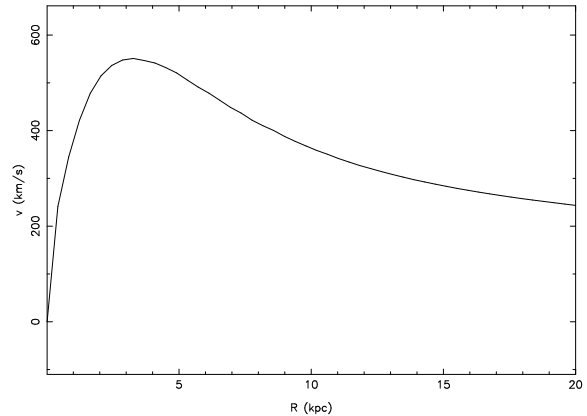}}
\caption{The orbital velocity (km s$^{-1}$) versus galactic radius
  (kpc) for a galaxy having the density distribution
  (\ref{deurdensity}) with total (baryonic) mass $M^\dagger = 4\pi h_z h^2_R
  \rho^\dagger_0 = 3\times 10^{11}$~M$_\odot$, radial scale length $h_R =
  1.5$~kpc and vertical scale length $h_z = 0.03 h_R$.
\label{fig:rotcurve}}
\end{figure}

It is of interest to compare the results obtained from the above
equations with those of~\gefc{Deur:2020wlg}; that work instead uses an approach
based on gravitational lensing, which we analyse separately in~\cref{lensing}. One could work directly with the
equations (\ref{deurv2}) and (\ref{deurphi}), but the integral in
(\ref{deurv2}) is numerically challenging to perform directly. As a
first step, however, one may evaluate more straightforwardly the
integral (\ref{deurphi}) and the corresponding expression for
$\delta\Phi(R,0)$, which is given by
\be
\begin{aligned}
\delta\Phi(R,0) & \approx 
36h_zG\rho^\dagger_0  \int_0^\infty \frac{R'}{R+R'}\,\Phi(R',0)\,
\\
&
\times\exp\left(-\frac{R'}{h_R}\right)
K\left(\frac{4RR'}{(R+R')^2}\right)\,\mathrm{d}R'.\label{deurdeltaphi}
\end{aligned}
\ee
The integrands in (\ref{deurphi}) and (\ref{deurdeltaphi}) each
contain an integrable singularity at $R'=R$, where the argument of $K$
becomes unity; this singularity occurs because the Green's function
must reproduce a delta function at this point. The
singularity is easily accommodated by breaking the integral into two
parts and using a standard one-sided open quadrature formula on either
side of the singularity. Following~\gefc{Deur:2020wlg}, we consider a galaxy
having the density distribution (\ref{deurdensity}) with total
(baryonic) mass $M^\dagger \equiv 4\pi h_z h^2_R \rho^\dagger_0 = 3\times
10^{11}$~M$_\odot$, radial scale length $h_R = 1.5$~kpc and vertical
scale length $h_z = 0.03 h_R$.  It is worth noting that our use of the
Laplace method to approximate integrals over $z'$ analytically means
that the expressions (\ref{deurphi}) and (\ref{deurdeltaphi}) are
independent of the value of $h_z$, and depend only on $M^\dagger$ and
$h_R$. Since the Laplace method is valid only in the thin-disc
approximation $h_z \ll h_R$, we expect this to be reasonably accurate
for our choice of $h_z = 0.03 h_R$. In Figure~\ref{fig:dphionphi}, we
plot the fractional correction $\delta\Phi(R,0)/\Phi(R,0)$ to the Newtonian
potential as a function of galactic radius (in kpc) that 
arises from the leading-order graviton self-interaction, as obtained
by performing the integrals
(\ref{deurphi}) and (\ref{deurdeltaphi}) numerically.
As one can see from the figure, $\delta\Phi(R,0)/\Phi(R,0) \sim {\cal
  O}\left(10^{-5}\right)$ over the entire range in galactic radius. Such a small
correction will lead to a similarly small fractional correction to the
orbital velocity $|\bm v(R,0)|$ and so we conclude that the leading-order
graviton self-interaction has a negligible effect on galaxy rotation
curves. This is in stark contrast to the findings in~\gefc{Deur:2020wlg}, derived
using a gravitational lensing approach. It will be of interest to
determine how the GEFC calculation leads to such a different conclusion, and we will explore this issue further in~\cref{postmortem}.
As a check of the numerical calculation, it is straightforward to
calculate the resulting rotation curve, which is plotted in
Figure~\ref{fig:rotcurve} and agrees well with that of~\gefc{Deur:2020wlg} in the
absence of the supposed self-interaction correction.
We reiterate in closing that the explicit rotation curve corrections obtained within this section are \emph{proxies} for the genuine $\ppn{2}$ effects implied by GR. However, as we anticipated in~\cref{newgauge}, nothing in the analysis suggests that the GEFC phenomena are preferentially hiding in the physics which is thrown out by radiative averageing. An attempt to obtain the genuine $\ppn{2}$ rotation curve would likely yield similar, uninteresting results, but without enjoying the simple perscription in~\eqref{electromaster}: if this claim is to be refuted, the calculation should be performed.

By this point we hope it is apparent that such calculations are not necessary. Particularly, the adjustments to the rotation curve introduced by the baryon profile clearly dwarf the nonlinear phenomena; our analysis makes it clear that one may rearrange the $\ppn{2}$ effects entirely by demanding that a given profile be represented by $\rho$, $\rho^*$ or $\rho^\dagger$ --- galactic baryon distributions, even if they can be measured, are not likely to be consistent to such precision across galaxies with flat and rising rotation curves~\cite{Li:2020iib}.

\section{Intragalactic lensing}\label{lensing}

In~\cref{gem2} we attempted to `steel-man' the case for GEFC near galactic discs, by discarding the GEFC scalar graviton of~\gefc{Deur:2016bwq,Deur:2014kya,Deur:2021ink,Deur:2009ya,Deur:2019kqi,Deur:2017aas} and studying the $\ppn{2}$ phenomena in the context of actual GR. However the main claims of~\gefc{Deur:2020wlg}, i.e. the most substantial exploration of galactic disc GEFC, are not directly based on either of these methods.
The methods used there instead concern the lensing of light rays in galaxies. These are meant to show how gravitational field lines are distorted in such a way that the (cylindrical) radial gravitational force near the edge of the galaxy declines like $1/R$ rather than the Newtonian $1/R^2$.

The key claim made in~\gefc{Deur:2020wlg} is that if one calculates the geodesic paths of photons emitted radially from the nucleus of an axisymmetric disc galaxy, then those emitted close to the disc are deflected such that they end up moving {\em parallel} to the disc by the time the edge of the galaxy is reached. These photon paths are meant to model gravitational field lines, meaning that the spreading of the field lines is just in one rather than two dimensions, leading to the $1/R$ force dependency.
We therefore devote this final section to exploring this claim, before concluding in~\cref{conclusions}.

\subsection{Exact lensing within the linearised background}

The calculations of the photon/graviton paths in~\gefc{Deur:2020wlg} are carried out by using the small angle deflection formula 
\be
\delta \upbeta = \frac{4G\mtot}{h},
\ee
for a `field line' passing near a point mass $\mtot $ with an impact parameter $h$, and with (in this case polar) angle of the ray tangent $\upbeta$. This deflection is then integrated along paths using a mass distribution model in which the galaxy is decomposed into slices in the form of concentric rings. We will return to consider these calculations later, but in the meantime we note that clearly, since the~\gefc{Deur:2020wlg} deflection effects are just added up along a path assuming this simple formula, it will be sufficient for our own comparison purposes to carry out the calculations for a photon path in a linearised gravitational background for an axisymmetric system. If we do the photon path calculations {\em exactly} in this linearised background, this must certainly capture any effects which the~\gefc{Deur:2020wlg} approach is able to capture. The non-linearity which is proposed in~\gefc{Deur:2020wlg} to be responsible for the rotation curve effects, would then come about from the gravitational field lines suffering distortion as they propagate within this background.

The way in which we carry out the lensing calculations merits some explanation in terms of the effective metric used. In the first instance, we employ our axisymmetric formulation from~\cref{axisy}. Working with~\eqref{eqn:next-h-function} and~\eqref{ch17:eqn17.16}, we derive the \emph{exact} equations for particle and photon motion. Thus this is {\em exact} lensing within what has the possibility, at least, of being an exact setup for a general static axisymmetric system.
We then insert $a_1$, $b_1$ etc.\ into the lensing equations, expand these to $\mathcal{O}\left(\varepsilon\right)$, and insert the results just given in~\cref{axisy} for the values of $a_1$ through $d_1$ into these equations, in order to calculate the lensing deflection for a given $a_1$. At the relevant order, clearly the equivalent metric which we can think of as giving rise to the lensing, is therefore
\begin{equation}
  \begin{aligned}
  \mathrm{d}s^2&=\left( 1-2a_1 \right)\mathrm{d}t^2\\
  &\ \ \ -\left( 1+2a_1 \right)\left( \mathrm{d}R^2+R^2\mathrm{d}\upvarphi^2+\mathrm{d}z^2 \right),
	\label{skipar}
  \end{aligned}
\end{equation}
in cylindrical polar coordinates.
We have concluded in~\cref{anmod} that exact static spacetimes corresponding to~\eqref{skipar} cannot be used, but the ansatz is nonetheless consistent with the line elements~\cref{pertan,linel}, or~\cref{ppn_pert,ppn_pert_2} to $\mathcal{  O}\left( \varepsilon^2 \right)$.
The coefficient $a_1$ is meant to be small, and to be a function of $R$ and $z$. The physical setup envisaged is a static axisymmetric mass distribution, with no pressure or rotation (i.e. $P=\bm v=0$). As discussed in~\cref{setupcon} this would of course collapse ordinarily, but we assume that the static distribution is possible since we are treating the density as small, of the same order as $a_1$, and that the pressure that would be needed for support in the absence of rotation (which~\gefc{Deur:2020wlg} assumes), would come in at the next order in both quantities.

We would like to use a density profile for the galaxy that is continuously differentiable so that there are no possible issues with lack of analycity in the calculations. Also we would like the density profile to result in an explicit analytic expression for the Newtonian potential, so that when we integrate the photon path numerically, we can evaluate the equations of motion for the photon without having to perform a numerical integral in order to get the potential and its gradients at the position where the photon is. To do this we will use a Miyamoto--Nagai (MN) profile for the density and potential. In this approach to density profiles~\cite{Miyamoto:1975zz}, one uses a fairly simple potential distribution given by
\be
a_1=\frac{GM}{\sqrt{R^2+\left(a+\sqrt{b^2+z^2}\right)^2}}.
\label{eqn:MN-potential}
\ee
Here $a$ and $b$ are characteristic scales in the $R$ and $z$ directions. Note that within~\cref{lensing} we will be using a system of units where length is measured in kiloparsecs, appropriate to galactic scales. 

The density profile which is implied by the Poisson equation~\eqref{ident} is then
\be
\begin{aligned}
\rho&(R, z)=\frac{M b^{2}}{4 \pi} 
\\
\times&\frac{a R^{2}+\left(a+3 \sqrt{b^{2}+z^{2}}\right)\left(a+\sqrt{b^{2}+z^{2}}\right)^{2}}{\left[R^{2}+\left(a+\sqrt{b^{2}+z^{2}}\right)^{2}\right]^{5 / 2}\left(b^{2}+z^{2}\right)^{3 / 2}},
\label{eqn:MN-density}
\end{aligned}
\ee
which agrees with the $\rho(R,z)$ given in~\cite{Ludwig:2021kea}. Recall that when integrated over space (without the determinant of the imposed spatial metric), the $\rho$ defined here yields the overall `mass' $M$ used in the potential.

Our choices here differ from those made in~\gefc{Deur:2020wlg}, which uses a density distribution which is the product of exponentials in the $R$ and $z$ directions. This leads to a cusp in density, and therefore a lack of analycity along the galactic plane. Furthermore, for a thick exponential disc we are likely to need a remaining integral to be done numerically in order to get the potential and its derivatives at a given point, whereas, as we have seen, the Miyamoto--Nagai density/potential pair are both simple analytic expressions. We will return below to any differences with the current analysis this causes, but we will aim to make the example galaxy we use as much like the one used in~\gefc{Deur:2020wlg} as possible in terms of its overall properties, such as mass, and the typical scales in the $R$ and $z$ directions.

\subsection{Gravitational lensing calculations}\label{intermex1}

As described above, we are going to carry out the calculations in the case where the gravitational fields are treated at $\ppn{1}$, and the `non-linearity' is brought in by considering the lensing of light rays, which act as a proxy for `gravitational field lines'.

So the aim is that we send out light rays from the origin and see how much they bend before heading off to infinity. Reverting to the Cartesian coordinates, we parameterise the photon momentum $\tensor{p}{^\mu}$ (with energy $E=p$) as
\begin{equation}
  \begin{gathered}
    \tensor{p}{^0}=p, \quad \tensor{p}{^1}=p\cos\upalpha \cos\upbeta,  \\
    \tensor{p}{^2}=-p\sin\upalpha\cos\upbeta, \quad \tensor{p}{^3}=p\sin\upbeta.
  \end{gathered}
\end{equation}
Then treating the lensing exactly, but within the linearised gravitational fields, one can demonstrate the following general results for motion in the $(R,z)$ plane, where $\upalpha=\upvarphi=0$. Introducing the affine parameter $\lam$ along the photon path, in place of the interval $s$
\be
\begin{aligned}
  \frac{\mathrm{d}\upbeta}{\mathrm{d}\lambda}&=2p \left(-\sin\upbeta\deriv{a_1}{R}+\cos\upbeta\deriv{a_1}{z}\right),\\
  \frac{\mathrm{d}p}{\mathrm{d}\lambda}&=p^2 \left(\cos\upbeta\deriv{a_1}{R}+\sin\upbeta\deriv{a_1}{z}\right),\\
  \frac{\mathrm{d}R}{\mathrm{d}\lambda}&=p\cos\upbeta, \quad
  \frac{\mathrm{d}z}{\mathrm{d}\lambda}=p\sin\upbeta, \quad
  \frac{\mathrm{d}\upvartheta}{\mathrm{d}\lambda}=\frac{p\cos(\upbeta+\upvartheta)}{\sqrt{R^2+z^2}},
\end{aligned}
\label{eqn:dots-on-path}
\ee
where $\upvartheta$ is the conventional polar angle of spherical coordinates. Note that $\upbeta$ and $\upvartheta$ are defined in opposite `senses', and respectively parameterise position and deflection, but are both polar in nature.
We first carry out a numerical evolution of these equations, and then seek to find an analytic approximation to the results, to help with understanding their physical meaning. 
We will do this using the Miyamoto--Nagai potential for $a_1$, since here there are no discontinuities in derivatives, no infinitesimal mass sheet in the $z=0$ plane, and everything concerning the potential and density distributions themselves is analytic.

\begin{figure}
\begin{center}
\includegraphics[width=\linewidth]{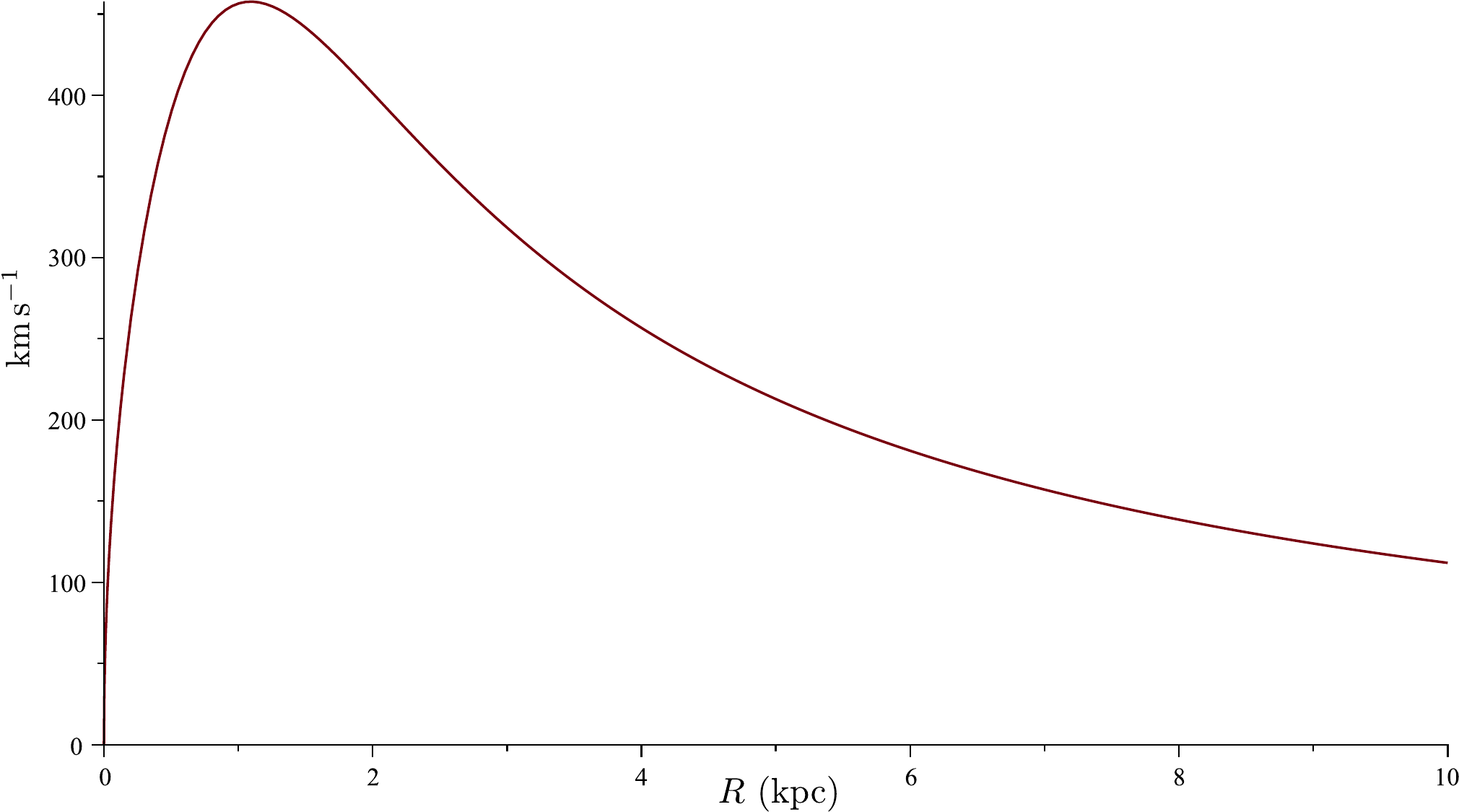}
\caption{Rotation curve, in $\kms$, for the example galaxy.\label{fig:deur-example-gal-rc}}
\end{center}
\end{figure}

\begin{figure*}
\begin{center}
\includegraphics[width=\textwidth]{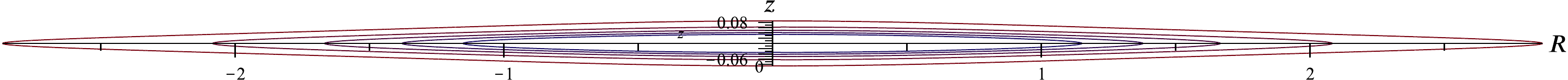}
	\caption{Isodensity contours for the example galaxy discussed in the text. This has a Miyamoto--Nagai profile with $M=3\times 10^{11} \msun$, $a=1.5 \kpc$ and $b=0.045\kpc$. The outer contour is 1/10\textsuperscript{th} of the central density. \label{fig:deur-example-gal-dens-conts}}
\end{center}
\end{figure*}

For the galaxy parameters, we will choose values yielding a similar ellipticity and overall dimensions as used in~\gefc{Deur:2020wlg}, but with them relating to the MN potential and density distribution, rather than one which has a product of exponentials in $R$ and $z$ for the density. So our values are $M=3\times 10^{11} \msun$, $a=1.5 \kpc$ and $b=0.045\kpc$. The rotation curve we would get for such a galaxy is shown in \cref{fig:deur-example-gal-rc}.
It can be seen that such a galaxy produces high rotation velocities, over $400 \kms$ at the peak. Nevertheless, it is not completely unreasonable, and we will use it as the example for our tests. The contours for density are shown in
\cref{fig:deur-example-gal-dens-conts}.
and we can see that the $a:b$ ratio of $~33$ has given a highly flattened galaxy. Again, this seems alright for a test, however, since it is best to look for effects in an object which stands the best chance of yielding something interesting, whilst not being impossible.

\begin{figure}
\begin{center}
\includegraphics[width=\linewidth]{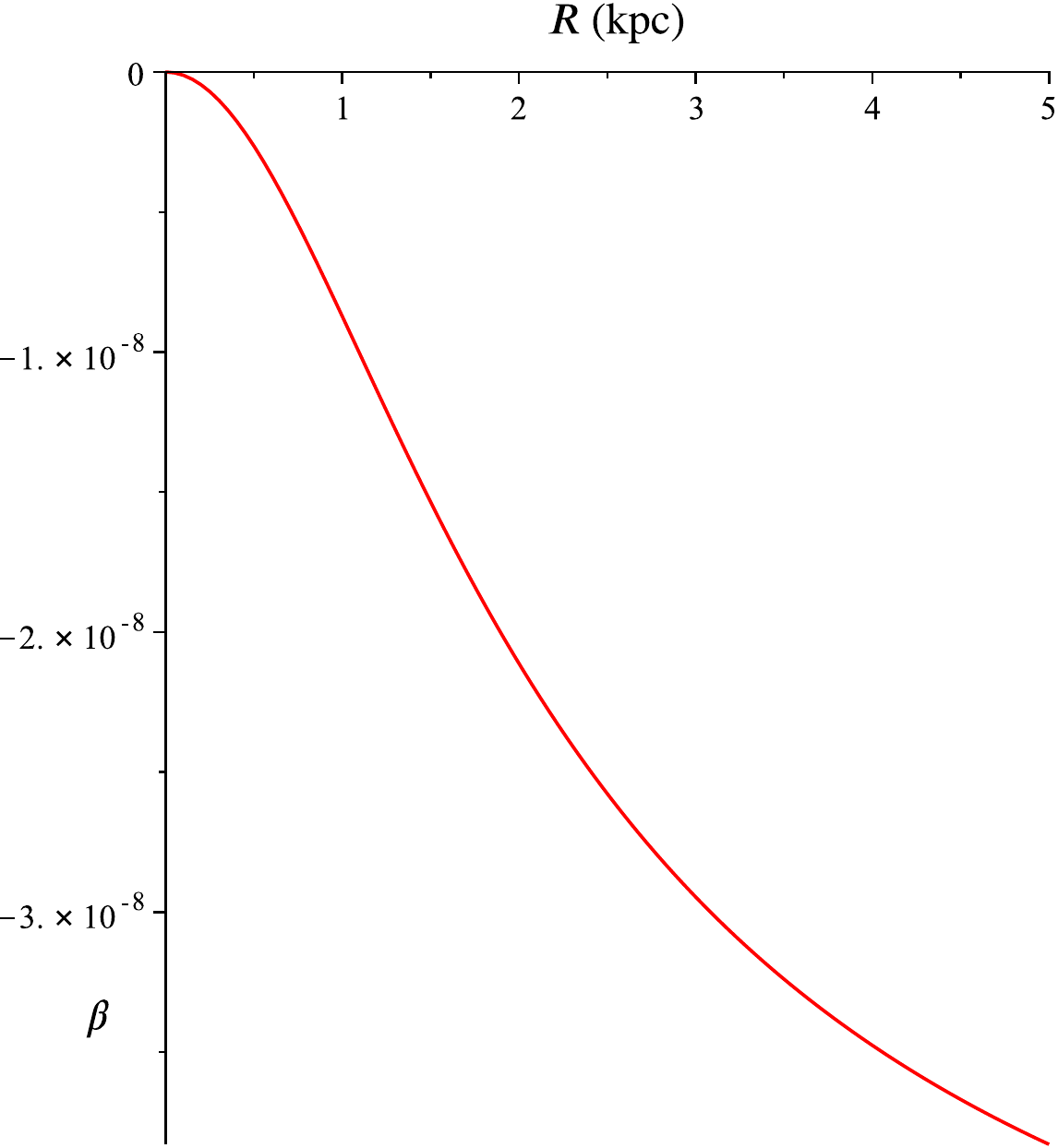}
\caption{Change in the $\upbeta$ angle of emitted photon as a function of affine parameter, in an exact numerical calculation. The units of the $\upbeta$ (vertical) axis are arcsec.\label{fig:my-exact-lensing-calcs-beta-change}}
\end{center}
\end{figure}

\cref{fig:my-exact-lensing-calcs-beta-change}
shows the change in the inclination angle $\upbeta$ when the photon is launched with a starting inclination (to the $x$-axis) of 18 arcsec. The vertical scale is in arcsec and shows that the `flattening' is by just $0.008''$ for this case. In terms of the trajectory, this is completely imperceptible, and we do not show the trajectory in the $(x,z)$ plane for this example, since it looks just like a radial straight line.

To explore the parameter space of this very small effect, and seek to find if we can get much larger values of the deflection, it makes sense to attempt to get an analytical approximation to the numerical results. We will then at least know the dependencies of the deflection on quantities such as the mass and two `principle radii' of the galaxy.

We can do this by inserting the MN results~\eqref{eqn:MN-potential} for the $a_1$ derivatives into the expression for $\mathrm{d}\upbeta/\mathrm{d}\lambda$ in equation \eqref{eqn:dots-on-path}, and then expanding in small quantities. Note we assume that both the deflection {\em and} the initial angle to the $x$-axis are small --- this matches the type of case we were looking at just now in the full numerical integration. Integrating along an affine path length $\lambda$ and assuming an initial angle of $\upbeta_0$ we find the following expression for the deflection in $\upbeta$:
\begin{equation}
\Delta \upbeta=-\frac{2 M \upbeta_0 a\left(\sqrt{a^{2}+2 a b+b^{2}+\lambda^{2}}-a-b\right)}{(a+b) b \sqrt{a^{2}+2 a b+b^{2}+\lambda^{2}}}.
\label{eqn:beta-deflection}
\end{equation}
To assess the quality of the approximation, we can can evaluate this expression for the same parameters as used to create \cref{fig:my-exact-lensing-calcs-beta-change}. In fact we do not show a fresh plot for this case, since the expression just given matches the full numerical result over the whole range better than the eye can discern the difference.

We should also get approximations for the $R$ and $z$ coordinates of the photon, since in principle it is these that measure the trajectory and from which we should derive the deflection (though we would expect this to match what we get from the momentum angle $\upbeta$ very closely in these cases). For $z$ we get the interesting expression
\begin{equation}
  \begin{aligned}
  z(\lambda)&=\upbeta_{0} \lambda-\frac{2a\upbeta_0 M}{(a+b)b}
\\
  \times&\Bigg[\ln\left(\frac{(a+b)}{\lambda+\sqrt{a^{2}+2 a b+b^{2}+\lambda^{2}}}\right)(a+b)+\lambda\Bigg].
\label{eqn:z-interesting}
\end{aligned}
\end{equation}

\begin{figure}
\begin{center}
\includegraphics[width=\linewidth]{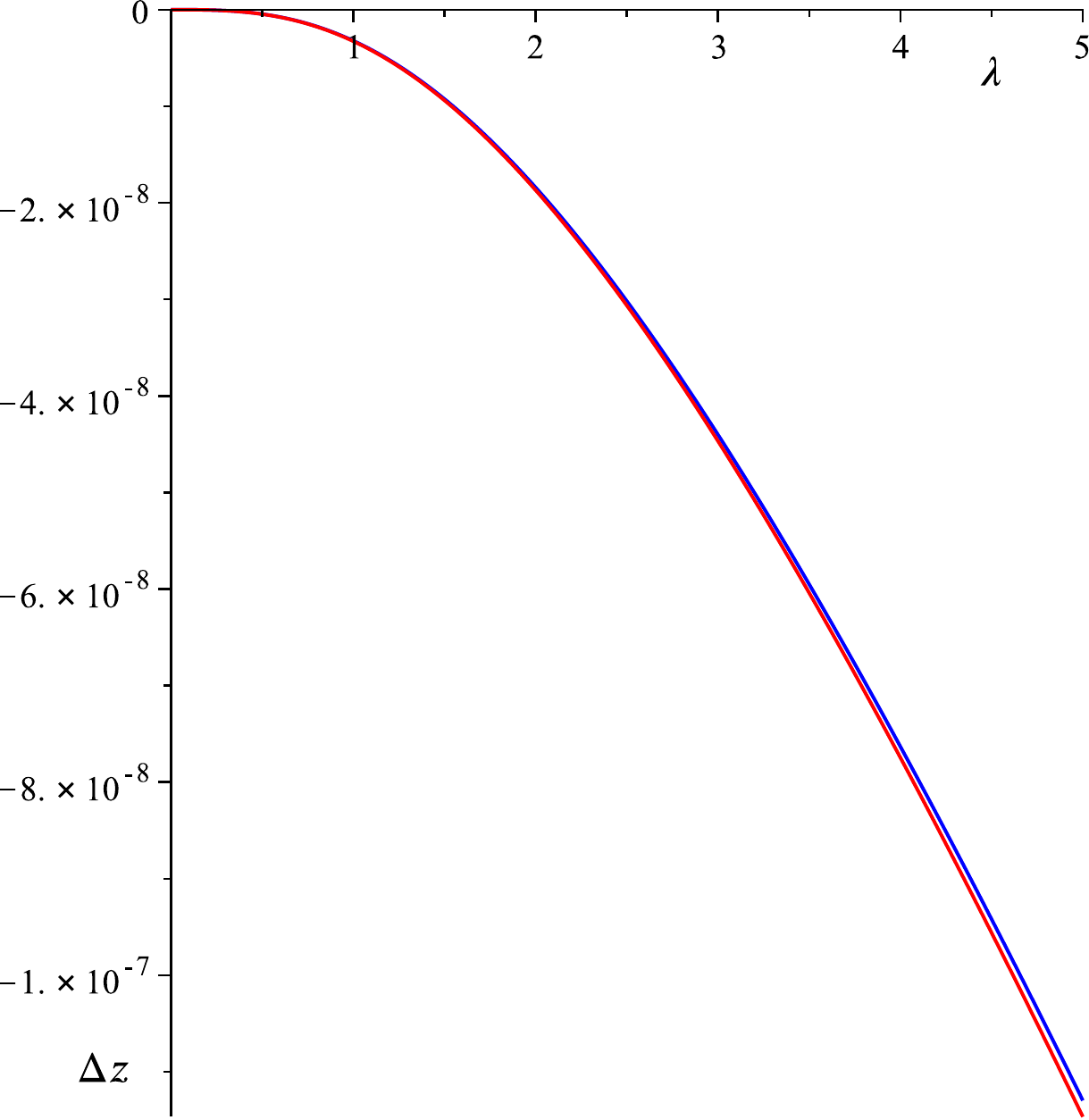}
\caption{Comparison of $z$ taken from the exact numerical integration (red) with the approximation given in equation \eqref{eqn:z-interesting} (blue), where $\lambda$ is the affine parameter for the photon path, covering the interval 0 to 5. Note that the undisturbed trajectory $z=\lambda\upbeta_0$ has been subtracted from each curve so that the residuals can be compared.\label{fig:z-deur-lensing}}
\end{center}
\end{figure}

We plot this against the exact numerical answer, for the same case as in \cref{fig:my-exact-lensing-calcs-beta-change}, in \cref{fig:z-deur-lensing}.

\begin{figure}
\begin{center}
\includegraphics[width=\linewidth]{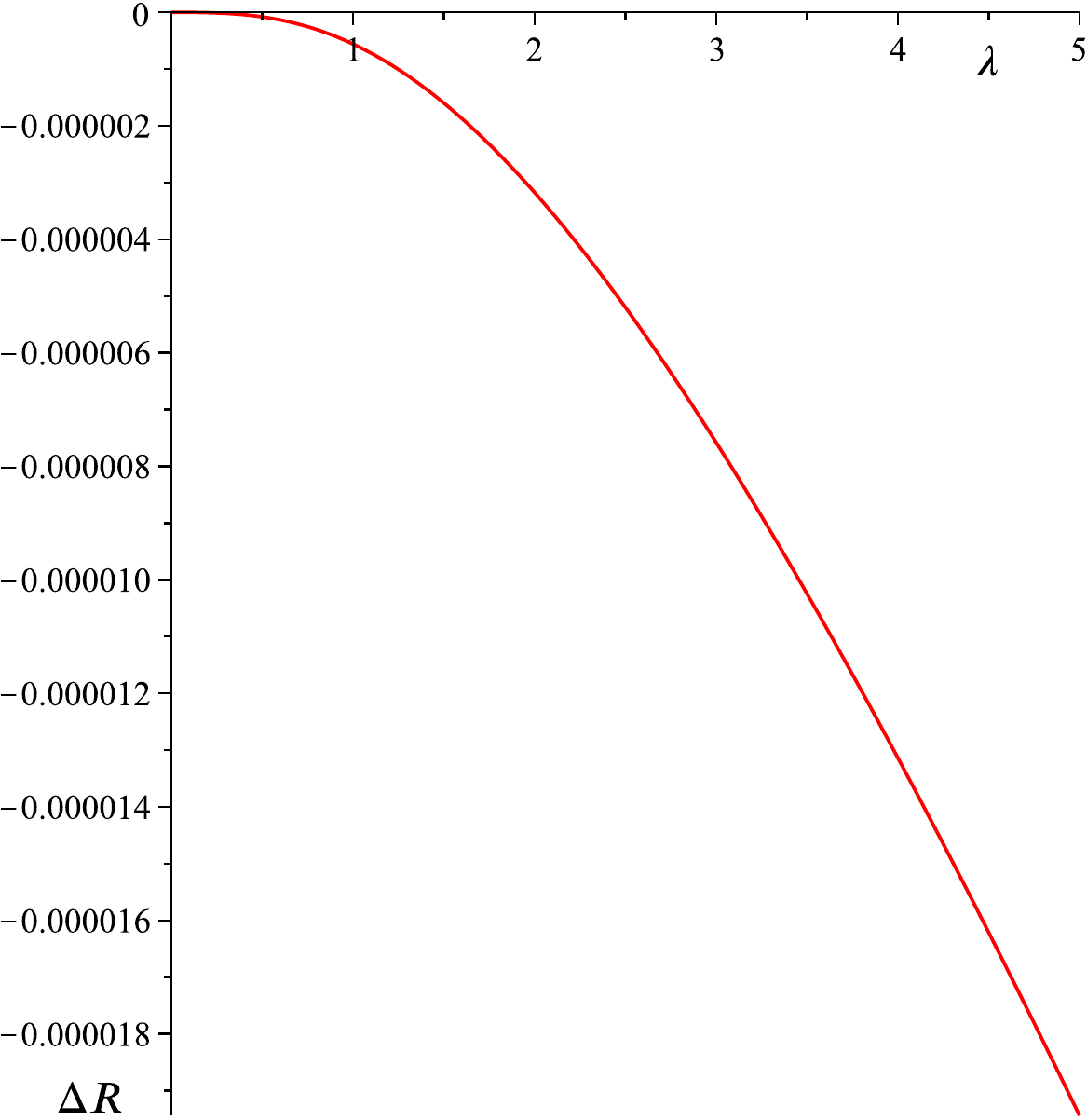}
\caption{Deviation of $R$ taken from the exact numerical integration with the approximation $R=\lambda$, where $\lambda$ is the affine parameter for the photon path, covering the interval 0 to 5. \label{fig:r-deur-lensing}}
\end{center}
\end{figure}

Here we can see a slight difference between exact result and the analytical approximation, but it is clearly small. To see whether this $z$-result ties in with the $\upbeta$ result of \cref{fig:my-exact-lensing-calcs-beta-change}, we need to understand also how $R$ evolves. In fact, at the accuracy being used here, we can take
\be
R=\lambda,
\ee
and this is verified in the following plot, \cref{fig:r-deur-lensing}
which shows the difference between the exact numerical $R$ and the affine parameter $\lambda$ as the latter goes over the range of integration used for the photon path, i.e.\ from 0 to 5. It can be seen that there is less than 1 part in $10^7$ deviation between the two over this range.

This means that the angle of motion, $\mathrm{d}z/\mathrm{d}R$ can be obtained just by differentiating equation \eqref{eqn:z-interesting} w.r.t.\ $\lambda$. Then evaluating this for the parameters of the galaxy, and for $\lambda=5$, we find that the difference from the initial $\upbeta_0$ is $\approx 0.008 {\rm \, arcsec}$, matching the result for $\upbeta$ shown in \cref{fig:my-exact-lensing-calcs-beta-change}. Thus the various approximations are all consistent for a case such as the present one.

\subsection{Parameter values for significant deflection}\label{intermex2}

In order to get a $1/R$ instead of $1/R^2$ behaviour for the force in the galactic plane,~\gefc{Deur:2020wlg} needs the photon paths (which are being used as proxies for `gravitational field lines') to be bent enough that they end up moving roughly parallel to the plane once the edge of the galaxy is reached, as in the top right plot of Fig.~3 in~\gefc{Deur:2020wlg}, for example. Although we have seen that for what appears to be a similar example galaxy to the one~\gefc{Deur:2020wlg} uses, the actual photon path deviation is many orders of magnitude below what~\gefc{Deur:2020wlg} proposes, it is of interest to see what sort of parameter values, specifically for $M$, $a$ and $b$, we would need in order to get this type of behaviour happening.

To do this, we can use our analytical approximations to get an estimate of what sort of values are required, and then check these out with the exact integrations, since it is likely that there will be some deviations between the two, given the extreme values of the parameters required.

To get motion parallel to the plane, we need the deflection angle $\Delta\upbeta$ in equation \eqref{eqn:beta-deflection} to equal minus the initial angle, $\upbeta_0$, itself. Solving this equation and assuming large $\lambda$, i.e.\ that this is happening for the eventual asymptotic motion of the photon, we find that we need
\be
M \to \frac{\left(a+b\right)b}{2a}.
\ee
For a highly flattened galaxy, with $a \gg b$, we have ${M\to b/2}$. This is very revealing. For a $b$ of $0.045 \kpc$ as above, this means the mass needs to be $\sim 4.7 \times 10^{14} \msun$, i.e.\ of the scale of a large cluster of galaxies!

\begin{figure}
\begin{center}
\includegraphics[width=\linewidth]{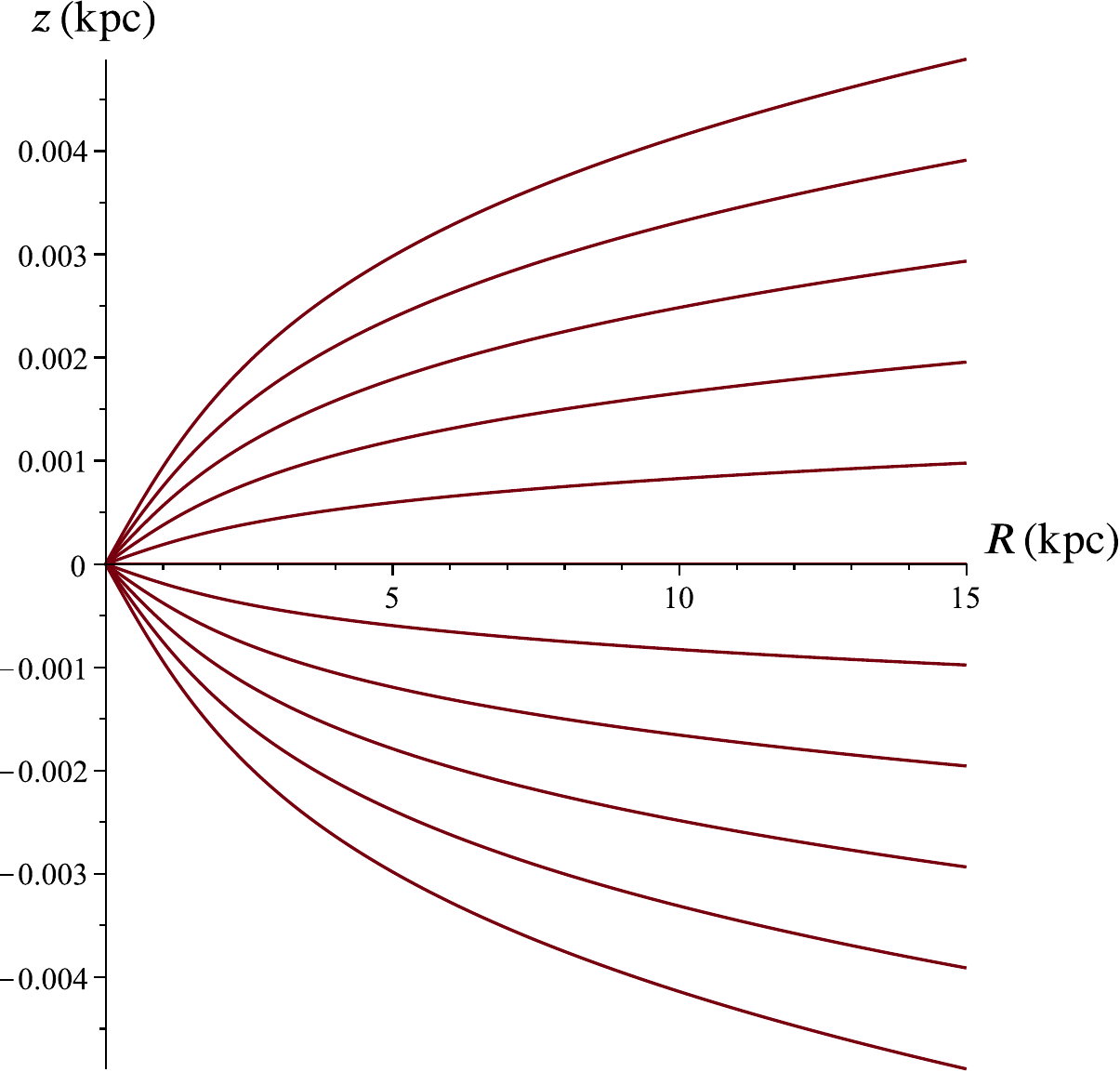}
\caption{Photon paths in the $(R,z)$ plane calculated using the analytical approximation for a range of initial $\upbeta_0$ in the case $M=b/2$. Note the two axes have been scaled independently. \label{fig:trajects-first-M}}
\end{center}
\end{figure}

\begin{figure}
\begin{center}
\includegraphics[width=\linewidth]{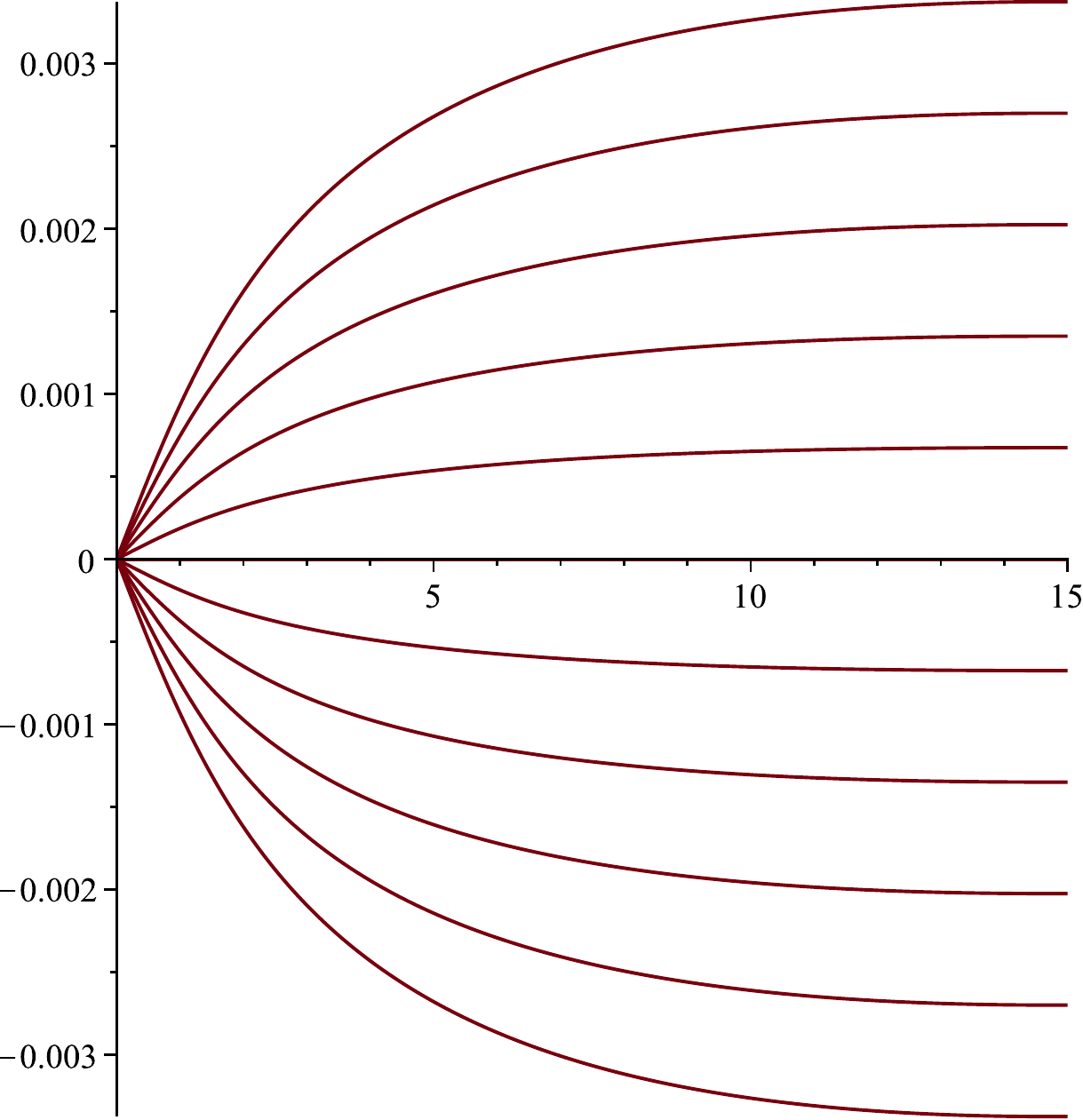}
\includegraphics[width=\linewidth]{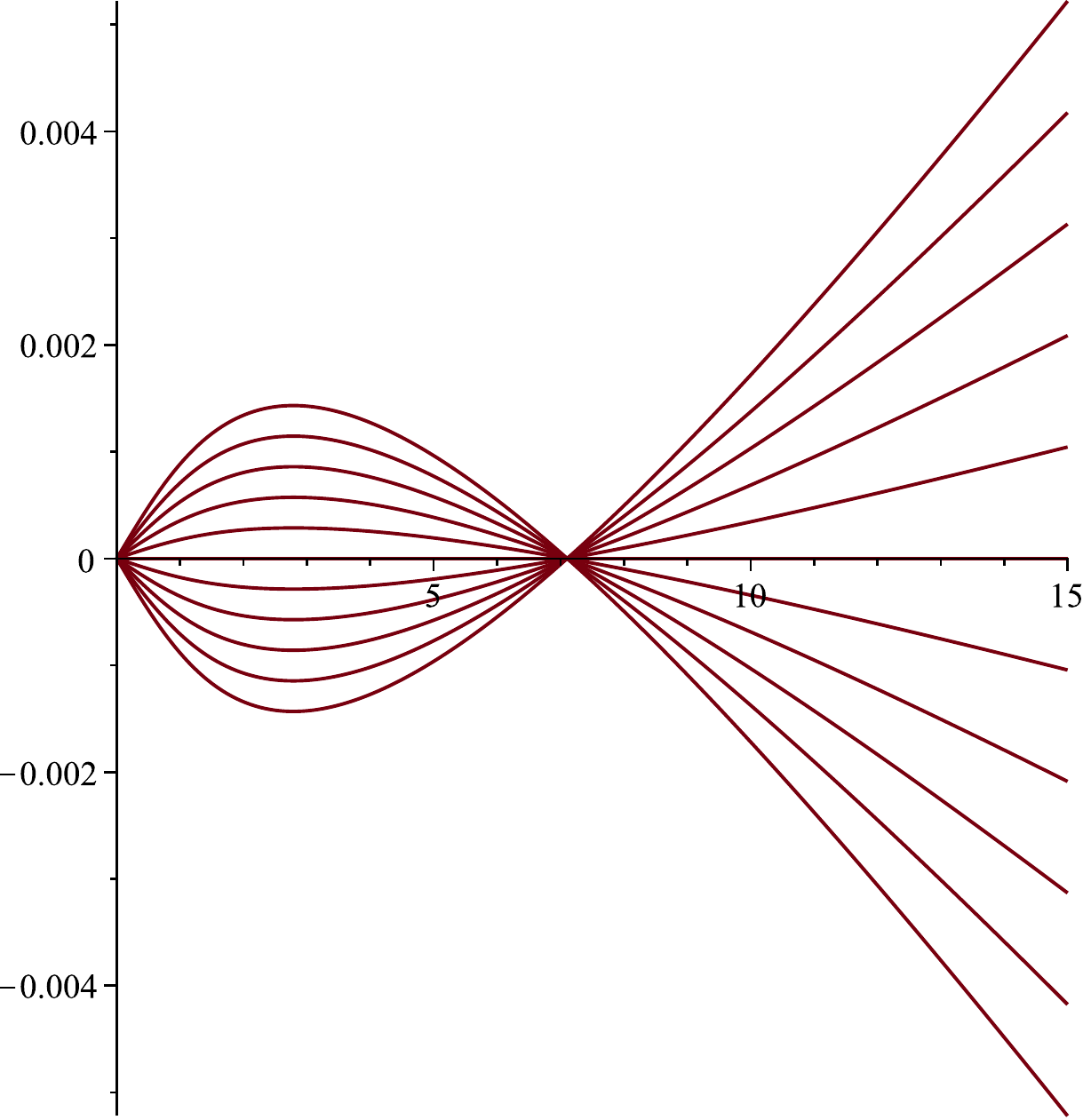}
\caption{Left: same as for \cref{fig:trajects-first-M} but for $M=1.15 (b/2)=0.575\,b$. This is just enough to flatten the trajectories at infinity. Right: same but for $M=b$, where an interesting `focussing' effect is visible. \label{fig:trajects-second-two-Ms}}
\end{center}
\end{figure}

Ignoring the obvious problems with this, let us see what plots of the photon paths look like for this case, using first our analytical approximation. In \cref{fig:trajects-first-M}
we show plots of paths in the $(z,R)$ plane for photons fired out at a range of initial $\upbeta$ angles, going in 11 steps between $\upbeta_0=-0.001$ and $\upbeta_0=+0.001$. (Note these angles are in radians, not arcsec.) We can see that indeed the paths become almost flattened. In the current analytical approximation, one finds that to get complete flattening, one needs $M \approx 1.15 \, b/2$. If one goes beyond this, then an interesting `focussing' effect becomes visible. These two cases are shown in
\cref{fig:trajects-second-two-Ms}.

\begin{figure}
\begin{center}
\includegraphics[width=\linewidth]{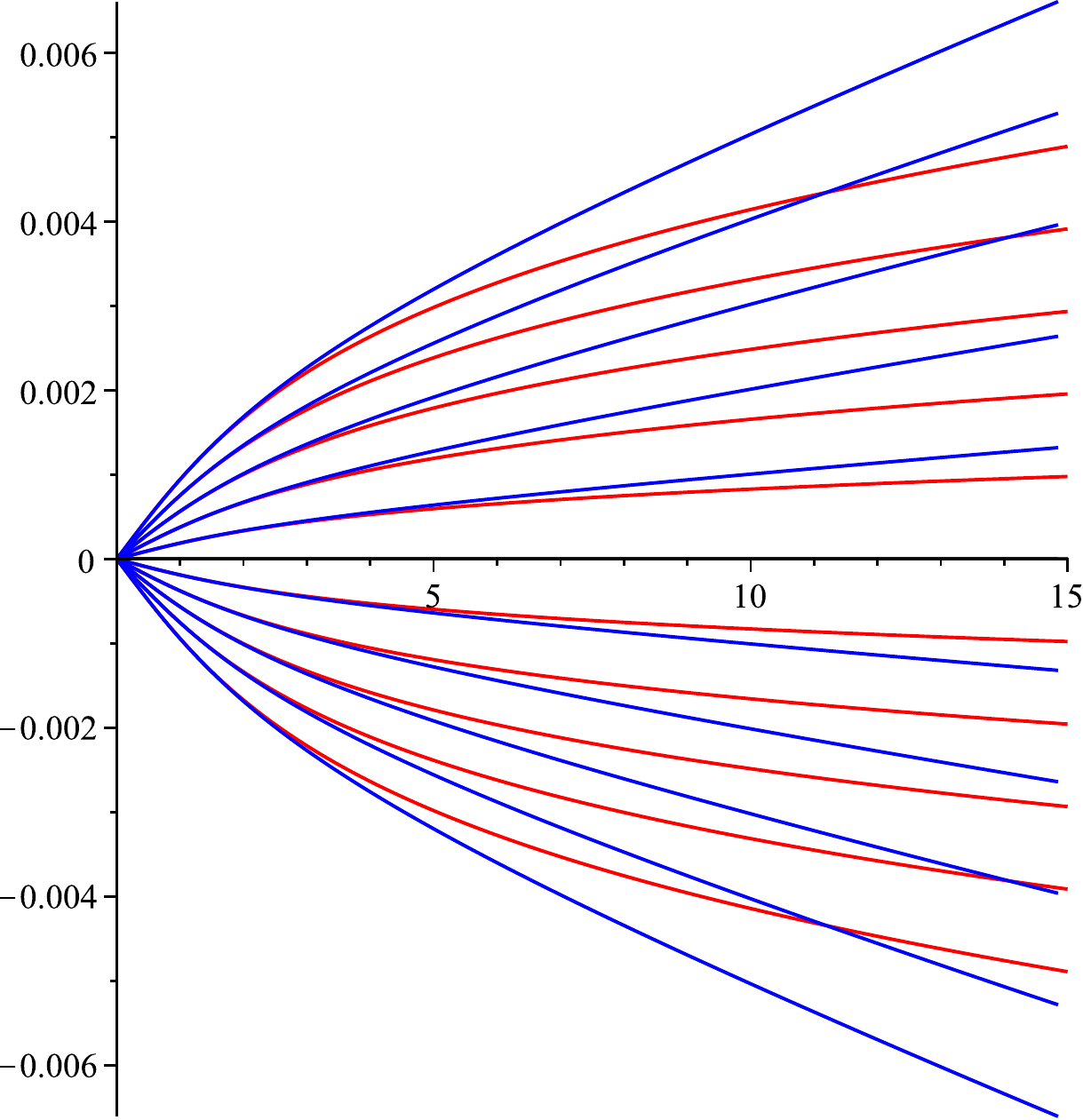}
\includegraphics[width=\linewidth]{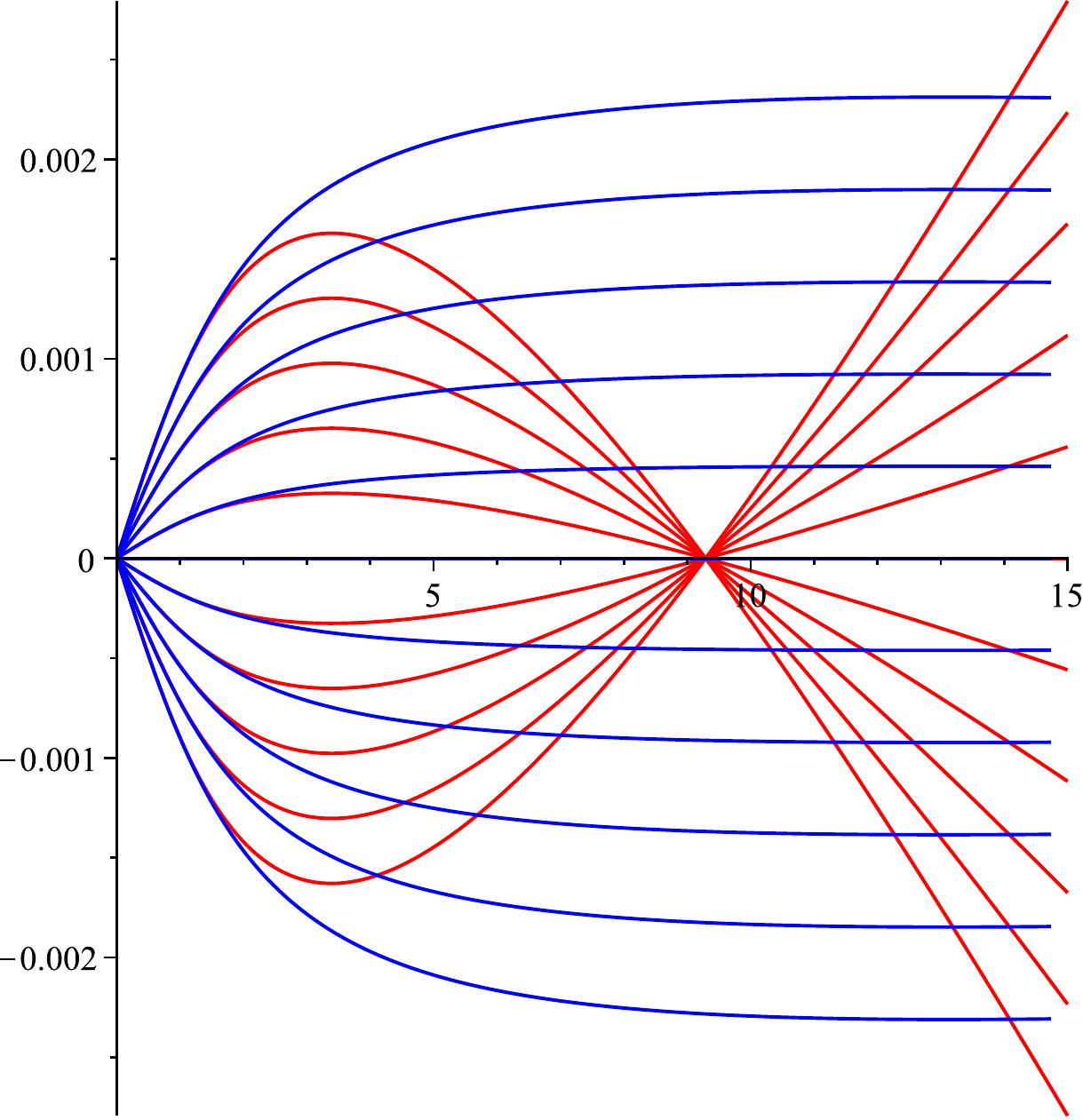}
\caption{Left: The blue curve shows the results of the exact numerical calculation for $M=b/2$, while the red curve is for the analytic approximation results for the same case. Right: same, but for the case $M=0.88 \, b$, which is just enough to give flattened field lines in the exact case. \label{fig:trajects-compare}}
\end{center}
\end{figure}

We now need to look at how the trajectories behave if we carry out exact numerical integration, rather than using our analytical approximation. In \cref{fig:trajects-compare}
we show the result of the exact numerical calculation in blue, and of the analytical approximation in red, for two different values of $M$. At the top we have the result for the initial case, corresponding to \cref{fig:trajects-first-M}, where $M$ was put to $b/2$. The red curves here are thus the same as in \cref{fig:trajects-first-M}. We can see that the exact calculation gives {\em less} deflection than the approximate one, although the two sets of curves are not wholly dissimilar.

In the bottom panel of \cref{fig:trajects-compare}, we show the equivalent but for $M=0.88\, b$. The point of choosing this $M$ is that for this value the exact curves (blue) become asymptotically parallel to the disc. Meanwhile, the larger deflection of the approximate curves (red) means that they turn round and refocus in this case. We have not shown it here, but if we continue increasing the mass, then the exact curves start to refocus as well, which is not surprising, and like the approximate curves, they appear to refocus exactly, i.e.\ all the curves go through the same point. This will be discussed further below.

\begin{figure}
\begin{center}
\includegraphics[width=\linewidth]{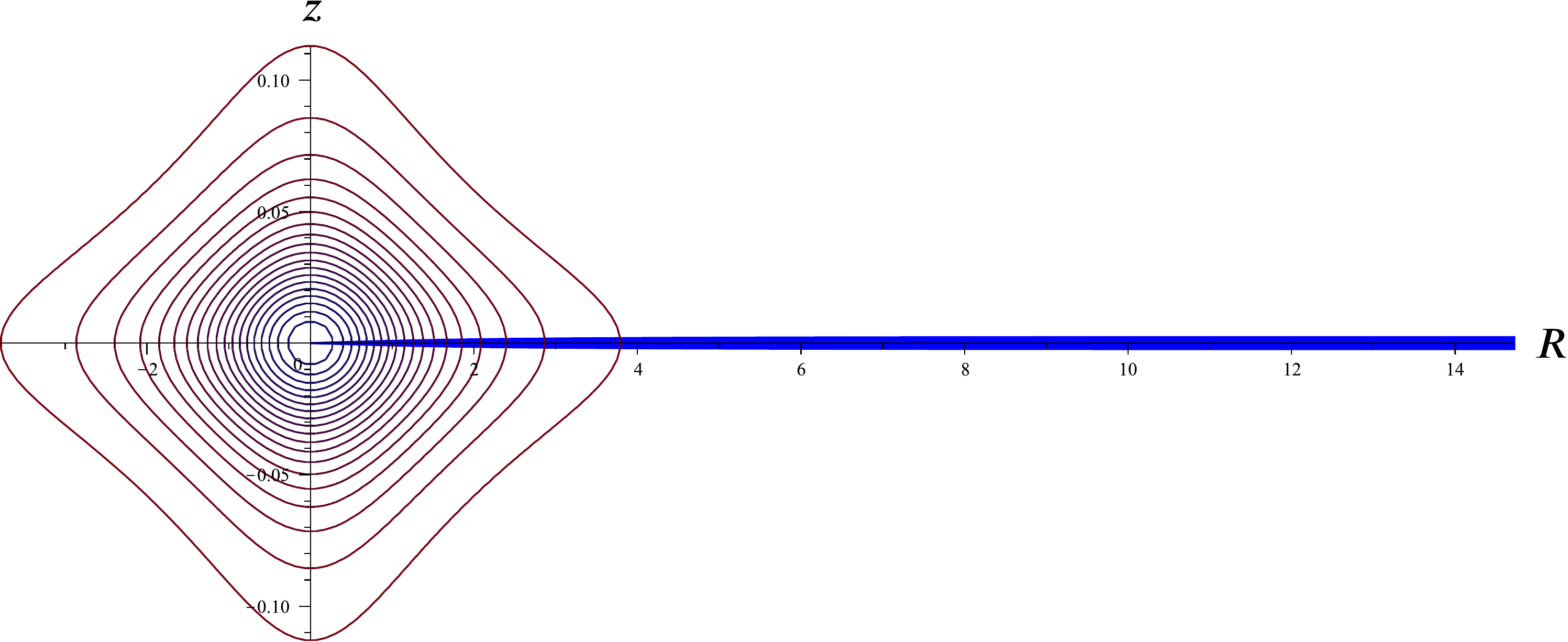}
\caption{This figure attempts to show some of the rays we have already discussed, in relation to the disc of the galaxy itself. The black lines are isodensity contours for the example galaxy, where the outermost contour corresponds to 1/20th of the central density. The blue lines show the same trajectories as the blue lines in \cref{fig:trajects-compare}, i.e.\ they are the exact numerical calculations for the case $M=0.88\, b$. The vertical scale is much larger than the horizontal scale, hence the galaxy no longer looks flattened, but even so it is difficult to see the individual trajectories. \label{fig:rays-plus-conts}}
\end{center}
\end{figure}

Another thing which it is useful to do at this point, is to illustrate where the types of trajectories we are plotting lie in relation to the disc of the galaxy itself. This is actually quite hard to show since the galaxy is very flattened, and the rays themselves are being emitted at angles very close to 0. Thus it is not possible to discern anything on a plot which has the same scaling for the $R$ and $z$ axes. In \cref{fig:rays-plus-conts},
we show a `compromise' plot, where the $z$-axis scale is sufficiently expanded that the galactic density contours are clearly visible --- the outer contour here represents 1/20th of the central density, so gives some feel for the extent of the galaxy on the plot. The blue curves are the same as those shown in the right panel of \cref{fig:trajects-compare}, i.e.\ they are the exact curves for the case where $M=0.88 \, b$, which leads to the trajectories just being flattened. We see that the photon paths we are looking at are indeed very close to the disc of the galaxy.

\begin{figure}
\begin{center}
\includegraphics[width=\linewidth]{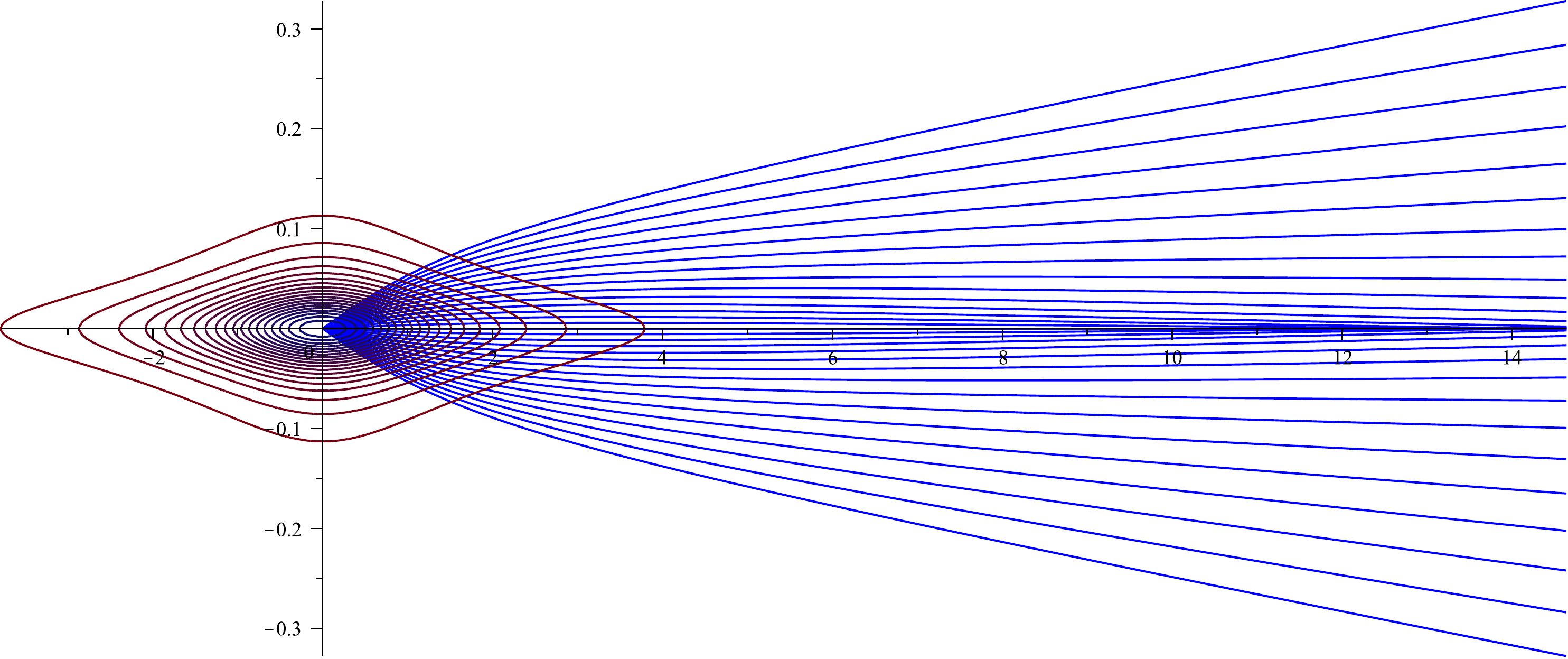}
\caption{Same as for \cref{fig:rays-plus-conts} but for $M=1.2\, b$, and for a wider range of initial $\upbeta$ values. Here the exact results show that the trajectories change shape as one moves outwards, with those closest to the $x$-axis refocusing, while those at the outside are defelected much less. \label{fig:rays-full-spectrum}}
\end{center}
\end{figure}

\begin{figure}
\begin{center}
\includegraphics[width=\linewidth]{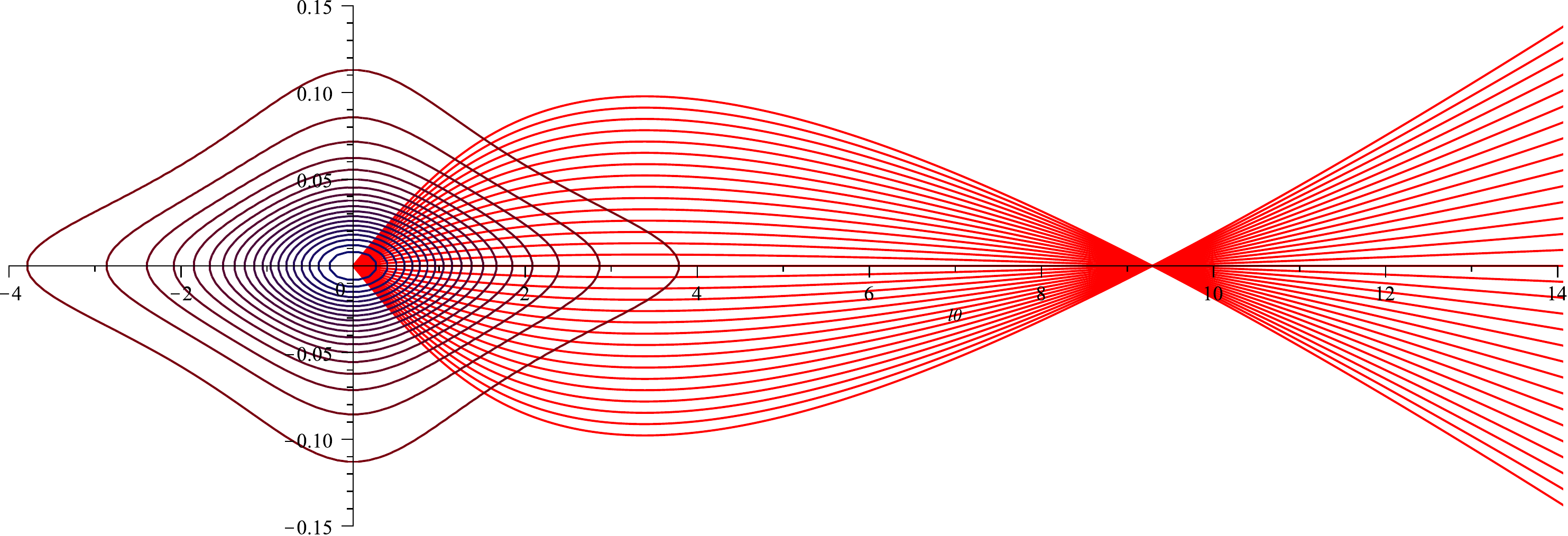}
\caption{Same as for \cref{fig:rays-full-spectrum}, but for the analytic approximation results for the case $M=0.88 \, b$. Here the curves do not change in shape, except for a vertical scaling, as one goes out to high initial $\upbeta$. \label{fig:rays-full-spectrum-from-approx}}
\end{center}
\end{figure}

Finally, in terms of the exact integrations, we show a plot for $M=1.2 \, b$, which leads to refocussing even in the exact case, but where we have covered a wider range of initial angles. This is shown in \cref{fig:rays-full-spectrum},
We can see here that the rays closest to the disc show refocussing, those slightly further out are `just flattened' and those further out still are relatively undeflected.This behaviour seems at first sight realistic, and is {\em not} what we obtain from the analytical approximation. We show this for the same range of initial $\upbeta$ but for a slightly smaller $M$ of $0.88 \,b$ (since otherwise the refocussing happens too quickly), in 
 \cref{fig:rays-full-spectrum-from-approx}.
Here it is clear that each trajectory has exactly the same `shape', with just a different vertical scaling. This is already evident from the form of approximation in equations \eqref{eqn:beta-deflection} and \eqref{eqn:z-interesting}, of course, which just scale directly proportional to the initial angle $\upbeta_0$.

\begin{figure}
\begin{center}
  \includegraphics[width=\linewidth]{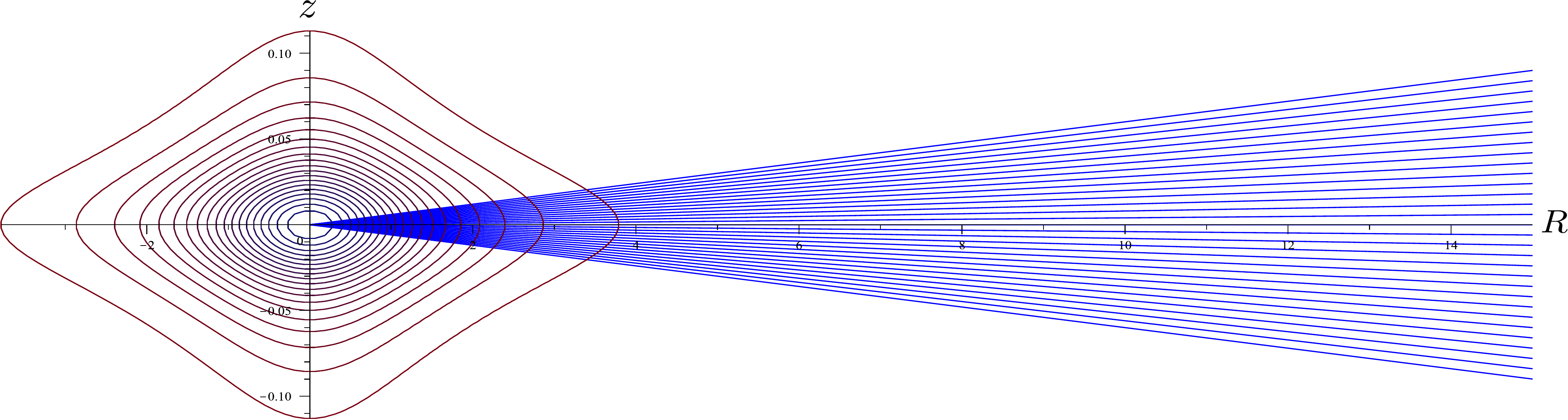}
\caption{Exact trajectories (blue) shown relative to the galactic disc (black contours) for realistic galaxy mass of $M=3 \times 10^{11} \msun$. Now no deflections at all are visible. \label{fig:rays-realistic-M}}
\end{center}
\end{figure}

So we might think that this is less realistic behaviour, and fails to capture what the exact results are telling us, and the type of behaviour we would need for the~\gefc{Deur:2020wlg} hypotheses to be true. However, the regime we are operating in for $M\sim b$ is completely unachievable in practice --- it would need, as already stated, an object of the mass of a rich cluster of galaxies confined to a region with typical scales of $1.5 \kpc$ horizontally, and $0.045 \kpc$ vertically. Basically, as we can already see from $M \sim b$, we are talking about something that is effectively a `black hole' in the $z$ direction, and it is only an object like this which can lead to any of the interesting effects seen here. If we were to plot the photon trajectories for realistic masses of a few times $10^{11}\msun$, then we would just get perfect looking radial lines for any initial $\upbeta_0$ for both the exact and analytical approximation cases, and no effects of the type that~\gefc{Deur:2020wlg} describe would be visible. We plot such a case in
\cref{fig:rays-realistic-M},
which is for the $M$ corresponding to our original example galaxy above, i.e.\ $M=3\times 10^{11} \msun$.This is for the exact numerical integrations, but exactly the same curves at the limit of resolution would be found for the analytical approximation here, and in particular there is no change in the `shape' as the disc is approached.

We promised earlier to discus the fact that the rays computed via the exact rather than approximate method, all appear to go through the same point when the refocusing occurs. This is not surprising from the approximate formula, since as already noted all the trajectories have the same shape here, but is perhaps surprising from the point of view of the exact calculations, since the curves are not all just vertically scaled versions of one another in this case. This would be worth investigating, except that of course this case, with an object approaching an effective black hole in the vertical direction, would need to be investigated using the fully non-linear Einstein equations, and not within the simple linearised ansatz for the fields~\eqref{skipar} which we have been using here.

\subsection{Postmortem of GEFC lensing}\label{postmortem}

In the context of our findings in~\cref{intermex1,intermex2} it is of interest to understand where the calculations of lensing in~\gefc{Deur:2020wlg} may have gone astray. The example calculation for which~\gefc{Deur:2020wlg} gives some details, and for which we can attempt to follow through what is happening, is for the effects of a annulus of matter on the path of a photon emitted radially from the centre of the galaxy. This is discussed in section II.B. of~\gefc{Deur:2020wlg}, which says that `the dominant bending comes from the rings with mid-planes at $z = 0$, henceforth referred to as ``central rings'''. The total effect of the galaxy is then found by adding the effects from all the different types of rings and slices together. What we will do here to compare, is to repeat our calculations above, but this time computing the lensing caused by an annulus of matter stretching from $R'$ to $R'+\Delta R$ in the $R$ direction, and effectively infinitesimally thin in the $z$ direction, since instead of a 3d density distribution $\rho(R,z)$ we will just ascribe to it a surface density distribution $\Sigma(R)$, with $R$ evaluated at $R'$ for the annulus of interest.

This in fact differs somewhat from the setup envisaged by~\gefc{Deur:2020wlg} for the annulus, which is shown as the blue object in Fig.~2 of~\gefc{Deur:2020wlg}. Here the vertical height of the annulus is given by the $z$ of the photon track at that point. However, we shall show below that the actual calculation was based on finding the Newtonian potential of the annulus assuming it is concentrated along $z=0$. Hence we shall follow this line in working out our results, and from these demonstrate that in fact it is allowable to take this approach for a non-infinitesimally thin annulus as well.

So we will work out the Newtonian potential for such a ring in the $z=0$ plane, and then use it (in the guise of $a_1$) in the formula for the rate of change of photon inclination $\upbeta$ given in equation \eqref{eqn:dots-on-path}. We can then do an exact numerical integration as above, to get an answer for the bending that will be suffered by a radially moving photon due to the annulus. Having established what the exact results are, we can then go through a process of approximation to get an explicit approximate answer, and check that this works to a sufficient level of accuracy. Finally, we can then check this approximate answer against what~\gefc{Deur:2020wlg} says the result is for the same case, and see how the answers compare.

The first step is to get the Newtonian potential of the annulus. For this we can use e.g.\ equation (34) of the paper by Cohl \& Tohline,~\cite{cohl1999compact}, which is for precisely this case. We find
\be
{a_1}_{\rm ring} (R,z)=\frac{2G}{\sqrt{R}}\, \Delta R \, \sqrt{R'}\,\Sigma(R')\, m  \, K( m )
+\mathcal{O}\left(\varepsilon^2\right),
\label{eqn:Phi-from-ring}
\ee
where
\be
 m \equiv\sqrt{\frac{4 R R'}{(R+R')^2+z^2}},
\label{eqn:mu-def}
\ee
and $K$ is a complete elliptic function of the first kind. Note one can verify by direct differentiation that this function satisfies $\bgrad^2 {a_1}_{\rm ring}=0$ away from the annulus.

\begin{figure}
\begin{center}
  \includegraphics[width=\linewidth]{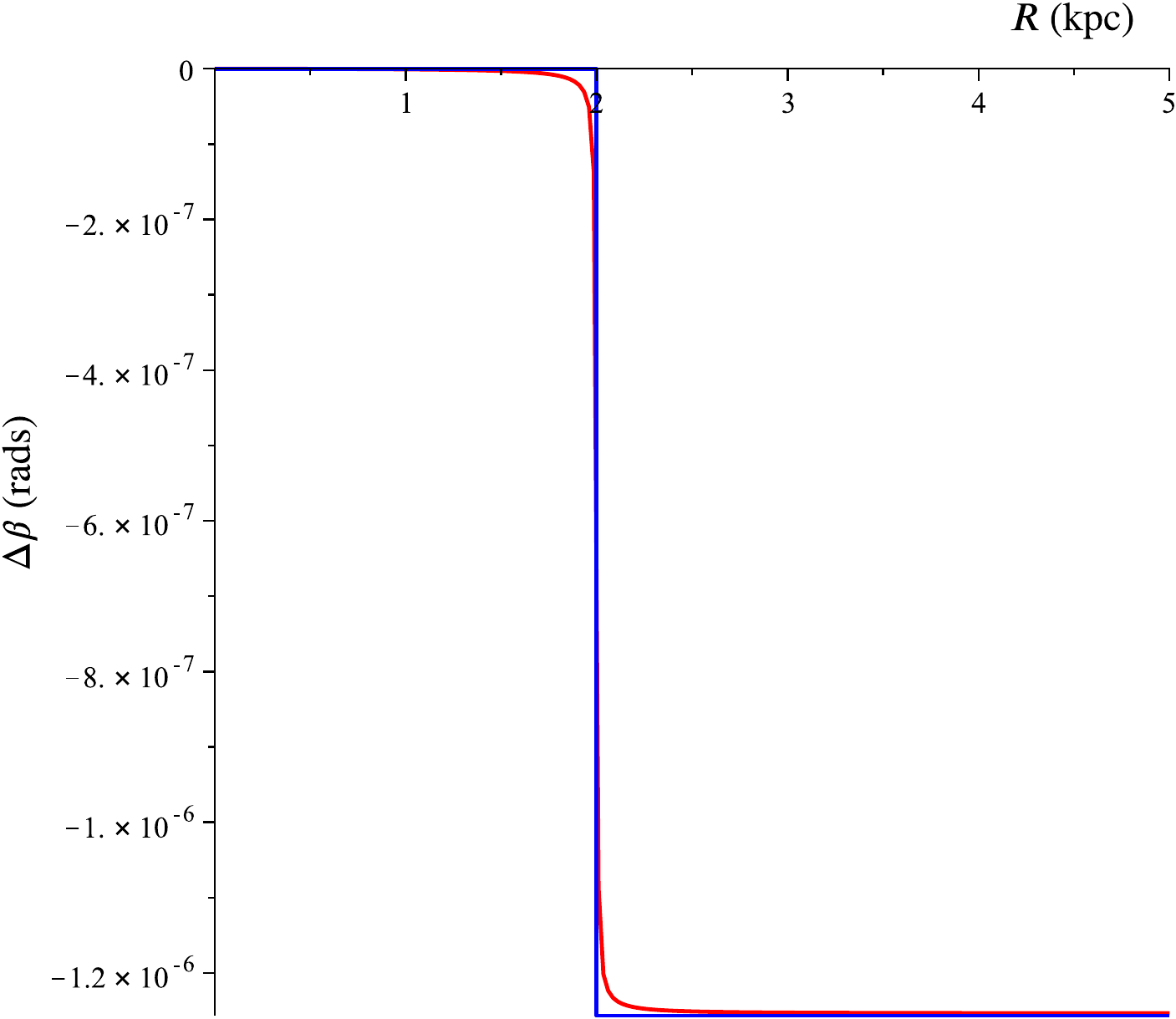}
\caption{In red, exact calculation of the deviation of photon inclination angle $\upbeta$ from its initial value, as a function of $R$, for a matter distribution consisting of an infinitesimally thin annulus at $R'=2 \kpc$. The initial $\upbeta$ angle is $0.0025 \, {\rm rads}$ for this example. In blue, a representation of  the leading term of the approximate answer for the total deflection, equation \eqref{eqn:ann-defl-ans}, is shown. \label{fig:annulus-comparison}}
\end{center}
\end{figure}

We now carry out a direct numerical integration using the equations in \eqref{eqn:dots-on-path}, with this new potential. \cref{fig:annulus-comparison}
shows how the photon inclination angle $\upbeta$ changes from its initial angle as it passes by the annulus, which is located at $R'=2 \kpc$. The parameters for the annulus used here are fairly arbitrary, since we just want to show indicative effects, but in detail they are that the width in the $R$ direction is $\Delta R=0.01$ and the surface density $\Sigma$ is $10^{-5}$ in the system where the unit of length (which therefore gives all the other units) is $1 \kpc$. (This corresponds to about $2 \times 10^{11} \msun {\rm \, kpc^{-2}}$.) The initial $\upbeta$ angle is $0.0025 \, {\rm rads}$ and we see that the total deflection happens more or less impulsively as the photon passes the annulus, and has a value of about $1.2 \times 10^{-6} \, {\rm rads}$.

We can now go about finding the (exact) deflection angles for a range of parameters, such as the ring radius $R'$ and the initial angle $\upbeta_0$ and use these as the `truth' in a comparison with an approximate solution which we would also like to find. To carry out the latter, quite an involved chain of approximations is necessary, starting from the exact formulae in which \eqref{eqn:Phi-from-ring} is inserted into \eqref{eqn:dots-on-path}. This requires being able to approximate the elliptic $K( m )$ function and the complete elliptic function of the second kind $E( m )$ that appears in its derivative, in the case where the argument $ m $  defined in \eqref{eqn:mu-def} is close to 1. This comes about since at the point where the photon is just passing the annulus, we can take $R\approx R'$ and $z$ small, hence $ m $ will be just below 1.
After this we need to be able to integrate the resulting expression for $\dot{\upbeta}$ over the photon path to get the total deflection. We omit these details here, and just give the result, to second order in the $z$ at closest approach. We get
\be
\Delta \upbeta = - \Delta R \,\, \Sigma(R')\left(4 \pi -4 \left(\ln 2 - \ln \theta\right) \theta - \frac{3\pi}{2}\theta^2 \right),
\label{eqn:ann-defl-ans}
\ee
where $\theta$ is used to denote the small quantity $z/R'$.

This deflection looks as though it might be singular as $z$ (and therefore $\theta$) tends to 0, but $\ln \theta$ is multiplied by $\theta$ and in fact the expression tends smoothly to the result $\Delta \upbeta = - 4 \pi \, \Delta R \,\, \Sigma(R')$. Moreover, the first term in the brackets in \eqref{eqn:ann-defl-ans} (i.e.\ $4\pi$) will in general be much larger than the others, hence we can predict from this there will be a relatively small dependence of the deflection angle on either the $z$ at closest approach or the $R'$ of the annulus location, and therefore also on the initial inclination angle $\upbeta_0$.

\begin{figure}
\begin{center}
\includegraphics[width=\linewidth]{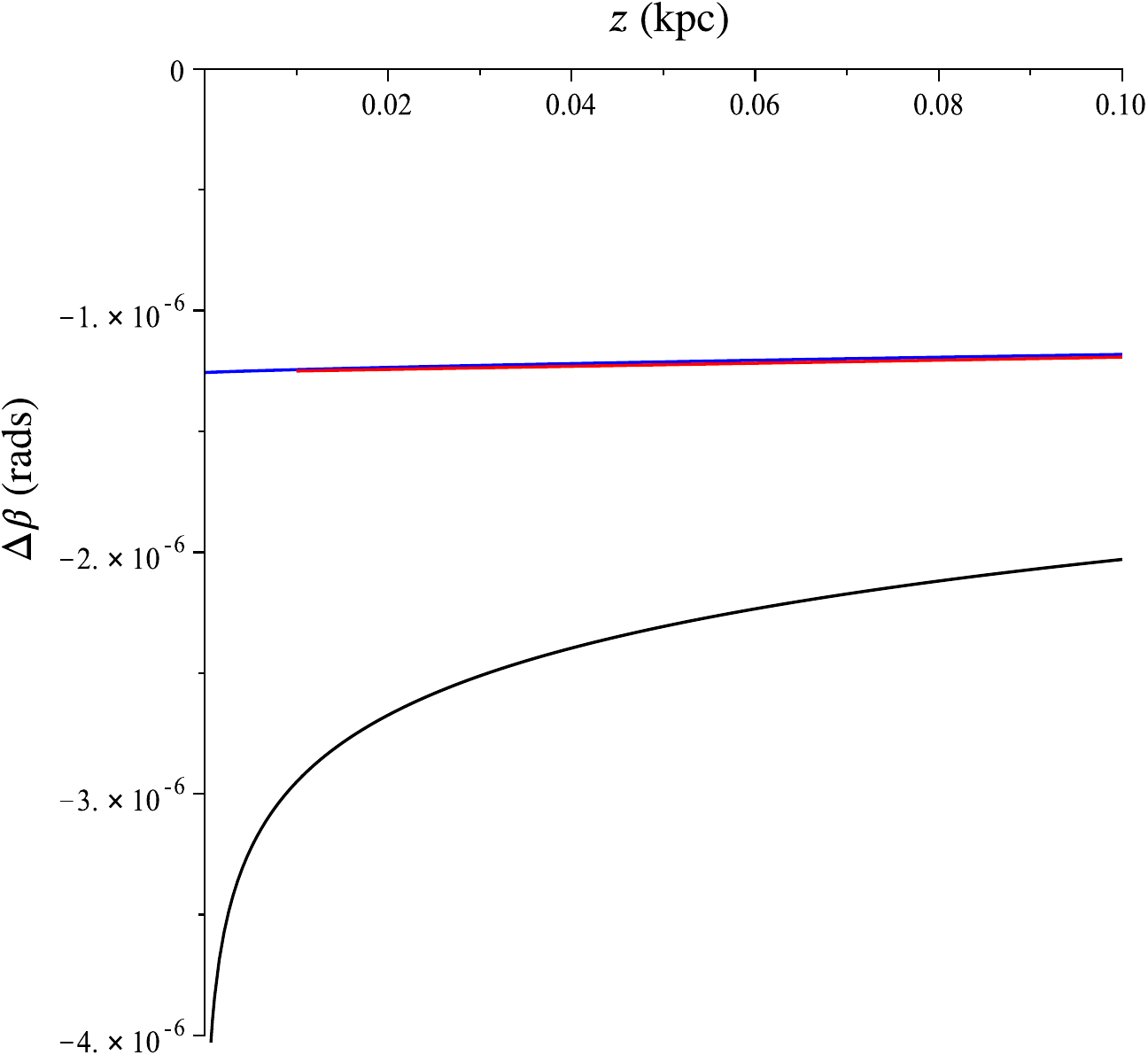}
\caption{In red, exact calculation of the final deviation of photon inclination angle $\upbeta$ from its initial value, where the initial values are varied so as to give the range of $z$'s at closest approach. (These $z$'s form the horizontal axis.) The example matter distribution consisting of an infinitesimally thin annulus at $R'=2 \kpc$ is being used. In blue, the approximate answer equation \eqref{eqn:ann-defl-ans} is shown. In black we show the result that follows from the equations given by~\gefc{Deur:2020wlg} for this case. \label{fig:deur-answer-comp}}
\end{center}
\end{figure}

This is borne out by what we find with the exact calculations. In
\cref{fig:deur-answer-comp}
we show exact calculations (red) for the final deflection angle, for a range of initial $\upbeta$'s chosen to give the range of $z$'s at closest approach as shown on the horizontal axis. The annulus is at a fixed radius of $R'=2\kpc$ but we get a very similar plot for other choices of $R'$. In blue we show our approximate answer \eqref{eqn:ann-defl-ans}, which clearly yields a good approximation. In particular both the exact and approximate answers confirm that the first (constant) term in \eqref{eqn:ann-defl-ans} dominates, and that the deflection goes smoothly to this value as the $z$ at closest approach goes to 0.

Given this agreement between the behaviour of the approximate and exact results, we can take it that this is what actually happens. Initially, at least, this behaviour may seem somewhat surprising. If we were considering the passage of photons past a point source, we know that the deflection would go reciprocally with the distance of closest approach, and therefore be singular. Extending the point source to an annulus, our intuitive guess for the result might be that the singularity is softened to become logarithmic in the closest approach distance, rather than reciprocal, but to still be singular. What we seem to have found is that instead there is no singularity, and the value of the deflection is roughly constant with the $z$ of closest approach. 

In fact such behaviour is widely known about already, for a case which initially looks totally different, but is in fact basically the same as we are seeing here. This case is that for lensing by {\em cosmic strings}. In~\cite{vilenkin1981gravitational}, Vilenkin showed how a line-like topological defect (which might be formed in an early universe phase transition) could cause lensing of light rays by an amount which (as long as the radius of curvature of the string was much bigger than the distance of closest approach) did not vary with impact parameter for the photon passing the string, but just changed with sign according to whether the photon passed one side or another of the string. The amount of the lensing was by a deflection angle $\Delta\upbeta$ given by 
\be
\Delta \upbeta = 4 \pi \mu,
\ee
where $\mu$ is the mass per unit length of the string. 

In the cosmic string literature, such behaviour is attributed to the string effectively causing a `wedge' to be taken out of an otherwise flat cylindrical spacetime surrounding the string, with a `defect angle' of the wedge of $2\Delta \upbeta$, and then the remaining spacetime having the cut edges glued back together in a form of spacetime surgery to form what is called a `conical' spacetime. One can picture that this could indeed give rise to the behaviour of light rays as described, since these would travel in straight lines in the still-flat remaining spacetime, but nevertheless, rays on opposite sides would appear to converge together after passage of the string. 

In our current case, we have a much more prosaic example of the same phenomenon. The `cosmic string' is now the thin annulus, and as long as our photon is on a path that takes it much closer to the annulus in terms of $z$ at closest approach than the annulus radius at that position, then we can expect the same logic to apply, and for the photon to be deflected by a fixed angle of $4\pi$ times the mass per unit length of the annulus. Since the latter is $\Delta R \Sigma(R')$ (remember $\Delta R$ is the annulus width and $\Sigma(R')$ its surface density at radius $R'$), then we expect a deflection of  $- 4\pi \Delta R \,\, \Sigma(R')$, exactly as found in the first term of \eqref{eqn:ann-defl-ans}. Of course in the current case we are not obliged to think in terms of `spacetime surgery' and topology --- just the weak field forces on the passing photons are enough to give us what we need, and indeed more generally one can give a full treatment of cosmic strings in terms of
gauge fields in flat space (as in electromagnetism), rather
than in terms of topological surgery upon spacetime --- see~\cite{doran1996physics} for a discussion of this approach.

Having got this satisfactory confirmation and justification for our result, we now turn to the answer that~\gefc{Deur:2020wlg} gets for this case, which is the only one for which an explicit answer is given. What we need to compare with is equation (6) in~\gefc{Deur:2020wlg}, namely
\be
\delta\upbeta(R,z) = \frac{GM}{\pi} E(R,z),
\label{eqn:deur-ans}
\ee
where the following definition is given for the quantity $E(R,z)$
\be
E(R,z) \equiv 2 \int_0^{\pi} \frac{d\psi}{\sqrt{\left(2 R \sin \left( \psi/2\right) \right)^2+z^2}},
\ee
which is described as `the complete elliptical integral of the first kind'. This seems like a non-standard designation (and we emphasise that it would conflict with our conventions both in~\cref{newgauge} and above in this section), but nevertheless, since an explicit expression for $E(r,z)$ is given, we can carry out the indicated integral and obtain
\be
E(R,z)= \frac{4 K\left(2 R/\sqrt{z^{2}+4 R^{2}}\right)}{\sqrt{z^{2}+4 R^{2}}},
\ee
where the $K$ here is the complete elliptic integral of the first kind.

Clearly what we are doing here with $E(R,z)$ is finding the average inverse distance from a point $(R,z,\upvarphi=0)$ in cylindrical coordinates, to the infinitesimally thin ring $(R,z=0,\upvarphi)$ as $\upvarphi$ varies over $0$ to $2\pi$. Indeed, comparing with equation \eqref{eqn:Phi-from-ring}, in which we need to set $R'=R$, we see that
$-E(R,z) GM/(2\pi)$, where $M$ is the mass of the ring, will be the Newtonian potential at the point $(R,z,\psi=0)$. The~\gefc{Deur:2020wlg} answer for the angular deflection \eqref{eqn:deur-ans} then appears to be twice this Newtonian potential.

It is not clear why~\gefc{Deur:2020wlg} believes that this is the way in which to get the deflection, and in particular it disagrees with our result \eqref{eqn:ann-defl-ans} by being {\em singular} as $z \rightarrow 0$. We can see this by expanding $E(R,z)$ in $z$, for which we get
\be
\begin{aligned}
	E(R,z)& \approx \frac{6\ln 2 - 2 \ln z + 2 \ln R}{R} 
	\\
	& \ \ \ +\frac{-3 \ln 2 + \ln z -\ln R +1}{8 R^3} z^2 + \ldots
\end{aligned}
\ee
which has a logarithmic singularity for small $z$. By contrast, our answer, backed up by the exact numerical calculations, tends to a constant for small $z$. To show the comparison between the GEFC and our answers, then in \cref{fig:deur-answer-comp} we have include a curve showing the prediction of the~\gefc{Deur:2020wlg} result \eqref{eqn:deur-ans}, calculated using the same ring mass. 

Now this looks like a big discrepancy, and a possible source of why~\gefc{Deur:2020wlg} says that the rays become parallel near the galaxy disc edge, whereas as we have seen, this would require densities about 1000 times larger than typical galactic densities. However, some caveats are in order.

As we have shown, the actual~\gefc{Deur:2020wlg} calculation seems to be assuming an infinitesimally thin ring, but that paper also contains a figure describing the setup and showing the ring vertical height being equal to the current $z$ of the photon path. Since we have shown that one can wind down the $z$ of the photon path to be as close as one likes to an infinitesimal ring, this is not a problem as long as $z$ is small compared to the $R'$ of the ring. Our answer should still apply.

More significantly, linked to this, is the fact that the mass of the ring as used in~\gefc{Deur:2020wlg} incorporates the height, i.e.\ for a fixed galactic density then in both that approach, and in applying ours to what is being done there, then (assuming a height $z$ over which the ring density does not vary much vertically as compared to its value at $z=0$), the ring mass should be taken as proportional to $z$. This will wipe out the singularity shown in \cref{fig:deur-answer-comp}, since the~\gefc{Deur:2020wlg} curve will now go like $z \ln z$ at small $z$, while ours will now go as $z$. These still differ in ratio by a factor of $\ln z$, but the absolute value of the discrepancy will not be large, and it seems difficult to understand how factors of order $10^3$ in the lensing could arise.

\section{Conclusions}\label{conclusions}

This paper has sought to demonstrate an effect which we term `\emph{gravitoelectric flux collapse}' (GEFC), and which is proposed in~\gefc{Deur:2009ya,Deur:2013baa,Deur:2014kya,Deur:2016bwq,Deur:2017aas,Deur:2019kqi,Deur:2020wlg,Deur:2021ink,Deur:2022ooc}. GEFC promises, among other things, to render galactic dark matter halos redundant by explaining flat and rising rotation curves via purely general-relativistic effects. To this end we have attempted to reproduce the remarkable results of~\gefc{Deur:2016bwq} and~\gefc{Deur:2020wlg}. We have enjoyed little success, and cannot conclude that the GEFC programme, in its current form, has a sound physical basis.

In particular we repeat certain observations which were made along the way:--
\begin{enumerate}
  \item We found in~\cref{matcou,nonrelsca} that the scalar gravity model which seems to
    underpin~\gefc{Deur:2016bwq,Deur:2014kya,Deur:2021ink,Deur:2009ya,Deur:2019kqi,Deur:2017aas,Deur:2022ooc}
    is essentially arbitrary, and not necessarily descriptive of GR.
    Superficially, the model would seem from~\cref{nonrel} to be \emph{inconsistent} with
    the nonlinear, static, vacuum EFEs. Its consistency with the
    Einstein--Infeld--Hoffman potential appears from~\cref{smokemirror} to be coincidental, and not too unlikely.
  \item The theoretical basis for the lattice techniques used to probe
    gravitational potentials in~\gefc{Deur:2009ya,Deur:2016bwq} does not appear
    fully watertight, as discussed in~\cref{ess}.
  \item At next-to-leading-order near a typical galactic baryon profile, usual tensorial GR does not appear to support substantial GEFC-type effects as proposed in~\gefc{Deur:2019kqi,Deur:2016bwq,Deur:2017aas,Deur:2020wlg}. This was verified throughout~\cref{gem2} using a variety of perturbation schemes and gauge choices.
  \item The lensing effects claimed in~\gefc{Deur:2020wlg}, which are used as a heuristic for GEFC-type phenomena, appear from~\cref{lensing} to have been overstated by three orders of magnitude.
\end{enumerate}

As mentioned in~\cref{introduction}, it is not clear how many of the other interesting effects promoted in~\gefc{Deur:2009ya,Deur:2014kya,Deur:2017aas,Deur:2019kqi,Deur:2021ink,Deur:2022ooc} can be salvaged if the points raised above are not adequately addressed.
In terms of mapping a road forwards, we are particularly interested in establishing clarity on the following question: `\emph{How are non-perturbative phenomena expected to emerge from closed, perturbative methods?}' In our calculations, for example, we encounter no warning that the perturbative approach is failing, such as divergent or unbounded quantities. It is then not too surprising that we recover only small corrections to the Newtonian phenomena. 

Despite this outlook, we recall that the above methods have, as a by-product, suggested a couple of interesting research avenues:--

\begin{enumerate}
	\item The result~\eqref{eqn:par_derivs_consist} in~\cref{axisy} may be of relevance to the fluid ball conjecture (Lichnerowicz's conjecture~\cite{2020arXiv200414240C}).
  \item The path integral separation in~\cref{ess} may orient a new gravitational energy localisation scheme.
\end{enumerate}

Finally, we distance ourselves from the previous analyses in~\gefc{Deur:2009ya,Deur:2013baa,Deur:2014kya,Deur:2016bwq,Deur:2017aas,Deur:2019kqi,Deur:2020wlg,Deur:2021ink,Deur:2022ooc} to emphasise that our `steel-man' approach precludes GEFC effects \emph{in degree but not in kind}. The nonlinear regime of gravity is very real, and doubtless still hides many unknown and exotic phenomena. 
Questions of utility and astrophysical realisation aside, a principled and considered correspondence between general relativity, nonlinear gravitoelectromagnetism and quantum chromodynamics --- should it exist --- would be a great asset in addressing the broader question of gravitational confinement.

\begin{acknowledgements}
	We are grateful to Alexandre Deur for rapid, thorough replies and vital clarifications at several junctures.
	We are also grateful to Craig Mackay and Amel Durakovi\'c for several useful discussions, and to John Donoghue and Subodh Patil for helpful correspondence. 

	W.~E.~V.~B. is grateful for the kind hospitality of Leiden University and the Lorentz Institute, and is supported by Girton College, Cambridge.
	
	The supplemental materials provided in~\cite{supp} incorporate elements of Cyril Pitrou's code from the repository at \href{https://github.com/xAct-contrib/examples}{www.github.com/xAct-contrib/examples}.

	During the course of this work, the literature advocating for GEFC was augmented by the preprint~\gefc{Deur:2022ooc}, which appears to be grounded in the same scalar gravity model addressed in the currect article.
\end{acknowledgements}

\bibliographystyle{apsrev4-1}
\bibliography{bibliography}

\appendix

\section{Two-point function as energy}\label{ess}

In this appendix we try to understand how and why the two-point function is being used to study gravitational potentials in~\gefc{Deur:2016bwq}. Although we have found in~\cref{anmod} that the scalar graviton underlying these calculations is not an appropriate model, for the sake of reproducibility we would still like to understand the lattice calculations which follow, and to weigh their physical significance. 

We firstly try to `steel-man' the theoretical basis for using a static two-point function as a proxy for potential energy in general. To this end, we consider the neutral Klein--Gordon theory $\phi$ in flat spacetime
\begin{equation}
  \mathcal{  L}_T=\frac{1}{2}\tensor{\partial}{_\mu}\phi\tensor{\partial}{^\mu}\phi-\frac{1}{2}m^2\phi^2+{J}\phi,
  \label{kleingordon}
\end{equation}
where $m$ is a mass and ${J}= J(x)$ is a source, such as a fermion current. The question of \emph{potential} in the case of staticity, i.e. $J(x)=J(\bm x)$, can be posed by asking `\emph{how much energy is associated with the imposition of the source}?' We will shortly make this question precise by combining techniques from~\cite{Coleman:2011xi} and~\cite{1995iqft.book.....P}, but first we recall the gravitational model to which these techniques will ultimately be applied.
Following the `steel-man' route, one can work with a unitary, relativistic alternative of the perturbative expansion of~\eqref{thoughttrain} in which the kinetic structure and d.o.f are preserved, but the coefficients are allowed to be arbitrary, i.e. $\mathcal{  L}_T\equiv\mathcal{  L}_G+\mathcal{  L}_M$ where
\begin{equation}
	\mathcal{L}_{{G}}
	\equiv\dalembertian\varphi\sum_{n=1}^{\infty}\tensor{a}{_n}\varphi^n,
	\quad
	\mathcal{L}_{{M}}
	\equiv
	J\sum_{n=1}^{\infty}\tensor{b}{_n}\varphi^n,
	\label{ansa}
\end{equation}
so that the $\tensor{a}{_n}$ control the nonlinear kinetic structure and the $\tensor{b}{_n}$ control the nonlinear coupling to matter. 
Following~\eqref{deur1} (with which we already have some physical concerns detailed in~\cref{matcou}), the matter stress-energy tensor $\tensor{\bar{T}}{^{\mu\nu}}$ is represented by $J$, indicating some scalar energy current.
Using rescalings of $\varphi$ and $J$ we can always set
\begin{equation}
	\tensor{a}{_1}\equiv-\frac{1}{2}, \quad \tensor{b}{_1}\equiv 1,
\end{equation}
just so that the Klein--Gordon conventions of~\eqref{kleingordon} are recovered.
The other couplings are features of the theory. Note that whilst~\eqref{ansa} generalises the gravity model used in~\gefc{Deur:2016bwq,Deur:2014kya,Deur:2021ink,Deur:2009ya,Deur:2019kqi,Deur:2017aas}, we are still not claiming that it is actually faithful to GR at any order.

\begin{figure*}[t!]
	\centering
	\includegraphics[width=\textwidth]{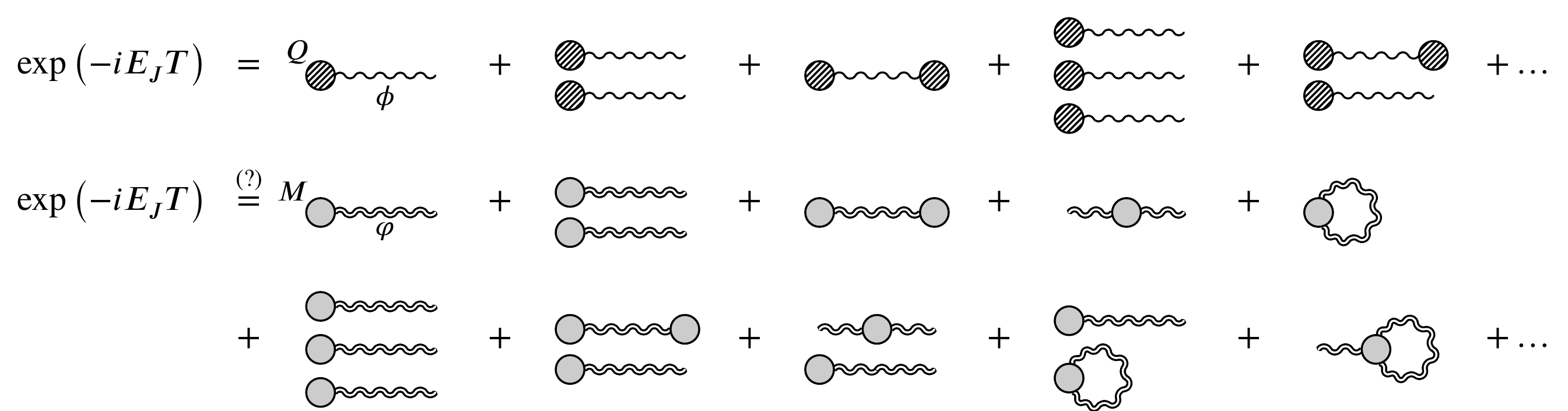}
	\caption{\label{diagrams}Some diagrams contributing to the vacuum--vacuum amplitude~\eqref{divphas} which encodes the potential energy associated with the Klein--Gordon field $\phi$ in the presence of an extended source charge $Q$, which is static for some long interval and may represent e.g. a pair of point sources. Below, a similar attempt may be made for the GEFC scalar graviton $\varphi$ in the presence of a mass $M$. That theory, however, inherits the ultraviolet divergences of GR, without inheriting the classical phenomenology. Moreover, we find that the nonlinearities destroy the separability of the path integral, so that an analytic expression for the two-point-mass potential is hard to obtain.}
\end{figure*}

For the moment we imagine that~\cref{kleingordon,ansa} are to be the bases for a pair of QFTs. 
In the case of~\eqref{kleingordon} the free part of the theory $H_0$ stems from the Lagrangian $\mathcal{  L}_0=-\frac{1}{2}\phi\left(\dalembertian\phi+m^2\phi\right)$. In the case of~\eqref{ansa} we do not take just the linear portion $\mathcal{L}_0=-\frac{1}{2}\varphi\dalembertian\varphi$, but rather use the full free gravity theory $\mathcal{L}_0=\mathcal{L}_G$, complete with its higher kinetic vertices and negative mass dimensions. 
Of course these features will technically be fatal to the QFT at high energies, but they are already infamous in quantum gravity, and an attempt to address them here falls well outside our scope.
In both theories the vertices mediated by $b_n$ and the matter $J$ form the interaction $H_I$. Denoting the ground state of the free theories by $|0\rangle$, the scattering of the vacuum into the vacuum $\langle 0| S|0\rangle$ can be given by the path integral representation
\begin{equation}
	\langle 0|U_I(-\infty,\infty)|0\rangle=\frac{Z[J]}{Z_0},
	\label{pintrep}
\end{equation}
where the generating functional becomes, in the case of~\eqref{ansa},
\begin{equation}
	\begin{aligned}
	Z[J]&\equiv\int\mathcal{D}\varphi\exp \left[i\int\mathrm{d}^4x\mathcal{L}_{{T}}\right]\\
		&=\int\mathcal{D}\varphi\exp \left[-\int\mathrm{d}^4x_{{E}}\left(\mathcal{L}_{{G}}+\mathcal{L}_{{M}}\right)\right],
	\label{pathint}
	\end{aligned}
\end{equation}
and we can (if needed) Wick rotate into the Euclidean action. The normalisation is given by $Z_0\equiv Z[0]$.
Note that the time evolution operator is the time-ordered exponent
\begin{equation}
	U_I(-\infty, \infty)\equiv
	\mathcal{T}\left\{\exp\left[-i\int_{-\infty}^{\infty}\mathrm{d}tH_I(t)\right] \right\}.
	\label{timeint}
\end{equation}
The field $\varphi$ as it appears on the RHS of~\eqref{pathint} is not an operator, but $H_I$ in~\eqref{timeint} will be constructed from the position-space representation $\varphi_{I}(x)$, of the field operator in the Heisenberg picture of the theory $\mathcal{L}_G$
\begin{equation}
	H_I(t)\equiv -\int\mathrm{d}^3xJ(x)\sum_{n=1}^\infty b_n{\varphi_I(x)}^n.
	\label{nullint}
\end{equation}
In the perturbative QFT, we would like to expand both within this interaction Hamiltonian, and in powers thereof stemming from the exponent, the relevant `small' quantity being powers of the reciprocal Planck mass in the couplings $b_n$. 
We give examples in~\cref{diagrams} of some diagrams which might then arise if ultraviolet considerations can somehow be overlooked.

How can this setup be related to energy? We will consider that a \emph{static} source $J(\bm x)$ switches on and off adiabatically within the function $J(x)$, and is essentially present for the comparatively long interval $T$, i.e. 
\begin{equation}
  J(x)\equiv j(t)J(\bm{x}), \quad -T/2< t < T/2\implies j(t)=1.
	\label{totsour}
\end{equation}
For the case of~\eqref{kleingordon}, it is shown in~\cite{Coleman:2011xi} that we can expect the transient imposition of the source in this way to `break' the QFT, in that the vacuum-vacuum scattering acquires a divergent phase as we send ${T\to\infty}$
\begin{equation}
	\langle 0| U_I(-\infty, \infty)|0\rangle=e^{-i(\gamma_-+\gamma_++E_JT)},
	\label{divphas}
\end{equation}
where $\gamma_+$ and $\gamma_-$ are finite shifts restulting from the adiabatic process. The QFT can be fixed by adding a constant counterterm to $H_I$, but one can keep it broken so that the constant $E_J$ is recovered through the angular velocity of the path integral in~\eqref{pintrep}, and equated with the \emph{potential} of $J(\bm x)$.

Considering still the case of~\eqref{kleingordon}, how can this path integral in~\eqref{pintrep} be efficiently evaluated? The elegant solution is presented in~\cite{1995iqft.book.....P}. One can ultralocally shift the field $\phi$ by the solution to its own (sourced) field equation, effectively completing the square in the exponent within $Z[J]$. The $\mathcal{  D}\phi$ integration then cancels over $Z_0$, and the energy $E_J$ from the logarithm is the autocorrelation of the source with the Klein--Gordon propagator as a kernel. It is very natural to consider a $J(\bm x)$ distribution comprising a pair of point sources, as set out for the case of mass-energy sources in~\eqref{manybod}. In that case regularised self-energies may be discarded from the autocorrelation: by retaining only the cross-terms one is left (once the time is integrated over) with the Yukawa (or Coulomb) potential at the source separation distance. By general considerations, this static portion of the Klein--Gordon propagator may also be obtained via the path integral representation of the static two-point function. We assume some version of this line of thinking to have motivated the lattice techniques employed in~\gefc{Deur:2016bwq}.

But is it safe to assume the above analysis holds for the (generalised) GEFC scalar in~\eqref{ansa}, as it does for~\eqref{kleingordon}? To find out, let us attempt to `separate' the path integral defined by~\cref{pathint,timeint,nullint} along lines similar to those used in~\cite{1995iqft.book.....P}. We ask if this is possible, under an ultralocal redefinition
\begin{equation}
	\varphi\to\varphi+\sum_{n=1}^{\infty}\sum_{m=0}^{n-1}c_{nm}\mathcal{J}^{n-m}\varphi^m.
	\label{locre}
\end{equation}
In~\eqref{locre} we introduce the \emph{nonlocal} source 
\begin{equation}
	\mathcal{J}(x)\equiv i\int\mathrm{d}^4y\tensor{D}{_{{F}}}(x-y)J(y),
\end{equation}
where $\feynman(x-y)$ is the (massless) Feynman propagator
\begin{equation}
  \feynman(x-y)\equiv\lim_{\epsilon\to 0}\frac{1}{(2\pi)^4}\int\mathrm{d}^4p\frac{e^{-i\tensor{p}{^\mu}(\tensor{x}{_\mu}-\tensor{y}{_\mu})}}{p^2+i\epsilon},
\end{equation}
which is a Green's function of the d'Alembertian with normalisation ${\dalembertian \feynman(x-y)=-i\delta^4(x-y)}$.
Note that~\eqref{locre} is \emph{induced} by the presence of the source, and connects smoothly to the identity as the source is switched off.

After some lengthy calculations (see supplemental material in~\cite{supp}), it turns out that that this is indeed possible for general $\mathcal{L}_G$, but \emph{only} if we impose some unique restrictions on the matter coupling in $\mathcal{L}_M$, which read at the lowest perturbative orders
\begin{subequations}
\begin{align}
	b_2&=-a_2,\label{res1}\\
	b_3&=-\frac{1}{3}(2a_2^2+3a_3),\\
	b_4&=-\frac{1}{2}(2a_2^3+3a_2a_3-2a_4).\label{res2}
\end{align}
\end{subequations}

This seems at first glance to be quite promising: notwithstanding the concerns raised in~\cref{anmod}, if some gauge choice, spacetime symmetry or perturbation scheme can be found in which~\crefrange{res1}{res2} hold (and in which the relativistic model~\eqref{ansa} is actually faithful to GR), then a transformation of the form~\eqref{locre} exists which would allow us to write
\begin{equation}
  Z[J]=Z_0\exp \left[i\int\mathrm{d}^4x\tilde{\mathcal{L}}_{{M}}\right].
\end{equation}
There would then be some natural functional counterpart $E[J]$ of Coleman's zero-point energy $E_J$ introduced in \eqref{divphas}
\begin{equation}
  \begin{aligned}
	E[J]&\equiv\lim_{T\to\infty}\frac{i}{T}\ln \left(\frac{Z[J]}{J_0}\right)
	\\
	&
	=
	-
	\lim_{T\to\infty}
	\frac{1}{T}\int\mathrm{d}^4x\tilde{\mathcal{L}}_M.
	\label{freeenergy}
  \end{aligned}
\end{equation}
The formula~\eqref{freeenergy} for $E[J]$, given in terms of~\eqref{mainexpansion}, is determined purely by the matter source $J(\bm x)$, making \emph{no reference} to the gravitational field $\varphi$. This would not only constitute a concrete theoretical basis on which to examine the validity of the lattice calculations in~\gefc{Deur:2016bwq}, it would also seem to suggest a new physically motivated programme for localising gravitational energy\footnote{We recall that covariant gravitational energy localisation is impossible~\cite{Barker:2018ilw}.}.

These hopes are quickly dashed. The separated Lagrangian can eventually be recovered after a long calculation
\begin{align}
	\tilde{\mathcal{L}}_{\text{M}}&\equiv\frac{1}{2}J\mathcal{J}
	-\frac{1}{2}\tensor{a}{_2}\mathcal{J}\left(2\tensor{\partial}{_\mu}\mathcal{J}\tensor{\partial}{^\mu}\mathcal{J}+J\mathcal{J}\right)
	\nonumber\\
	 -&\frac{7}{3}\left(\tensor{a}{_2}^2+3\tensor{a}{_3}\right)\mathcal{J}^2\left(3\tensor{\partial}{_\mu}\mathcal{J}\tensor{\partial}{^\mu}\mathcal{J}+J\mathcal{J}\right)
	 \nonumber\\
	 -&\frac{1}{12}\left(140\tensor{a}{_2}^3+105\tensor{a}{_2}\tensor{a}{_3}+18\tensor{a}{_4}\right)\mathcal{J}^3\left(4\tensor{\partial}{_\mu}\mathcal{J}\tensor{\partial}{^\mu}\mathcal{J}+J\mathcal{J}\right)
	\nonumber\\
	&+\cdots.
	 \label{mainexpansion}
\end{align}
The first term in~\eqref{mainexpansion} will tell us that the energy of two static point masses is given by the Coulomb potential, and encodes the linear, Newtonian part of~\eqref{ansa}. All the interesting corrections are pure surface terms, and it is easy to verify in hindsight that the conditions~\crefrange{res1}{res2} are acting to ensure that the theory~\eqref{ansa} is merely a polynomial reparameterisation of~\eqref{kleingordon} in the massless limit. No other solutions are to be found.

It may be interesting to consider more serious attempts at gravitational energy localisation along the lines set out in this appendix, in which the gauge-invariant, tensorial (and low-energy-effective-field-theoretic) nature of GR is properly accounted for. We leave this somewhat daunting task to future work, and concede that, for the time being, the physical meaning of the lattice calculations in~\gefc{Deur:2016bwq} has not been fully resolved.

\section{Tracking the GEFC vs PPN corrections}\label{tracking}

  The expansion of $\mathcal{L}_G$ as it is defined in~\eqref{gravmat} is more challenging than that of $\mathcal{L}_M$, since the Riemann curvature introduces two derivatives whose action on the PPN potentials must be interpreted in powers of the virial velocity. In order to extract, by means of these gradients, the factor of $\rho^*$ in common with~\eqref{matpert}, the quantity $\sqrt{-g}R$ is expanded to $\mathcal{O}\left(\varepsilon^3\right)$, producing in a semi-covariant notation
\begin{align}
    \mathcal{L}_G&=\frac{1}{8\kappa}\Big[
      4\ddddot X
      +2\tensor{\partial}{_\mu}\ddot X\tensor{\partial}{^\mu}\ddot X
      -2\dddot X\dalembertian\dot X
      -\tensor{\partial}{_\mu}\tensor{\partial}{_\nu}\dot X\tensor{\partial}{^\mu}\tensor{\partial}{^\nu}\dot X
      \nonumber\\
      &
      +\left(\dalembertian\dot X\right)^2
	  -8 \left\{\ddot\Phi_2+2\ddot U U+2\dot U^2\right\}
      +8\dalembertian \left\{\Phi_2+U^2\right\}
      \nonumber\\
      &
      +16\ddot U
      +8\dalembertian U
      +32\tensor{\partial}{_\alpha}\tensor{\dot V}{^\alpha}
      +16\ddot U \ddot X
      +16\tensor{\partial}{_\alpha}\tensor{\dot V}{^\alpha}\ddot X
      \nonumber\\
      &
      -16\tensor{\partial}{_\mu}\tensor{\partial}{^\mu}\dot X
      -16\tensor{\ddot V}{_\alpha}\tensor{\partial}{^\alpha}\dot X
      -16\tensor{\partial}{_\mu}\tensor{\partial}{_\alpha}\tensor{V}{^\alpha} U\tensor{\partial}{^\mu}\dot X
      \nonumber\\
      &
      +16\dalembertian\tensor{V}{_\alpha}\tensor{\partial}{^\alpha}\dot X
      +16\dddot X \dot U
      -12\dalembertian \dot X \dot U
      \nonumber\\
      &
      -48  \left\{\dot\Phi_2+2\dot U U\right\} \dot U
      +64\dot U^2
      -4\tensor{\partial}{_\mu}\ddot X\tensor{\partial}{^\mu}U
      \nonumber\\
      &
      +24\tensor{\partial}{_\mu}\left\{\Phi_2+U^2\right\}\tensor{\partial}{^\mu}U
      -40\tensor{\partial}{_\mu}U\tensor{\partial}{^\mu}U
      -16\tensor{\partial}{_\alpha}\ddot X\tensor{\dot V}{^\alpha}
      \nonumber\\
      &
      +32\tensor{\partial}{_\alpha}U\tensor{\dot V}{^\alpha}
      -96\tensor{\dot V}{_\alpha}\tensor{\dot V}{^\alpha}
	+16\dddot X \tensor{\partial}{_\alpha}\tensor{V}{^\alpha}
	  -16\dalembertian \dot X\tensor{\partial}{_\alpha}\tensor{V}{^\alpha}
      \nonumber\\
      &
	  +96\dot U\tensor{\partial}{_\alpha}\tensor{V}{^\alpha}
	  +16\tensor{\partial}{_\mu}\tensor{\partial}{_\alpha}\dot X\tensor{\partial}{^\mu}\tensor{V}{^\alpha}
      -96\tensor{\partial}{_\mu}\tensor{V}{_\alpha}\tensor{\partial}{^\mu}\tensor{V}{^\alpha}
      \nonumber\\
      &
      +32\tensor{\partial}{_\beta}\tensor{V}{_\alpha}\tensor{\partial}{^\alpha}\tensor{V}{^\beta}
	  +64\left(\tensor{\partial}{_\alpha}\tensor{V}{^\alpha}\right)^2
      -48\ddot U \left\{\Phi_2+U^2\right\}
      \nonumber\\
      &
      +24\dalembertian U \left\{\Phi_2+U^2\right\}
      +8\ddddot X U
      \nonumber\\
      &
      -16 \left\{\ddot\Phi_2+2\ddot U U+2\dot U^2\right\} U
      +16\dalembertian \left\{\Phi_2+U^2\right\} U
      \nonumber\\
      &
      +96\ddot U U
      -48 \dalembertian U U
      +64\tensor{\partial}{_\alpha}\tensor{\dot V}{^\alpha}U
      +32\dot U^2 U
      \nonumber\\
      &
      +64\tensor{\partial}{_\mu}U\tensor{\partial}{^\mu}U U
      +32 \ddot U U^2
      +16\dalembertian U U^2
      \nonumber\\
      &
      +4\dalembertian \ddot X\left(1+2U\right)
      +128\tensor{\partial}{_\alpha}\dot U\tensor{V}{^\alpha}
      +128\tensor{\ddot V}{_\alpha}\tensor{V}{^\alpha}
      \nonumber\\
      &
      +128\tensor{\partial}{_\alpha}\tensor{\partial}{_\beta}\tensor{V}{^\beta}\tensor{V}{^\alpha}
      -128\dalembertian\tensor{V}{_\alpha}\tensor{V}{^\alpha}
    \Big]
+\mathcal{O}\left(\varepsilon^4\right)
    .
    \label{smallexpr}
  \end{align}
  Once again, we keep track of the $\mathcal{O}\left(\varepsilon^2\right)$ correction to $\tensor{h}{_{00}}$ via braces.
  The expansion~\eqref{smallexpr} can be simplified by a careful addition of surface terms, with the help of the identities
  \begin{subequations}
\begin{gather}
	\dot U\equiv-\vect{\nabla}\cdot\bm{V}, \quad
	\vect{\nabla}^2X\equiv-2U, \label{superpot}\\
  \vect{\nabla}^2 U\equiv -\frac{\kappa}{2}\rho^*, \quad
  \vect{\nabla}^2 \bm{V}\equiv -\frac{\kappa}{2}\rho^*\bm{v},
  \label{ident}
\end{gather}
  \end{subequations}
  which can be obtained from~\eqref{ppnpot}. Here,~\eqref{superpot} encode the conserved matter current and superpotential while~\eqref{ident} recover the matter and momentum sources.
  Suitable use of~\cref{superpot,ident} then yields the far simpler form~\eqref{ehpert}, and to accelerate this manipulation (see supplemental material in~\cite{supp}) we make use of the \texttt{xAct}, \texttt{xPert} and \texttt{xTras} tensor manipulation, perturbation and field theory software~\cite{Brizuela:2008ra,Nutma:2013zea}. Note that \texttt{xAct} already has an advanced PPN implementation in \texttt{xPPN}~\cite{Hohmann:2020muq}, but we happen not to use these tools in the current analysis.

\section{Potential of a uniform density sphere}\label{axisph}

Following on from our discussion in~\cref{axisy}, we now calculate the potential of a uniform density sphere. The purpose of this calculation will be to confirm --- for a case where the exact solution to the EFEs in~\eqref{efe} is known --- that the approximaion in~\eqref{eqn:2nd-order-pot} is actually \emph{corrective}, i.e. reducing the error in the $\ppn{1}$ result.
By `potential' here, we mean that we are going to calculate the $\ppn{2}$ correction to the $a_1$ metric coefficient for a uniform sphere. Since in the axisymmetric work we have already adopted an ansatz at $\ppn{1}$ which has a spatial metric which is conformally equivalent to flat space (although the conformal coefficient is generally a general function of both $R$ and $z$), this means that the metric it is most convenient to work with here in the spherical case should be the one appropriate to {\em isotropic} space. We can write this in the exact case as
\be
\begin{aligned}
	\tensor{g}{^{00}}  = \left(1+a_1\right)^2,  \quad
	\tensor{g}{^{11}}  = \left(1+b_1\right)^2,  \\
	\tensor{g}{^{22}}  = \left(1+b_1\right)^2,  \quad
	\tensor{g}{^{33}}  = \left(1+b_1\right)^2,  \\
\end{aligned}
\label{eqn:iso-spher-h-function}
\ee
where $a_1$ and $b_1$ are functions of the spherical coordinate $r$.

So the problem we are dealing with in exact terms is finding the Schwarzschild interior and exterior solutions written in isotropic spatial coordinates. This is discussed in e.g.\ Section III.E of~\cite{Barker:2018ilw} and was first solved by Wyman in 1946~\cite{wyman1946schwarzschild}. Since the solutions are not well known, we give the results, transferred to our metric coefficients, as
\begin{subequations}
	\begin{align}
a_1 & =\frac{2 \mtot \left(3 a^{3}-r^{2} a+\mtot r^{2}\right)}{4 a^{4}-4 \mtot a^{3}+4 \mtot r^{2} a-\mtot^{2} r^{2}}, 
\\
b_1 & = -\frac{\mtot\left(12 a^{2}+6 a \mtot+\mtot^{2}-4 r^{2}\right)}{(2 a+\mtot)^{3}}.
	\end{align}
\end{subequations}
Here $\mtot$ is the `gravitational mass' of the object, which has radius $a$, and these are related to the (constant) density $\rho$ via
\be
\rho=\rho_0=\frac{48 \mtot a^3}{\pi \left( 2 a + \mtot \right)^6}.
\label{eqn:rho0-from-ma}
\ee
(See~\cite{Barker:2018ilw} for a discussion of the different types of mass definition possible in this context.) The other quantity of interest is the pressure, $P$, which is not given by Wyman, but has the expression
\be
\begin{aligned}
&P=\\
&\frac{96 a^{4} \mtot^{2}(a-r)(a+r)}{\pi(2 a+\mtot)^{6}\left(4 a^{4}-4 \mtot a^{3}+4 \mtot r^{2} a-\mtot^{2} r^{2}\right)}.
\end{aligned}
\ee
We can see this vanishes at the edge of the object, as we would expect. 

So these quantities define the exact solution, and we now wish to see if our expression \eqref{eqn:2nd-order-pot} works correctly in terms of moving us towards the exact solution, if we start from a $\ppn{1}$ GEM type solution. Note we will only be applying this to the $a_1$ coefficient. In the spherical case, it is possible to do the same for the $b_1$ function which controls the spatially isotropic part of the metric, but while we are able to have a similar setup to this at $\ppn{1}$ in the axisymmetric case, this does not persist to $\ppn{2}$ since one soon finds that if the pressure is non-zero, then $b_1$ has to be different to $c_1$, and a spatially isotropic metric is not possible.

The order in which we carry out solution is that we first solve \eqref{eqn:a1f-poisson} for $a_1^f$ given the constant $\rho$, and then use this to get the pressure $P$ at $\ppn{2}$ from \eqref{eqn:P-so}. These are then substituted in \eqref{eqn:2nd-order-pot}, enabling us to get $a_1^s$. The only boundary conditions that are perhaps non-obvious are that the derivatives of $a_1^f$ and $a_1^s$ are taken to match individually either side of the boundary of the object, at $r=a$. This enables us to say that the boundary conditions, in which the total $a_1$ certainly has to have matching derivatives either side of $r=a$, to match the `singularity' in $\rho$, which is in the form of a step, apply to each order in the expansion.

Carrying out this process, the solutions we find are
\begin{subequations}
	\begin{align}
		a_1^f &= \begin{cases}
{ \frac{2 \pi \rho_0\left(-r^{2}+3 a^{2}\right)}{3}} & r<a, \\
{ \frac{4\pi \rho_0 a^3}{3 r}} & r>a,
\end{cases}\\
		P &= \begin{cases}
			{ \frac{2 \pi \rho_0^{2}(a-r)(a+r)}{3}+\ppn{3}} & r<a, \\
{ 0+\ppn{3}} & r>a,
\end{cases}\\
		a_1^s &= \begin{cases}
{ \frac{8 \pi^{2} \rho_0^{2}\left(r^{4}-6 a^{2} r^{2}+12 a^{4}\right)}{9}
			 +\ppn{3}} & r<a, \\
{ \frac{8 \pi^{2} \rho_0^{2} a^{5}(a+6 r)}{9 r^{2}}+\ppn{3}} & r>a.
\end{cases}
	\end{align}
\end{subequations}
The results for $a_1^f$ and $P$ are the expected Newtonian values, and we now want to see if the addition of the $a_1^s$ corrections to the $a_1^f$ is successful in bringing them closer to the exact results. We can do this in two ways. First by plotting the results and making a visual comparison, and secondly, and more accurately, by expanding to $\ppn{2}$ and checking that this successfully brings the $a_1^f+a_1^s$ and exact $a_1$ results into alignment.

\begin{figure}
\begin{center}
\includegraphics[width=\linewidth]{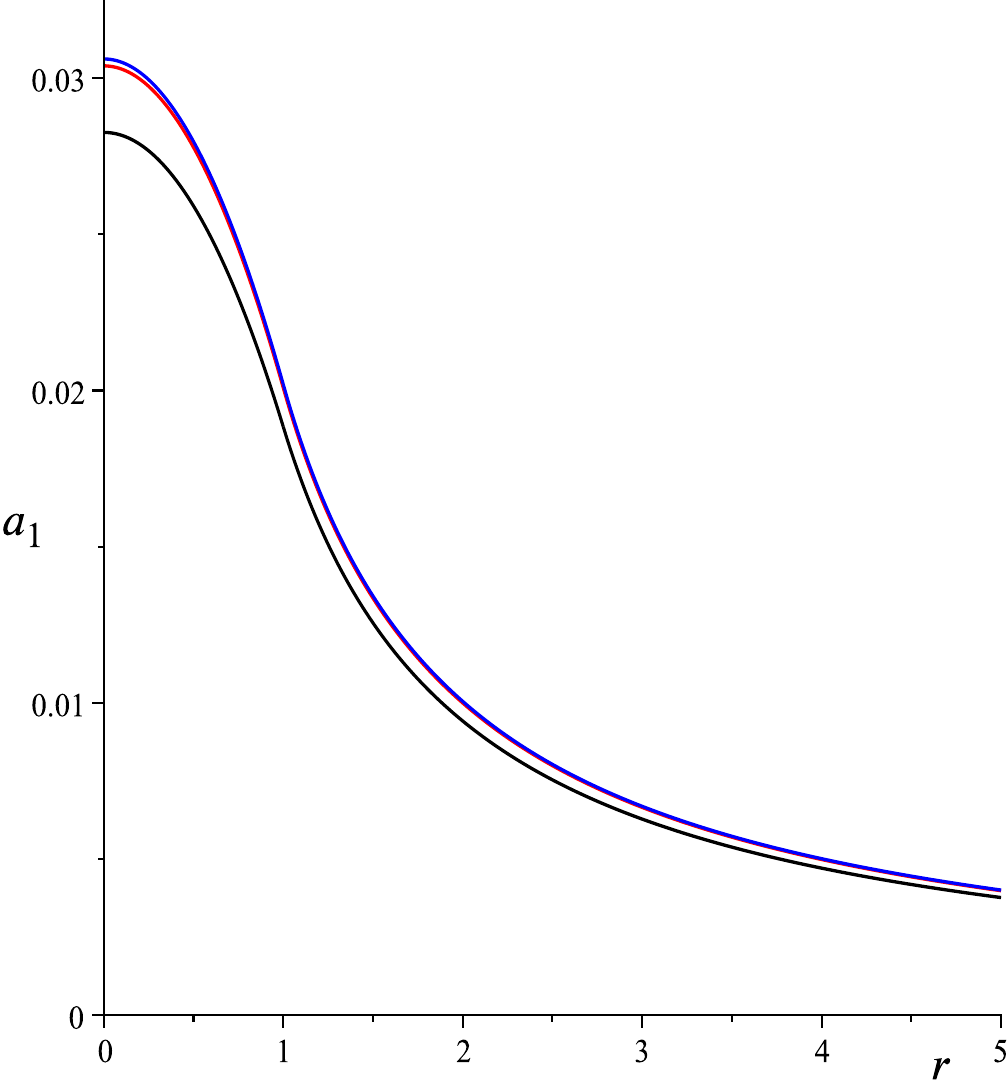}
	\caption{Comparison of the $\ppn{1}$, $\ppn{2}$ and exact results for the metric coefficient $a_1$ (which is minus the potential at $\ppn{1}$), for a spherically symmetric case with $m=1/50$, and where the boundary of the object is at $r=a=1$. Black shows $\ppn{1}$, red shows the combination of $\ppn{1}$ and $\ppn{2}$, and blue is exact. \label{fig:a1-comp}}
\end{center}
\end{figure}

In
\cref{fig:a1-comp}
we show a comparison of the $\ppn{1}$, $\ppn{2}$ and exact results for a case with $\mtot=1/50$ and $a=1$. (Note we do not need to specify $\rho_0$ as well, since this is given by \eqref{eqn:rho0-from-ma}.) It can be seen that the $\ppn{2}$ results are definitely different from the $\ppn{1}$ result, but virtually indistinguishable from the exact result over nearly the entire range plotted, so the method appears to be working.

To carry out a series comparison, we need to decide on the appropriate variable to expand in. Since $r$ can in principle be small, the variable we should work with is $\mtot /a$, which we will denote $x$. Similarly we will denote $r/a$ by $y$, since this simplifies the expressions, but we are not expanding in this. We find the following:
\begin{widetext}
\begin{subequations}
	\begin{align}
a_1^f &= \begin{cases}
{ \left(-\frac{y^{2}}{2}+\frac{3}{2}\right) x 
	 +\left(\frac{3 y^{2}}{2}-\frac{9}{2}\right) x^{2}} + \ldots & r<a, \\
{ \frac{x}{y}-\frac{3 x^{2}}{y}} + \ldots & r>a,
\end{cases}\\
a_1^f +a_1^s &= \begin{cases}
{ \left(-\frac{y^{2}}{2}+\frac{3}{2}\right) x
	 +\Big(-\frac{3}{2} y^{2}
	 +\frac{3}{2}+\frac{1}{2} y^{4}\Big) x^{2}} + \ldots & r<a, \\
{ \frac{x}{y}+\frac{x^{2}}{2 y^{2}}} + \ldots & r>a,
	\label{axisphr1}
\end{cases}\\
a_1^{\rm exact} &= \begin{cases}
{ \left(-\frac{y^{2}}{2}+\frac{3}{2}\right) x
	 +\left(-\frac{3}{2} y^{2}+\frac{3}{2}+\frac{1}{2} y^{4}\right) x^{2}} + \ldots & r<a, \\
{ \frac{x}{y}+\frac{x^{2}}{2 y^{2}}} + \ldots & r>a.
	\label{axisphr2}
\end{cases}
	\end{align}
\end{subequations}
\end{widetext}
We see that indeed the exact and $\ppn{2}$ results agree at $\ppn{2}$, so the method has been successful. Of course, these particular results are only applicable to the spherically symmetric case, and this is the case in which~\gefc{Deur:2020wlg} would say there is no effect of the kind desired, since there is no flattening. 

Turning now from the expression
\eqref{eqn:2nd-order-pot} and instead considering the
radiative GEM
  proxy in~\eqref{electromaster}, as a first example let us try something very similar to the above analysis by calculating the $\ppn{2}$ gravitational
potential of a sphere of uniform density $\tensor*{\rho}{^\dagger_0}$ and radius $a$. For any
spherically symmetric system, the integral solution in the first equation
in (\ref{ppnpot}) may be written in the simple form
\be
\begin{aligned}
  \Phi(r) = -\frac{\kappa}{2}\bigg(&\frac{1}{r}\int_0^r \tensor*{\rho}{^\dagger_0}(r')r^{\prime 2}\,\mathrm{d}r' 
\\
&\ \ \
  +
\int_r^\infty \tensor*{\rho}{^\dagger_0}(r')r'\,\mathrm{d}r'\bigg).
\label{phisolsphere}
\end{aligned}
\ee
By analogy, one may write (\ref{deltaphisol}) for any
spherically-symmetric system as
\be 
\begin{aligned}
\delta\Phi(r) = \frac{9\kappa}{4}\bigg(&\frac{1}{r}\int_0^r
  \Phi(r')\tensor*{\rho}{^\dagger_0}(r')r^{\prime 2}\,\mathrm{d}r' 
\\
&\ \ \
+ \int_r^\infty
\Phi(r')\tensor*{\rho}{^\dagger_0}(r')r'\,\mathrm{d}r'\bigg)+\mathcal{  O}\left(\varepsilon^3\right).
\label{deltaphisolsphere}
\end{aligned}
\ee
Evaluating the expression (\ref{phisolsphere}) for a sphere of uniform
density $\tensor*{\rho}{^\dagger_0}$ and radius $a$, with `mass' $M^\dagger\equiv\frac{4}{3}\pi\tensor*{\rho}{^\dagger_0} a^3$, and substituting into
(\ref{deltaphisolsphere}) yields
\begin{widetext}
  \begin{subequations}
  \begin{align}
\Phi(r) &=
\left\{
\begin{array}{llll}
-\frac{\kappa}{12}\tensor*{\rho}{^\dagger_0}(3a^2-r^2) & = & -GM^\dagger(3a^2-r^2)/2a^3 &\mbox{for $r \le a$}, \\[1mm]
-\frac{\kappa}{6}\tensor*{\rho}{^\dagger_0} a^3/r & = & -GM^\dagger/r & \mbox{for $r > a$},
\end{array}
\right.
\\
\delta\Phi(r) &= 
\left\{
\begin{array}{llll}
3\left(\frac{\kappa\tensor*{\rho}{^\dagger_0}}{4}\right)^2
\left(\frac{5}{4}a^4-\frac{1}{2}a^2r^2+\frac{1}{20}r^4\right)+\mathcal{  O}\left(\varepsilon^3\right) &
 = & 3\left(\frac{3
  GM^\dagger}{2a^3}\right)^2
\left(\frac{5}{4}a^4-\frac{1}{2}a^2r^2+\frac{1}{20}r^4\right)+\mathcal{  O}\left(\varepsilon^3\right)
& \mbox{for $r \le a$}, \\ 
-\frac{12}{5}\left(\frac{\kappa\tensor*{\rho}{^\dagger_0}}{4}\right)^2a^5/r+\mathcal{  O}\left(\varepsilon^3\right) & = & -\frac{12}{5}\left(\frac{3
  GM^\dagger}{2a^3}\right)^2a^5/r+\mathcal{  O}\left(\varepsilon^3\right) & \mbox{for $r > a$}.
\end{array}
\right.
\end{align}
  \end{subequations}
\end{widetext}
In particular, for $r > a$, one may write the gravitational potential
up to $\ppn{2}$ as
\be
\Phi(r) + \delta\Phi(r) = -\frac{GM^\dagger}{r}\left(1 + \frac{27}{5}\frac{GM^\dagger}{a}\right)+\mathcal{  O}\left(\varepsilon^3\right).
\label{cerdis}
\ee
As discussed in~\cref{secgem}, a consequence of the radiative average is that we do not expect the correction in~\eqref{cerdis} to be strictly faithful to the exact result which would follow from~\cref{axisphr1,axisphr2}, even if the difference between $\tensor*{\rho}{^\dagger_0}$ and $\tensor*{\rho}{_0}$ is factored in: it is, however, straightforward to calculate, and perfectly comparable in magnitude.

\section{The `bipartite' harmonic gauge}\label{bipartite}

To understand why the nonlinear gravitoelectric correction obtained in~\cref{secgem} differs from the $\mathcal{  O}\left(\varepsilon^2\right)$ PPN correction, we must understand how the gauge choice which is developed over~\cref{oldgauge,newgauge} departs from the exact harmonic gauge
\begin{equation}
  \tensor{\partial}{_\mu}\tensor{\mathfrak{g}}{^{\mu\nu}}= 0, 
  \quad
  \tensor{\mathfrak{g}}{^{\mu\nu}}\equiv \sqrt{-g}\tensor{g}{^{\mu\nu}},
  \label{harmonicgauge}
\end{equation}
at the same PN order. 
From the outset, it is clear from~\cref{oldgauge} that all gauges under consideration take $\tensor{h}{_{\mu\nu}}=\mathcal{  O}\left( \varepsilon \right)$, and so the nonlinear harmonic gauge from~\eqref{harmonicgauge} demands
\begin{equation}
  \begin{aligned}
  -\tensor{\partial}{_{\mu}}\tensor{\bar h}{^{\mu\nu}}&=
  \frac{1}{2}\tensor{\partial}{_\mu}\left(\bar h\tensor{\bar h}{^{\mu\nu}}\right)
  +\frac{1}{2}\tensor{\bar h}{^{\mu\sigma}}\tensor{\partial}{^\nu}\tensor{\bar h}{_{\mu\sigma}}
  -\frac{1}{4}\bar h \tensor{\partial}{^\nu}\bar h
  \\
  &\ \ \ 
  -\tensor{\partial}{_\mu}\left( \tensor{\bar h}{^{\nu\sigma}}\tensor*{\bar h}{^\mu_\sigma} \right)
  +\mathcal{  O}\left( \varepsilon^3 \right),
  \label{harmonicgaugenonlinear}
  \end{aligned}
\end{equation}
which for linear gravity becomes $\tensor{\partial}{_{\mu}}\tensor{\bar h}{^{\mu\nu}}=\mathcal{  O}\left( \varepsilon^2 \right)$. Now we recall from~\cref{newgauge} that ${\tensor{h}{_{\mu\nu}}\equiv\tensor{\ell}{_{\mu\nu}}+\tensor{\delta h}{_{\mu\nu}}}$ is exact, where ${\tensor{\ell}{_{\mu\nu}}=\mathcal{  O}\left( \varepsilon \right)}$ and ${\tensor{\delta h}{_{\mu\nu}}=\mathcal{  O}\left( \varepsilon^2 \right)}$. The condition ${\tensor{\partial}{_{\mu}}\tensor{\bar \ell}{^{\mu\nu}}= 0}$ need not be exact, but it can only be relaxed to ${\tensor{\partial}{_{\mu}}\tensor{\bar \ell}{^{\mu\nu}}= \mathcal{  O}\left( \varepsilon^3 \right)}$ if~\eqref{master} is still to hold. Similarly, we may relax the second part of this `bipartite' harmonic gauge to ${\tensor{\partial}{_{\mu}}\tensor{\delta\bar h}{^{\mu\nu}}= \mathcal{  O}\left( \varepsilon^3 \right)}$ while still keeping~\eqref{masterfinal}, but even this relaxation violates~\eqref{harmonicgaugenonlinear} which has by now become
\begin{equation}
  -\tensor{\partial}{_{\mu}}\tensor{\delta\bar h}{^{\mu\nu}}=
  \frac{1}{2}\tensor{\ell}{^{\mu\sigma}}\tensor{\partial}{^\nu}\tensor{\ell}{_{\mu\sigma}}
  -\tensor*{\ell}{^\mu_\sigma}\tensor{\partial}{_\mu}\tensor{\ell}{^{\nu\sigma}}
  +\mathcal{  O}\left( \varepsilon^3 \right).
  \label{departure}
\end{equation}
We thus identify in~\eqref{departure} the $\mathcal{  O}\left( \varepsilon^2 \right)$ departure from the harmonic gauge.

If~\eqref{harmonicgauge} holds, it is well known (see e.g.~\cite{landau}) that the EFEs in~\eqref{efe} adopt the form
\begin{equation}
  -\dalembertian \tensor{\mathfrak{g}}{^{\mu\nu}}= 2\kappa g\left( \tensor{T}{^{\mu\nu}}+\tensor*{t}{^{\mu\nu}}+\tensor*{\tau}{^{\mu\nu}} \right),
  \label{harmonicefe}
\end{equation}
where we define a pseudotensor
\begin{equation}
-2\kappa g\tensor*{\tau}{^{\mu\nu}}\equiv\tensor{\partial}{_\rho}\tensor{\mathfrak{g}}{^{\mu\sigma}}\tensor{\partial}{_\sigma}\tensor{\mathfrak{g}}{^{\mu\rho}}+\left(\tensor{\mathfrak{g}}{^{\rho\sigma}}\tensor{\partial}{_\rho}\tensor{\partial}{_\sigma}-\dalembertian\right)\tensor{\mathfrak{g}}{^{\mu\nu}},
\end{equation}
and $\tensor{t}{^{\mu\nu}}$ is the Landau--Lifshitz pseudotensor whose formula is given elsewhere. Now it is easy to verify that by substituting $h_{\mu\nu} = \ell_{\mu\nu} + \delta h_{\mu\nu}$ into~\eqref{harmonicefe} and imposing the linear harmonic gauge that the $\tensor{\ell}{_{\mu\nu}}$ solutions are as found previously to $\mathcal{O}\left(\varepsilon^{3/2}\right)$. 
In fact we can make things more precise by splitting the stress-energy tensor into $\mathcal{O}\left(\varepsilon\right)$ and $\mathcal{O}\left(\varepsilon^2\right)$ parts
\begin{equation}
	\begin{gathered}
	\tensor{T}{^{\mu\nu}}\equiv\tensor{\mathcal{T}}{^{\mu\nu}}+\tensor{\delta T}{^{\mu\nu}}, \quad 
	\tensor{\mathcal{T}}{^{\mu\nu}}\equiv \rho^*\tensor{\bar u}{^\mu}\tensor{\bar u}{^\nu},
	\\
	\tensor{\delta T}{^{\mu\nu}}\equiv -\rho^*U\tensor{\bar u}{^\mu}\tensor{\bar u}{^\nu}+\mathcal{O}\left(\varepsilon^3\right),
		\label{splitting}
	\end{gathered}
\end{equation}
so that $\tensor{\ell}{_{00}}=-2U$, etc., and this approach is similar in spirit to our earlier taking $\tensor{h}{_{\mu\nu}}\equiv\tensor{\ell}{_{\mu\nu}}+\tensor{\delta h}{_{\mu\nu}}$.
If we now try to solve for $\tensor{\delta h}{_{\mu\nu}}$ we will be obliged to impose~\eqref{departure}, and after some lengthy calculations we find that~\eqref{harmonicefe} can then be rearranged to yield
\begin{equation}
	\begin{aligned}
		\dalembertian & \tensor{\delta\bar h}{^{\mu\nu}}=
		\kappa \ell \tensor{{\mathcal{T}}}{^{\mu\nu}}+\frac{1}{2}\kappa\tensor{{\delta{T}}}{^{\mu\nu}}
		+\frac{1}{4}\Big[
			-2\tensor{\eta}{^{\mu\nu}}\tensor{\ell}{_{\rho\sigma}}\tensor{\partial}{^{\rho\sigma}}\ell
			\\
			&
			-2\Big(
			\tensor{\partial}{^\mu}\tensor{\ell}{^{\rho\sigma}}
			\tensor{\partial}{^\nu}\tensor{\ell}{_{\rho\sigma}}
			+2
			\ell	
			\dalembertian\tensor{\ell}{^{\mu\nu}}
			+
			\tensor{\ell}{^{\mu\nu}}
			\dalembertian
			\ell	
			\\
			&
			-
			\tensor{\eta}{^{\mu\nu}}
			\ell
			\dalembertian
			\ell	
			-2\Big(
			\tensor{\ell}{^{\rho\sigma}}
			\tensor{\partial}{_\rho}\tensor{\partial}{_\sigma}\tensor{\ell}{^{\mu\nu}}
			+
			2\tensor{\ell}{^{(\nu|\rho}}
			\dalembertian\tensor*{\ell}{^{|\mu)}_\rho}
			\\
			&
			+
			\Big(
			\tensor{\partial}{^\nu}\tensor{\ell}{^{\rho\sigma}}
			-2\tensor{\partial}{_{[\rho|}}\tensor*{\ell}{^\nu_{|\sigma]}}
			\Big)
			\tensor{\partial}{^\sigma}\tensor{\ell}{^{\mu\rho}}
			+
			\tensor{\partial}{^\mu}\tensor{\ell}{_{\rho\sigma}}
			\tensor{\partial}{^\sigma}\tensor{\ell}{^{\nu\rho}}
			\Big)
			\\
			&
			+
			\tensor{\eta}{^{\mu\nu}}
			\tensor{\ell}{^{\rho\sigma}}
			\dalembertian\tensor{\ell}{_{\rho\sigma}}
			\Big)
			-
			\tensor{\eta}{^{\mu\nu}}
			\Big(
			2\tensor{\partial}{_\sigma}\tensor{\ell}{_{\rho\lambda}}
			+\tensor{\partial}{_{\lambda}}\tensor{\ell}{_{\sigma\rho}}
			\Big)
			\tensor{\partial}{^\lambda}\tensor{\ell}{^{\sigma\rho}}
			\Big]
		\\
		&+\mathcal{O}\left(\varepsilon^3\right).
		\label{harmonicefelinear}
	\end{aligned}
\end{equation}
We initially sought to recover the $\mathcal{O}\left(\varepsilon^2\right)$ correction to $\tensor{h}{_{00}}$ in~\eqref{ppn_pert}, and by substituting the various quantities into the RHS of~\eqref{harmonicefelinear} we obtain after some work
\begin{equation}
	\begin{aligned}
		\dalembertian\tensor{\delta h}{_{00}}&\equiv \dalembertian\left(\tensor{\delta\bar h}{_{00}}-\frac{1}{2}\delta\bar h\right)
\\
		&
		=
	-3\kappa\rho^* U-12U{\vect\nabla}^2U-4|\vect{\nabla}U|^2+\mathcal{O}\left(\varepsilon^3\right).
		\label{finallandau}
	\end{aligned}
\end{equation}
One can then proceed immediately from~\eqref{finallandau} to~\eqref{ppn_pert} by means of~\eqref{ident} and its corollary 
\begin{equation}
	\int\frac{|\vect{\nabla}U|^2\mathrm{d}^3x'}{|\bm{x}-\bm{x}'|}\equiv
	4\pi\Phi_2-2\pi U^2.
	\label{coroll}
\end{equation}

By contrast to~\eqref{finallandau}, we obtain from~\eqref{masterfinal} that
\begin{equation}
		\dalembertian\tensor{\delta h}{_{00}}		=
	-3\kappa\rho^* U-2|\vect{\nabla}U|^2+\mathcal{O}\left(\varepsilon^3\right).
		\label{finalmike}
\end{equation}
In computing the overall $\mathcal{O}\left(\varepsilon^2\right)$ correction implied by the methods of~\cref{newgauge} we must recall that the relevant $\tensor{\ell}{_{\mu\nu}}$ solutions \emph{also} differ from those of~\eqref{harmonicefe} by $\mathcal{O}\left(\varepsilon^2\right)$, since $\Phi=-U+\mathcal{O}\left(\varepsilon^2\right)$ in~\eqref{GEMtoPPN}. Even taking this into account, we still find from~\eqref{finalmike} that
\begin{equation}
	\tensor{h}{_{00}}=-2U-6\Phi_2+U^2+\mathcal{O}\left(\varepsilon^3\right),
	\label{finalmikeclear}
\end{equation}
which is in contradiction with~\eqref{ppn_pert}.

How to understand the discrepancy between~\eqref{finalmikeclear} and~\eqref{ppn_pert}? 
This could be a feature of (i) the bipartite gauge, or (ii) the use of the radiative formula, or (iii) some admixture of the two. To investigate, we repeat the whole analysis of~\cref{newgauge} while correcting both aspects. Rather than defining $\tensor{\ell}{_{\mu\nu}}$ via GEM potentials as in~\cref{newgauge}, we stick to the PPN potentials and corresponding definition of the stress-energy tensor from the outset --- thus we have exactly $\tensor{\ell}{_{00}}=-2U$, etc. After some tedious calculations, we obtain in place of~\eqref{masterfinal}
  \begin{align}
	  \dalembertian &\tensor{\delta\bar h}{_{\mu\nu}}=
  \frac{1}{2}\Big[
    2\tensor{\partial}{^\rho}\tensor*{\ell}{_\nu^\sigma}\tensor{\partial}{_\mu}\tensor{\ell}{_{\rho\sigma}}
    +2\left( 
      \tensor{\partial}{^\rho}\tensor*{\ell}{_\mu^\sigma}
      -\tensor{\partial}{_\mu}\tensor{\ell}{^{\rho\sigma}}
    \right)
      \tensor{\partial}{_\nu}\tensor{\ell}{_{\rho\sigma}}
      \nonumber\\
      &
      +\tensor{\ell}{^{\rho\sigma}}
      \left(
	4\tensor{\partial}{_{(\mu|}}\tensor{\partial}{_\sigma}\tensor{\ell}{_{|\nu)\rho}}
	-2\tensor{\partial}{_{\mu}}\tensor{\partial}{_\nu}\tensor{\ell}{_{\rho\sigma}}
      \right)
      +\tensor{\eta}{_{\mu\nu}}
      \Big( 
	\tensor{\ell}{^{\rho\sigma}}\dalembertian \tensor{\ell}{_{\rho\sigma}}
      \nonumber\\
      &
	-\tensor{\ell}{_{\rho\sigma}}\tensor{\partial}{^\rho}\tensor{\partial}{^\sigma}\ell
	+\left( \tensor{\partial}{_\lambda}\tensor{\ell}{_{\rho\sigma}}-2\tensor{\partial}{_\sigma}\tensor{\ell}{_{\rho\lambda}} \right)
      \Big)
      \tensor{\partial}{^\lambda}\tensor{\ell}{^{\rho\sigma}}
  \Big]
      \nonumber\\
      &
      -8\kappa\tensor{\ell}{_{(\nu|\sigma}}\tensor*{\mathcal T}{^\sigma_{|\mu)}}
      -2\kappa\tensor{\delta T}{_{\mu\nu}}+2\tensor*{G}{^{(2)}_{\mu\nu}}\left( \ell \right)
		+\mathcal{O}\left(\varepsilon^3\right).
      \label{correctexp}
\end{align}
In~\eqref{correctexp}, the initial collection of terms in square brackets shifts the nonlinear harmonic guage in~\eqref{masterfinal} from \emph{bipartite} to \emph{exact}: it stems from the action of~\eqref{departure} in $\tensor*{G}{^{(1)}_{\mu\nu}}\left( \delta h \right)$. The next two terms account for index-lowering and potential corrections to the fluid energy density, and ultimately occur because $\tensor{\ell}{_{\mu\nu}}$ is redefined to $\mathcal{O}\left(\varepsilon^2\right)$. The final term in $\tensor*{G}{^{(2)}_{\mu\nu}}\left( \ell \right)$ must be evaluated strictly according to~\cref{ch17:eqn17.61,ch17:eqn17.59} (in which it is safe to replace $\tensor{T}{^{\mu\nu}}$ with $\tensor{\mathcal T}{^{\mu\nu}}$), rather than via the radiative formula in~\eqref{ch17:eqn17.64}.
Of these three corrections to~\eqref{masterfinal}, we find that only the latter two are relevant to gravitostatics, while the gauge correction \emph{cancels internally} as may be verified by direct calculation. After substituting the various quantities on the RHS of~\eqref{correctexp} and reversing the trace, we obtain
\begin{equation}
		\dalembertian\tensor{\delta h}{_{00}}=	5\kappa\rho^* U+4U{\vect\nabla}^2U-4|\vect{\nabla}U|^2+\mathcal{O}\left(\varepsilon^3\right),
		\label{finalcorrected}
\end{equation}
and we immediately see that~\eqref{finalcorrected}, as with~\eqref{finallandau}, is consistent with~\eqref{ppn_pert} as required.

\end{document}